\documentclass[11pt]{article}
\usepackage{jheppub}
\usepackage{bm,bbm}
\usepackage{booktabs}
\usepackage{accents}
\usepackage{multirow}
\usepackage{graphics}
\usepackage{braket}
\usepackage{mathtools}
\usepackage{comment}
\usepackage{subcaption}
\usepackage{float}
\usepackage{xcolor}
\usepackage{tikz}

\definecolor{deepblue}{RGB}{25,70,130}
\definecolor{deepred}{RGB}{140,40,40}
\definecolor{deepgreen}{RGB}{35,115,60}
\definecolor{softgray}{RGB}{245,245,245}

\usepackage{enumerate}
\usepackage{blkarray}
\usepackage{longtable}
\usepackage{enumitem}
\usepackage{soul}
\usepackage{ascmac}
\usepackage{mathrsfs}
\usepackage{graphicx}
\usepackage{booktabs}
\usepackage{array}
\usepackage{hyperref}
\usepackage{physics}
\usepackage{xcolor}
\usepackage{listings}
\usepackage{caption}
\usepackage{placeins}
\usepackage{microtype}
\usepackage{graphicx}
\usepackage{float}
\usepackage{comment}
\usepackage{booktabs}
\usepackage{tabularx}
\usepackage{array}
\usepackage{amsmath,amssymb,amsfonts,mathtools,bm}
\usepackage{arydshln}
\usepackage{mathtools}
\usepackage{tikz}

\usetikzlibrary{patterns}
\usetikzlibrary{arrows,shapes,positioning}
\usetikzlibrary{decorations.markings}
\usetikzlibrary{decorations.pathmorphing}
\tikzset{snake it/.style={decorate, decoration=snake}}

\hypersetup{
    colorlinks=true,
    linkcolor=blue,
    filecolor=magenta,      
    urlcolor=cyan,
    pdfpagemode=FullScreen,
}

\edef\restoreparindent{\parindent=\the\parindent\relax}
\usepackage{parskip}
\restoreparindent

\hypersetup{colorlinks=true,linkcolor=blue,citecolor=blue,urlcolor=blue}
\graphicspath{{./}}

\renewcommand{\dd}{\mathrm{d}}
\newcommand{\ii}{\mathrm{i}}

\newcommand{\AdS}{\mathrm{AdS}}

\newcommand{\BTZ}{\mathrm{BTZ}}

\newcommand{\Tcal}{\mathcal{T}}

\newcommand{\Om}{\Omega}
\newcommand{\Del}{\Delta}
\newcommand{\kap}{\kappa}

\newcommand{\calC}{\mathcal{C}}

\newcommand{\calA}{\mathcal{A}}
\newcommand{\calE}{\mathcal{E}}

\renewcommand{\arraystretch}{1.25}

\usepackage{amsthm}
\newtheoremstyle{break}
  {\topsep}{\topsep}%
  {\upshape}{}%
  {\bfseries}{}%
  {\newline}{}%
\theoremstyle{break}

\def\Tr{{\rm Tr}}
\def\d{{\rm d}}

\def\d{\mathrm{d}}

\title{A Holographic Open Quantum System}

\preprint{UT-WI-31-2026}

\author[a,b]{Ignacio Quiroz Vargas} 
\author[a]{Merna Youssef}

\affiliation[a]{Physics Department, University of Texas, Austin,\\
Austin TX 78712, USA}

\affiliation[b]{Department of Physics and Astronomy, Rice University,\\
Houston, TX 77005, USA}

\emailAdd{iaq1@rice.edu, myoussef@utexas.edu}

\abstract{We construct a holographic open quantum system by coupling two CFTs through a double trace deformation and tracing out the bath degrees of freedom within the Schwinger–Keldysh formalism. This procedure yields an effective description of the nonunitary reduced dynamics of the remaining sector, whose linear response is encoded in its retarded Green’s function. Working perturbatively in the double trace coupling, we derive closed form expressions for the quasinormal modes of several coupled geometries involving pure AdS and rotating or nonrotating BTZ black holes, considering both possible system–bath assignments. For the nonrotating configurations previously investigated in \cite{Karch:2025hof}, our analytic expressions agree with the numerical results, while our treatment extends the analysis to configurations involving rotating BTZ black holes. Finally, moving beyond perturbation theory, we derive a relation between dissipation and transmission in the general AdS–BTZ setup.}

\begin{document} 
\maketitle
\section{Introduction}
In AdS/CFT correspondence, a quantum gravity theory living in AdS spacetime is dual to a conformal field theory. More generally, any theory of quantum gravity with an asymptotic boundary is dual to a strongly coupled field theory uniquely defined through its correlation functions.  By translating otherwise intractable many body problems into questions about classical or semiclassical geometry, holographic methods have provided qualitative and quantitative insights into the quark–gluon plasma, relativistic hydrodynamics, confinement and deconfinement, quantum critical matter, thermalization, and entanglement.

Recent advances in quantum information science and condensed-matter physics have renewed interest in the study of open quantum systems. Unitary evolution governed by the Schrödinger equation provides a complete description of an isolated quantum system. In practice, however, no physical system is perfectly isolated: it continually exchanges energy and information with its surroundings. Upon tracing out environment degrees of freedom, the system generally exhibits dissipation and decoherence, and its reduced state can become mixed even though the combined system and environment evolve unitarily. The resulting reduced dynamics depends sensitively on the dynamics of the environment, as well as on the form of the coupling. Consequently, unlike in the isolated system case, there is no universal equation of motion that applies without additional assumptions. 

Interaction between a system and an environment can nevertheless be turned into a diagnostic tool. In quantum probing, for example, an atom playing the role of a probe, may be coupled to a many body system whose properties are then inferred from the response of the probe. In this setting, the probe constitutes the open system, while the many body sector plays the role of the environment. The effects of dissipation and decoherence on the probe therefore encode information about the spectrum and dynamics of the surrounding environment.

For high energy theory, one of the most important applications of open systems is the evaporation of black holes in anti-de Sitter (AdS) spacetime. Although holographic duality provides a powerful framework for studying black holes, the standard reflecting boundary conditions of AdS prevent radiation from escaping through the asymptotic boundary. Consequently, a sufficiently large AdS black hole can equilibrate with its Hawking radiation rather than evaporate completely. To model evaporation, one may couple the dual conformal field theory (CFT) to an auxiliary thermal bath, thereby allowing energy and excitations to leave the holographic system and rendering the CFT an open subsystem. The aim of the present work is to construct a concrete model of such a holographic open quantum system and to identify its gravitational signatures through the quasi-normal mode spectrum.

In previous work \cite{Karch:2025hof}, the authors studied open holography through double trace deformation by coupling a d dimensional CFT, with a gravitational dual AdS$_{d+1}$, to a second  d dimensional CFT that possesses its own AdS$_{d+1}$ dual and plays the role of an environment. Double-trace deformations are a particularly tractable class of multi-trace deformations, originally introduced in \cite{Aharony:2001pa,Klebanov_1999} and subsequently developed in \cite{Witten:2001ua,Aharony:2005sh,Aharony:2006hz,Kiritsis:2006hy,Karch:2023wui}. On the bulk side, the deformation imposes mixed boundary conditions that relate the near boundary data of fields in the two AdS spacetimes. These boundary conditions effectively render the two sectors transparent to one another, permitting the exchange of energy and a nonvanishing flux between the two geometries while preserving the unitary evolution of the combined system.

In the present work, we extend this construction to combine the Schwinger–Keldysh formalism with the influence-functional framework. Schwinger–Keldysh methods for open quantum systems coupled to holographic environments were studied in \cite{Jana:2020vyx,Loganayagam_2023}, where a quantum field theory serving as the system was coupled to a holographic bath. Related works \cite{Pelliconi:2023ojb,Arenas-Henriquez:2025rpt} developed the influence functional description of open quantum field theories possessing holographic duals. From a complementary perspective, \cite{ishii2025lindbladdynamicsholography} constructed a holographic description of a system governed by Lindblad evolution. Here, we integrate out the bath degrees of freedom and construct the resulting influence functional for the system. From this effective description, we obtain the retarded Green’s function of the open system. At leading nontrivial order in the coupling, the poles of this Green’s function yield a closed analytic expression for the corresponding quasi-normal modes, allowing us to determine how coupling to the environment modifies the dissipative dynamics of the bulk theory.

In this manuscript, we provide an analytical treatment for the equation:
\begin{equation}
  \alpha_L\alpha_R
  -
  h^2(2\Delta_L-d)(2\Delta_R-d)\beta_L\beta_R
  =0.
  \label{eq:QNM-master}
\end{equation}
previously solved numerically \cite{Karch:2025hof}. We discuss multiple scenarios where we first couple a Schwarzschild BTZ to AdS and recover the same quasi normal modes in \cite{Karch:2025hof} and then we extend it to coupling black holes at different temperatures. The analytic form for the quasi normal modes to leading orders in the coupling are summarized in Table \ref{tab:perturbative-poles}. 

Each scenario requires a somewhat distinct mathematical treatment. We derive analytical expressions for finite-volume spectral density of different baths: AdS, Schwarzschild BTZ and Spinning BTZ. We show how we can use the perturbative framework to identify qualitatively and quantitatively the characteristic properties of dissipation: Dissipation widths (shift in the imaginary part of a given mode induced by the coupling, namely the difference between its value in the coupled and the uncoupled system), and Real shifts (Oscillatory shifts). We show how by tracing out one system, we obtain a good agreement with the quasinormal modes obtained by solving numerically the full system as it was done before. The general lesson is small coupling dissipation comes from the system's residues and the bath's retarded green's function contributions. 

Further, we derived a non-perturbative expression to compute dissipation in the high-$n$/high frequency limit of the AdS-BTZ system. Such expression links together dissipation and the transmission coefficient that depends only on boundary conditions and the coupling. Surprisingly, the expression resembles the structure of an optical depth law in optics, and allows for an interpretation of the high-n limit of open holography in AdS (the system) and BTZ (the environment) as a wave scattering through an optical material. 

This paper is organized as follows. Section \ref{sec:review_dble_trace} reviews the basics of double-trace deformations and their holographic realization as transparent boundary conditions. In Section \ref{sec:tracing_bath}, we formulate the system–bath setup within the Schwinger–Keldysh formalism, trace out the bath degrees of freedom, and recover the mixed boundary conditions describing the coupled system. Section \ref{sec:dissipation_quants} derives the retarded Green’s functions for all configurations considered in the subsequent sections and introduces several quantities used to characterize dissipation. In Section \ref{sec:QNM_AdS_schld_BTZ}, we derive the quasinormal mode spectrum, to leading order in the double trace coupling, for multiple scenarios: 
\begin{itemize} 
\item Empty AdS coupled to a nonrotating BTZ black hole,
\item Two coupled nonrotating BTZ black holes, and
\item Rotating BTZ black hole coupled to empty AdS.
\end{itemize}
Section \ref{sec:optical_depth}, we provide a nonperturbative analysis relating dissipation to transmission in the general AdS–BTZ setup. We conclude with a summary of our results and a discussion of future directions for applying CFT deformations to the study of open quantum systems. 

\section{Review of Double Trace Coupled Bath} 
\label{sec:review_dble_trace}
In this work, we model an open quantum system through deformations of CFTs. We explore two weakly coupled CFTs where one CFT plays the role of the system and the other we think of as the bath. The deformation from the perspective of one of them will look like:
\begin{equation}
    S_{\text{CF}T_L} \to S_{\text{CFT}_L}+h\int d^dx \,\mathcal{O}_L \mathcal{O}_R
    \label{eq:def}
\end{equation}
Where $\mathcal{O}_L$ and $\mathcal{O}_R$ are scalar primaries of the CFT on the left and the CFT on the right. These operators on the field theory are dual to scalar fields on the gravity side. The scalar field profile near the boundary is given by:
\begin{equation}
    \phi = \alpha z^{d-\Delta}+\beta z^\Delta
\end{equation}
where $\Delta_{\pm} = d/2 \pm  \sqrt{d^2+ 4 m^2}/2$. For $\frac{d}{2}<\Delta<d$, that is standard quantization, both coefficients converge to 0 but $\alpha$ converges much slower than $\beta$. Therefore, in the case of a decoupled CFT we take $\alpha$ to be the non normalizable mode, and set it 0. It plays the role of the source and $\beta$, being the normalizable mode, plays the role of the vev \cite{Klebanov_1999}. In the Breitenloner-Freedman regime $\tfrac{d-2}{2}<\Delta<\tfrac{d+2}{2}$, there is a second quantization scheme possible \cite{Breitenlohner:1982bm} which instead assigns the operator a dimension $\Delta$ with $\tfrac{d-2}{2}<\Delta<\tfrac{d}{2}$ dubbed alternate quantization. In this regime, the speed of convergence of $\alpha$ and $\beta$ are swapped  
but we still consider $\alpha$ to be the non-normalizable mode due to divergent energy flux across the boundary. For a complete review and explicit calculations of this point, see section 3 of \cite{Karch:2025hof}. When $\alpha =0$, the boundary CFT is undeformed. 

When $\alpha \neq 0$, however, the corresponding CFT is deformed and the source in the boundary field theory is exactly characterized by this $\alpha$.
\begin{equation}
    S_{\text{CF}T_L} \to S_{\text{CFT}_L}+h\int d^dx \,\alpha(x) \mathcal{O}_R
\end{equation}
For our setup, we consider a marginal deformation \ref{eq:def} where $\Delta_L+\Delta_R =d$. In this case, one can see that the variation of the action with respect to one operator will couple the source of that operator to the vacuum expectation value of the other operator giving us the mixed boundary conditions allowing for interaction and exchange of energy from one side to the other. 
\begin{equation}
    \alpha_L(\vec{x}) = h (2 \Delta_R-d)\beta_R(\vec{x}), \qquad \alpha_R(\vec{x}) = h (2 \Delta_L-d)\beta_L(\vec{x}).
\end{equation}
with the prefactor coming from the relationship between the coefficient $\beta$ and the one point function \cite{Klebanov_1999}. 
\section{Tracing Out a Holographic Bath}
\label{sec:tracing_bath}
We consider two theories, one will be the system denoted by $S$ and the other is the bath, $B$, coupled through the double trace interaction \cite{Karch:2025hof,Klebanov_1999,Witten:2001ua,Berkooz:2002ug,Hartman:2006dy}
\begin{equation}
 S_{\rm tot}=S_S[\phi]+S_B[\chi]-h\int \d^d x\,O_S(x)O_B(x).
 \label{eq:total-action}
\end{equation}
We take the coupling to be real and $h\geq0$. The same microscopic theory may of course be studied with the two assignments interchanged. The reduced density matrix of the system after tracing out the bath is $\rho_S=\Tr_B \rho_{SB}$. Since the evolution is non-unitary, its real-time generating functional lives on a Schwinger-Keldysh contour \cite{Sieberer:2015svu, Haehl:2025zfn}
\\\\
We use $+$ and $-$ for the forward and backward branches of the contour and define
\begin{equation}
 O_{c}=\frac{O_++O_-}{2},
 \qquad
 O_{q}=O_+-O_-.
 \label{eq:cq-definition}
\end{equation}
where a stands for Left(L) or Right(R) system. A derivation of the closed-time-path functional from the Lindblad equation is included in \cite{Sieberer:2015svu}. A review of the Schwinger-Keldysh formalism is included in \cite{Sieberer:2015svu}, \cite{Haehl:2025zfn}. 
The Keldysh partition function is:
\begin{equation}
Z[J_c,J_q]
  =
  \int_\mathcal{C}
  \mathcal D[\Phi_c,\Phi_q]\,
  \exp\left\{
    iS
    +
    i\int_{t,\mathbf x}
    \left(
      J_q^\dagger\Phi_c
      +
      J_c^\dagger\Phi_q
    \right)
  \right\}
  \nonumber\\
  =
  \left\langle
  \exp\left[
    i\int_{t,\mathbf x}
    \left(
      J_q^\dagger\Phi_c
      +
      J_c^\dagger\Phi_q
    \right)
  \right]
  \right\rangle
\end{equation}
where $c$ and $q$ stand for classical and quantum fields after rotating the $+$ and $-$fields. In the above equation, we used the shorthand
\begin{equation}
 \int_x\equiv\int\d^d x,
 \qquad
 \int_{x,x'}\equiv\int\d^d x\,\d^d x'.
 \label{eq:integration-shorthand}
\end{equation}

\subsection{Schwinger-Keldysh for Double Trace Coupling}
We start from the microscopic theory containing two coupled CFTs,
\begin{equation}
S_{\rm tot}
=
S_L[O_L]+S_R[O_R]+S_{\rm int}[O_L,O_R],
\end{equation}
where $O_L$ is dual to the scalar field $\phi$ and $O_R$ is dual to the scalar field $\chi$. The double-trace interaction is
\begin{equation}
S_{\rm int}
=
-h\int d^dx\,O_L[\phi](x)\,O_R[\chi](x).
\end{equation}
In the open-system interpretation, we will interpret the left CFT as the system and the right CFT as the bath. The reduced left density matrix is obtained by tracing over the right degrees of freedom:
\begin{equation}
\rho_L(t)
=
\mathrm{Tr}_R\,\rho_{LR}(t).
\end{equation}
\\
To compute real-time evolution of a density matrix, we use the Schwinger--Keldysh contour. Thus every field is doubled:
\begin{equation}
\phi \longrightarrow (\phi_+,\phi_-),
\qquad
\chi \longrightarrow (\chi_+,\chi_-),
\end{equation}
where \(+\) denotes the forward branch and \(-\) denotes the backward branch. The full microscopic Schwinger--Keldysh generating functional is therefore
\begin{align}
Z[J_+,J_-]
&=
\int \mathcal D\phi_+\,\mathcal D\phi_-\,
\mathcal D\chi_+\,\mathcal D\chi_-\,
\exp\Bigg\{
iS_L[\phi_+]-iS_L[\phi_-]
+iS_R[\chi_+]-iS_R[\chi_-]
\nonumber\\
&
-ih\int_C d^dx\,O_L(x)O_R(x)
+i\int d^dx\,
\left(
J_+(x)O_{L,+}(x)-J_-(x)O_{L,-}(x)
\right)
\Bigg\}.
\label{eq:full_LR_SK_generating}
\end{align}
Here
\begin{equation}
\int_C d^dx\,O_LO_R = \int d^dx\,O_{L,+}O_{R,+} - \int d^dx\,O_{L,-}O_{R,-}.
\end{equation}
The minus sign on the \(-\) branch comes from the reversed orientation of the backward part of the contour. 
\\\\
Now we separate the left and right path integrals. Treating the left histories \(O_{L,+}\) and \(O_{L,-}\) as fixed external functions from the point of view of the bath, we write
\begin{align}
Z[J_+,J_-]
&=
\int \mathcal D\phi_+\,\mathcal D\phi_-\,
\exp\left\{
iS_L[\phi_+]-iS_L[\phi_-]
+i\int d^dx\,
\left(
J_+O_{L,+}-J_-O_{L,-}
\right)
\right\}
\nonumber\\
&\hspace{1.5cm}
\times
\left[
\int \mathcal D\chi_+\,\mathcal D\chi_-\,
\exp\left\{
iS_R[\chi_+]-iS_R[\chi_-]
-ih\int_C d^dx\,O_L(x)O_R(x)
\right\}
\right].
\label{eq:separate_bath_integral}
\end{align}
The standard Schwinger--Keldysh generating functional for a theory with operator \(O_R\) and external source \(\tilde{J}\) \cite{Haehl_2017}:
\begin{equation}
Z_R[\tilde{J}_+,\tilde{J}_-]
=
\mathrm{Tr}_R\!\left[
U_R[
\tilde{J}_+]\,\rho_R\,U_R^\dagger[\tilde{J}_-]
\right].
\label{eq:bath_SK_trace}
\end{equation}
and equivalently, as a contour-ordered expression:
\begin{equation}
Z_R[\tilde{J}_+,\tilde{J}_-]
=
\left\langle
T_C
\exp\left[
i\int_C d^dx\,\tilde{J}(x)O_R(x)
\right]
\right\rangle_R,
\label{eq:bath_SK_contour}
\end{equation}
\\
\paragraph{Tracing out the bath.} For further details, see \cite{Haehl_2017} and
\cite[Sec.~II.D--E]{FEYNMAN1963118}. 
\begin{align}
Z[J_+,J_-]
&=
\int D\phi_+D\phi_-\,
\exp\Bigg\{
iS_L[\phi_+]-iS_L[\phi_-]
+i\int d^d x\,
\Big(
J_+O_{L,+}-J_-O_{L,-}
\Big)
\Bigg\}
\nonumber\\
&\times
\left[
\int D\chi_+D\chi_-\,
\exp\Bigg\{
iS_R[\chi_+]-iS_R[\chi_-]
-ih\int d^d x\,
\Big(
O_{L,+}O_{R,+}
-
O_{L,-}O_{R,-}
\Big)
\Bigg\}
\right].
\label{eq:partitionfn_coupled}
\end{align}
The Schwinger--Keldysh generating functional of the isolated bath is
\begin{align}
Z_R[J_{R,+},J_{R,-}]
&=
\int D\chi_+D\chi_-\,
\exp\Bigg\{
iS_R[\chi_+]-iS_R[\chi_-]
+i\int d^d x\,
\Big(
J_{R,+}O_{R,+}
-
J_{R,-}O_{R,-}
\Big)
\Bigg\}.
\end{align}
Therefore the square bracket above in \ref{eq:partitionfn_coupled} is exactly the bath generating functional evaluated at the left-history-dependent source
\begin{align}
J_{R,+}(x)=-hO_{L,+}(x),
\qquad
J_{R,-}(x)=-hO_{L,-}(x).
\end{align}
Hence
\begin{align}
\left[
\int D\chi_+D\chi_-\,
e^{iS_R[\chi_+]-iS_R[\chi_-]
-ih\int d^d x\,
(
O_{L,+}O_{R,+}
-
O_{L,-}O_{R,-}
)}
\right]
=
Z_R[-hO_{L,+},-hO_{L,-}].
\end{align} 
If the bath Schwinger--Keldysh functional is normalized by \(Z_R[0,0]\), then
\begin{align}
\frac{
Z_R[-hO_{L,+},-hO_{L,-}]
}{
Z_R[0,0]}
&\equiv
e^{iS_{\rm IF}[O_{L,+},O_{L,-}]}.
\end{align}
Equivalently,
\begin{align}
iS_{\rm IF}[O_{L,+},O_{L,-}]
=
\log
\frac{
Z_R[-hO_{L,+},-hO_{L,-}]
}{
Z_R[0,0]
}.
\end{align}
Thus \(e^{iS_{\rm IF}}\) is the bath
path integral left over after the \(\chi_\pm\) fields have been integrated out.
\\\\
The reduced, normalized open-system generating functional is therefore
\begin{align}
Z_{\rm open}[J_+,J_-]
&=
\frac{Z[J_+,J_-]}{Z_R[0,0]}
\nonumber\\
&=
\int D\phi_+D\phi_-\,
\exp\Bigg\{
iS_L[\phi_+]-iS_L[\phi_-]
+iS_{\rm IF}[O_{L,+},O_{L,-}]
\nonumber\\
&\hspace{3.5cm}
+i\int d^d x\,
\Big(
J_+O_{L,+}
-
J_-O_{L,-}
\Big)
\Bigg\}.
\end{align}

\begin{equation}
\begin{split}
Z_{\rm open}[J_+,J_-]
=
\int \mathcal D\phi_+\,\mathcal D\phi_-\,
\exp \left\{
\begin{aligned}
&iS_L[\phi_+]-iS_L[\phi_-]
+iS_{\rm IF}[O_{L,+},O_{L,-}]
\\
&\quad
+i\int d^dx\,
\left(
J_+O_{L,+}-J_-O_{L,-}
\right)
\end{aligned}
\right\}
\label{eq:open_SK_generating}
\end{split}
\end{equation}
More explicitly,
\begin{equation}
e^{iS_{\mathrm{IF}}[O_{L,+},O_{L,-}]}
=
\left\langle
T_C
\exp\left[
-ih\int d^d x\,
\left(
O_{L,+}(x)O_{R,+}(x)
-
O_{L,-}(x)O_{R,-}(x)
\right)
\right]
\right\rangle_R
\end{equation}

\subsection{Recovered Transparent Boundary Conditions for QNM from $G^R_{L,open}$}
In the AdS/CFT dictionary \cite{Son:2002sd}, the retarded Green's function is obtained from the ratio of response to source. Recall, the near boundary behavior of a scalar with mass $m$ is: $\phi_i=\alpha_i z^{d-\Delta_i}+\beta_iz^\Delta+\dots$, where $\Delta_\pm=\frac{d}{2}\pm\sqrt{d^2+4m^2}/2$. For a scalar dual to $O_i$,
\begin{equation}
    G^R_i(\omega,k) = (2\Delta_i-d)\frac{\beta_i(\omega,k)}{\alpha_i(\omega,k)}
\end{equation}
The poles of $
G^R_{L,\rm open}(\omega,k)
=
\frac{1}
{
(G_L^R(\omega,k))^{-1}
-
h^2G_B^R(\omega,k)
}$ derived in Appendix \ref{appendix_G_open} gives
\begin{equation}
  (G_L^R)^{-1}-h^2G_B^R=0.
\end{equation}
Using the above AdS/CFT relation,
\begin{equation}
  \frac{\alpha_L}{(2\Delta_L-d)\beta_L}
  -
  h^2(2\Delta_R-d)\frac{\beta_R}{\alpha_R}
  =0.
\end{equation}
Multiplying by $(2\Delta_L-d)\beta_L\alpha_R$, we find
\begin{equation}
  \alpha_L\alpha_R
  -
  h^2(2\Delta_L-d)(2\Delta_R-d)\beta_L\beta_R
  =0.
  \label{eq:QNM-master}
\end{equation}
\textbf{This is exactly the structure of the coupled quasinormal mode equation previously derived in \cite{Karch:2025hof}}.

\section{Spectral Baths, Green's functions, and Dissipation Quantities}
\label{sec:dissipation_quants}
\subsection{Some Definitions of and Spectral Baths for reference}
We set the AdS length to $\ell=1$ and the BTZ $l=1$. For our general BTZ black hole, the angular quantum number is denoted by $m\in \mathbb{Z}$. The BTZ mass and angular momentum are encoded in $r_\pm$, with
\begin{equation}
    M=r^2_++r^2_-\quad J=2r_+r_-
\end{equation}
We define the \textit{chiral inverse length} as
\begin{equation}
    \kappa_\sigma=r_++\sigma r_-\quad\sigma=\pm
\end{equation}
For a nonrotating BTZ Black Hole, $J=0$ then $r_-=0$, so $\kappa_+=\kappa_-=\sqrt{M}$. Define
\begin{equation}
    \nu(\Delta)=(2\Delta-2)\frac{\Gamma(2-\Delta)}{\Gamma(\Delta)}
\end{equation}
A pole contributes in time as $e^{-i\omega t}$. If $\omega=\omega_R-i\gamma$. We define the \textit{damping parameter} as 
\begin{equation}
    \gamma=-\Im\omega
\end{equation}
The term \textit{Liouvillian Gap} or \textit{Spectral Gap} often referes to the \textbf{smallest} $\gamma$ since this mode dominates late-time behavior.
\begin{equation}
    \gamma_{gap}=-\min[\Im\omega_j]
\end{equation}
for a given pole branch $j$. This index is needed for the case when we have multiple poles for the retarded Green's function.
\\\\
Denote unperturbed/original poles as $\Omega_n$ and perturbed/shifted poles as $\omega_n$. In the literature it is conventional to define the real and imaginary shifts as
\begin{equation}
\omega_n=\Omega_n+\delta\Omega_n-\frac{i}{2}\Gamma_n
\end{equation}
Then, 
\begin{equation}
    \Gamma_n=-2\Im(\omega_n-\Omega_n)\quad\delta\Omega_n=\Re(\omega_n-\Omega_n)
\end{equation}
For simple poles, we define the following object to get rid of the 2 factor
\begin{equation}
    \delta\gamma_n=\Gamma_n/2
\end{equation}
For multiple poles, there are several, we define for each pole branch $j$
\begin{equation}
    \Gamma_j=-\Im(\omega_j-\Omega) \quad \delta\Omega_j=\Re(\omega_j-\Omega)
\end{equation}
\subsection{AdS General Spectral Bath}
For global empty AdS\(_3\) with the quantum number \(\ell=0\), we introduce the \textit{quantization dimension }\(\delta\).  Standard quantization has \(\delta=\Del\); the alternate quantization has \(\delta=2-\Del\). This parameter helps us keep the quantization scheme general.
\begin{equation}
 G_{\AdS}^R(\omega;\delta)=
 (2\delta-2)\frac{\Gamma(1-\delta)}{\Gamma(\delta-1)}
 \frac{\Gamma\!\left(\frac{\delta-\omega}{2}\right)
       \Gamma\!\left(\frac{\delta+\omega}{2}\right)}
      {\Gamma\!\left(\frac{2-\delta-\omega}{2}\right)
       \Gamma\!\left(\frac{2-\delta+\omega}{2}\right)}.
 \label{eq:EAdS-G}
\end{equation}
The positive-frequency poles are
\begin{equation}
 \Om^{\AdS}_n=\delta+2n,
 \qquad n=0,1,2,\ldots.
\end{equation}
The residue at \(\Om_n\) is
\begin{equation}
R_n(\delta)=-Z_n=-4(\delta-1)^2\left[\frac{(\delta)_n}{n!}\right]^2.
 \label{eq:EAdS-Z}
\end{equation}
For the alternate quantization convention \(\delta=2-\Del\), this is
\begin{equation}
 R_n=-4(\Del-1)^2\left[\frac{(2-\Del)_n}{n!}\right]^2.
\end{equation}
\textbf{High Dissipation limit of the bath}
\\\\
For large imaginary frequency $\Im\omega\gg1$, set
\begin{equation}
 \omega=-\ii\lambda,
 \qquad \lambda>0.
\end{equation}
The two gamma ratios in \eqref{eq:EAdS-G} have arguments
\begin{equation}
 \frac{\delta\pm\omega}{2}=\frac{\delta}{2}\mp \frac{\ii\lambda}{2},
 \qquad
 \frac{2-\delta\pm\omega}{2}=1-\frac{\delta}{2}\mp \frac{\ii\lambda}{2}.
\end{equation}
Using
\begin{equation}
 \frac{\Gamma(z+a)}{\Gamma(z+b)}=z^{a-b}\left[1+O(z^{-1})\right],
 \qquad |z|\to\infty,
\end{equation}
and multiplying the conjugate large-\(\lambda\) ratios gives
\begin{equation}
 G_{\AdS}^R(\Im\omega\gg1;\delta)
 =\calE_\delta\left(\frac{\lambda}{2}\right)^{2\delta-2}
 \left[1+O(\lambda^{-2})\right]=\calE_\delta i^{2\delta-2}\left(\frac{\omega}{2}\right)^{2\delta-2}
 \label{eq:EAdS-asymp}
\end{equation}
\begin{equation}
 \calE_\delta=(2\delta-2)\frac{\Gamma(1-\delta)}{\Gamma(\delta-1)}.
 \label{eq:Edelta}
\end{equation}
This coincides with characteristics of scale invariant open quantum systems where the environment acts as an unparticle bath with a universal scaling for the spectral density $\rho \sim \omega^{2 \Delta-d}$ as discussed in \cite{Arguelles:2026bah,Karch:2026scinv} where if we set $\delta=\Delta$, then we get
\begin{equation}
    G^R_{AdS}(\Im\omega\gg1)=\calE_\Delta (\frac{i}{2})^{2\Delta-2}\omega^{2\Delta-2}
    \label{eq:high_dissipation}
\end{equation}
Recall $i^\Delta=e^{i\pi\Delta/2}$, then $i^{2\Delta-2}=e^{i(\pi(\Delta-1))}$
Then,
\begin{equation}
    \rho_{AdS}(\omega)=2\Im G^R_{AdS}=\calE_\Delta 2^{3-2\Delta}\sin(\pi(\Delta-1))\omega^{2\Delta-2}
\end{equation}
Provides another check between the unparticle baths at finite volume, and the high dissipation limit of AdS \eqref{eq:high_dissipation}.

\subsection{Spinning BTZ Green's function and Spectral density}

For spinning BTZ,
\begin{equation}
    \alpha(\omega)=-\frac{\pi\csc
    (\pi\Delta)\Gamma(\frac{il(lr_+\omega-mr_-)}{r_-^2-r+^2}+1)}{\Gamma(2-\Delta)\Gamma[\frac{1}{2}(\Delta+\frac{il(l\omega-m)}{r_-r_+})]\Gamma[\frac{1}{2}(\Delta-\frac{il(m+l\omega)}{r_-+r_+})]}
\end{equation}
\begin{equation}
    \beta(\omega)=-\frac{\pi\csc
    (\pi\Delta)\Gamma(\frac{il(lr_+\omega-mr_-)}{r_-^2-r+^2}+1)}{\Gamma(\Delta)\Gamma[\frac{1}{2}(-\Delta+\frac{il(l\omega-m)}{r_-r_+}+2)]\Gamma[\frac{1}{2}(-\Delta-\frac{il(m+l\omega)}{r_-+r_+}+2)]}
\end{equation}
The Green's function is obtained by the usual holographic relation. Defining,
\begin{equation}
 A_\sigma(\omega,m)=\frac{\Del}{2}-\frac{\ii(\omega+\sigma m)}{2\kap_\sigma},
 \qquad
 B_\sigma(\omega,m)=1-\frac{\Del}{2}-\frac{\ii(\omega+\sigma m)}{2\kap_\sigma}.
\end{equation}
The retarded Green function is
\begin{equation}
 G_{\BTZ}^R(\omega,m)=-\nu(\Del)
 \prod_{\sigma=\pm}\frac{\Gamma(A_\sigma)}{\Gamma(B_\sigma)}.
 \label{eq:BTZ-G}
\end{equation}
The QNM poles are
\begin{equation}
 \Om_{n,\sigma}=-\sigma m-\ii\kap_\sigma(2n+\Del),
 \qquad n=0,1,2,\ldots.
 \label{eq:BTZ-poles}
\end{equation}
For real frequency define the quantity 
\begin{equation}
 y_\sigma=\frac{\omega+\sigma m}{2\kap_\sigma}
\end{equation}
Using the reflection identity,
\begin{align}
 \frac{\Gamma(\Del/2-\ii y_\sigma)}{\Gamma(1-\Del/2-\ii y_\sigma)}
 &=\frac{\left|\Gamma(\Del/2+\ii y_\sigma)\right|^2}{\pi}
 \sin\left[\pi\left(\frac{\Del}{2}+\ii y_\sigma\right)\right].
\end{align}
Writing
\begin{equation}
 \sin\left[\pi\left(\frac{\Del}{2}+\ii y\right)\right]
 =\sin\frac{\pi\Del}{2}\cosh(\pi y)+\ii\cos\frac{\pi\Del}{2}\sinh(\pi y),
\end{equation}
one finds
\begin{equation}
 \rho_{\BTZ}(\omega,m)=2\operatorname{Im}G_{\BTZ}^R(\omega,m)
 =-\nu(\Del)\frac{\sin(\pi\Del)}{\pi^2}
 \sinh\!\left[\pi(y_++y_-)\right]
 \prod_{\sigma=\pm}\left|\Gamma\!\left(\frac{\Del}{2}+\ii y_\sigma\right)\right|^2.
 \label{eq:BTZ-rho}
\end{equation}
For \(1<\Del<2\) and positive frequency, \(\sin(\pi\Del)<0\), so \(\rho_{\BTZ}\) is positive.
\\\\
For non-rotating BTZ with \(m=0\),
\begin{equation}
 y_+=y_-\equiv y=\frac{\omega}{2\sqrt M},
\end{equation}
and
\begin{equation}
 \rho_{\BTZ}^{(0)}(\omega)=
 -\nu(\Del)\frac{\sin(\pi\Del)}{\pi^2}
 \sinh\!\left(\frac{\pi\omega}{\sqrt M}\right)
 \left|\Gamma\!\left(\frac{\Del}{2}+\frac{\ii\omega}{2\sqrt M}\right)\right|^4.
 \label{eq:BTZ-rho-nonrot}
\end{equation}
The real part for the same non-rotating bath is
\begin{equation}
 \operatorname{Re}G_{\BTZ}^{R,(0)}(\omega)=
 -\nu(\Del)\frac{|\Gamma(\Del/2+\ii y)|^4}{\pi^2}
 \left[\sin^2\frac{\pi\Del}{2}\cosh^2(\pi y)
 -\cos^2\frac{\pi\Del}{2}\sinh^2(\pi y)\right].
 \label{eq:BTZ-Re-nonrot}
\end{equation}
\textbf{Large real-frequency BTZ spectral density}
\\\\
For \(y\to+\infty\),
\begin{equation}
 |\Gamma(a+\ii y)|^2=2\pi e^{-\pi y}y^{2a-1}\left[1+O(y^{-2})\right].
\end{equation}
For non-rotating BTZ,
\begin{align}
 \rho_{\BTZ}^{(0)}(\omega)
 &=-\nu\frac{\sin(\pi\Del)}{\pi^2}
 \left(\frac12e^{2\pi y}\right)(2\pi)^2e^{-2\pi y}y^{2\Del-2}
 \left[1+O(y^{-2})\right]\nonumber\\
 &=-2\nu(\Del)\sin(\pi\Del)y^{2\Del-2}
 \left[1+O(y^{-2})\right].
 \label{eq:BTZ-rho-asymp-nonrot}
\end{align}

For spinning BTZ at large positive \(\omega\),
\begin{equation}
 \rho_{\BTZ}(\omega,m)
 =-2\nu(\Del)\sin(\pi\Del)
 \left[\frac{(\omega+m)(\omega-m)}{4\kap_+\kap_-}\right]^{\Del-1}
 \left[1+O(\omega^{-1})\right].
 \label{eq:BTZ-rho-asymp-spin}
\end{equation}
For fixed quantum angular number \(m\), $\rho_{\BTZ}$ is proportional to \(\omega^{2\Del-2}\) as expected from a scale invariant open quantum system \cite{Arguelles:2026bah}.

\begin{table*}[t]
    \centering
    \renewcommand{\arraystretch}{1.6}
    \setlength{\tabcolsep}{8pt}
    \begin{tabular}
    {p{0.20\textwidth} p{0.60\textwidth} c}
        \toprule
        \textbf{Pole structure} &
        \textbf{Perturbed frequency} &
        \textbf{Leading shift} \\
        \midrule

        Simple pole &
        \[   \omega_{n,\pm}=\Omega_{n,\pm}+h^2R_{n,\pm}G^R_B(\Omega_{n,\pm})+O(h^4)\]
        & $O(h^2)$ \\

        Double pole &
        \[
        \omega_{n,\pm}
        =
        \Omega_{S,n}
        \pm h\sqrt{A_{S,n}g_n}
        +\frac{h^2}{2}
        \left(
            A_{S,n}g_n'
            +B_{S,n}g_n
        \right)
        +O(h^3)
        \]
        & $O(h)$ \\

        Double-pole resonance &
        \[
        \omega_{n,m,j}
        =
        \Omega
        +h^{1/2}q_j
        +h\,
        \frac{
            A_{S,n}B_{B,m}
            +B_{S,n}A_{B,m}
        }{4q_j^2}
        +O(h^{3/2}),
        \qquad
        q_j^4=A_{S,n}A_{B,m}
        \]
        & $O(h^{1/2})$ \\

        \bottomrule
    \end{tabular}

    \caption{
        Leading perturbative corrections to the open-system poles.
        Here $g_n=G_B^R(\Omega_{S,n})$ and
        $g_n'=\partial_\omega G_B^R(\Omega_{S,n})$.
        At resonance, $\Omega=\Omega_{S,n}=\Omega_{B,m}$ and
        $j=1,\ldots,4$ labels the four branches.
    }
    \label{tab:perturbative-poles}
\end{table*}

\subsection{Dictionary of Dissipation Quantities}
There are several quantities that are useful for the purpose of quantifying dissipation of the system due to a bath. Here we collect the main quantities used to describe the open system
quasi-normal mode spectrum after tracing out a bath. These are summarized in Table \ref{tab:dissipation-quantities}. We use the time dependence
\[
\Phi(t)\sim e^{-i\omega t},
\]
so a mode with
\[
\omega=\omega_R+i\gamma
\]
decays as
\[
e^{-i\omega t}=e^{-i\omega_R t}e^{-\gamma t}.
\]
Therefore the physical damping rate, or gap, of a pole is
\begin{equation}
\gamma_j=-\operatorname{Im}\omega_j .
\label{eq:def_gamma_j}
\end{equation}

\begin{table}[t]
    \centering
    \small
    \renewcommand{\arraystretch}{1.35}
    
    \begin{tabular}{@{}p{0.47\textwidth}p{0.47\textwidth}@{}}
        \toprule
        \textbf{Definition} & \textbf{Interpretation} \\
        \midrule

        $\displaystyle \gamma_j=-\operatorname{Im}\omega_j$
        &
        Damping rate of pole $j$.
        \\

        $\displaystyle \gamma_{\mathrm{gap}}=\min_j\gamma_j$
        &
        Smallest damping rate. 
        \\

        \midrule

        $\displaystyle
        \delta\Omega_n=\operatorname{Re}(\omega_n-\Omega_n)$
        &
        Real-frequency shift of a simple pole.
        \\

        $\displaystyle
        \Gamma_n=-2\operatorname{Im}(\omega_n-\Omega_n)$
        &
        
        Midpoint drift of the damping branches
        \\

        $\displaystyle
        \Gamma_{n,\pm}=\operatorname{Im}(\omega_{n,\pm}-\Omega_n)$
        &
        Damping width of a simple pole.
        \\

        \midrule

        $\displaystyle
        \gamma_{n,\pm}=-\operatorname{Im}\omega_{n,\pm}$
        &
        Damping rates of the two branches of a split double pole.
        \\

        $\displaystyle
        \delta\gamma_{\mathrm{cent},n}
        =
        \frac{\gamma_{n,+}+\gamma_{n,-}}{2}
        -\gamma_n^{(0)}$
        &
        Shift of the midpoint of the damping branches.
        \\

        $\displaystyle
        \Delta\gamma_n=\gamma_{n,+}-\gamma_{n,-}$
        &
        Vertical separation of the two branches.
        \\

        $\displaystyle
        \Delta\Omega_n=
        \operatorname{Re}(\omega_{n,+}-\omega_{n,-})$
        &
        Horizontal separation of the two branches.
        \\

        \bottomrule
    \end{tabular}

    \caption{
        Main quantities used to characterize dissipation.
        Here $\Omega_n$ and $\omega_n$ are the uncoupled and
        coupled poles, respectively, and
        $\gamma_n^{(0)}=-\operatorname{Im}\Omega_n$.
    }
    \label{tab:dissipation-quantities}
\end{table}

\subsubsection{Liouvillian gap}

The open-system Liouvillian gap is the smallest positive damping rate among the
quasinormal poles:
\begin{equation}
\gamma_{\rm gap}
=
\min_j\bigl(-\operatorname{Im}\omega_j\bigr).
\label{eq:def_liouvillian_gap}
\end{equation}
Equivalently, in a plot of poles in the complex $\omega$-plane, $\gamma_{\rm gap}$ is
the vertical distance from the real axis to the pole closest to it. This is the
mode that controls the slowest late-time decay. 

\subsubsection{Simple-pole frequency shift and damping width}

When the unperturbed system pole is simple, the open pole can be written as
\begin{equation}
\omega_n
=
\Omega_n+\delta\Omega_n-\frac{i}{2}\Gamma_n ,
\label{eq:simple_pole_parametrization}
\end{equation}
where $\Omega_n$ is the unperturbed normal-mode frequency, $\delta\Omega_n$ is the
real frequency shift, and $\Gamma_n$ is the damping width. Thus
\begin{equation}
\delta\Omega_n
=
\operatorname{Re}\bigl(\omega_n-\Omega_n\bigr),
\qquad
\Gamma_n
=
-2\,\operatorname{Im}\bigl(\omega_n-\Omega_n\bigr).
\label{eq:def_simple_shift_width}
\end{equation}
At weak double-trace coupling $h$, the simple-pole damping width takes the
Fermi golden-rule form (interaction strength x density of states)
\begin{equation}
\Gamma_n
=
h^2 Z_n \rho_B(\Omega_n),
\label{eq:golden_rule_width}
\end{equation}
where $Z_n$ is the residue strength of the system pole and
\begin{equation}
\rho_B(\omega)
=
2\,\operatorname{Im}G_B^R(\omega)
\label{eq:def_spectral_density}
\end{equation}
is the bath spectral density. Eq \eqref{eq:golden_rule_width}  applies when the system pole is simple $\Omega_n$. 

\subsubsection{Double Pole dissipation branches}

For a BTZ system, the relevant unperturbed pole can be a double pole. In that
case one cannot describe the open spectrum by a single width $\Gamma_n$. Instead,
one double pole splits into two branches (Here we define $g_n = G_{bath}^R(\Omega))$
\begin{equation}
\omega_{n,\pm}
=
\Omega_{S,n}
\pm h\sqrt{A_{S,n}g_n}
+
\frac{h^2}{2}
\left(
A_{S,n}g'_n+B_{S,n}g_n
\right)
+O(h^3).
\label{eq:double_pole_branches}
\end{equation}
For each branch of the split double pole, define the branch damping gap
by
\begin{equation}
\gamma_{n,\pm}
=
-\operatorname{Im}\omega_{n,\pm}.
\label{eq:def_branch_resolved_gap}
\end{equation}
The dissipation widths are defined as
\begin{equation}
\Gamma_{n,\pm}=-\Im(\omega_{n,\pm}-\Omega_n)
\end{equation}
\textbf{Central damping shift:}
The central damping shift is the motion of the midpoint of the two damping
branches relative to the original unperturbed damping:
\begin{equation}
\delta\Gamma_{{\rm cent},n}
=
\frac{\gamma_{n,+}+\gamma_{n,-}}{2}
-
\gamma_{n}^{(0)},
\label{eq:def_central_damping_shift}
\end{equation}
where
\begin{equation}
\gamma_{n}^{(0)}
=
-\operatorname{Im}\Omega_{n}.
\label{eq:def_unperturbed_system_gap}
\end{equation}
\textbf{Damping splitting}:
The damping splitting is the difference between the two branch gaps:
\begin{equation}
\Delta\gamma_n
=
\gamma_{n,+}-\gamma_{n,-}.
\label{eq:def_damping_splitting}
\end{equation}
Therefore $\Delta\gamma_n$ measures whether the two branches acquire different
lifetimes. In a plot of $\gamma_{n,+}$ and $\gamma_{n,-}$, it is the vertical separation
between the two branches
\\\\
\textbf{Real frequency splitting}:
The real splitting is the horizontal separation of the two branches in the complex
$\omega$-plane:
\begin{equation}
\Delta\Omega_n
=
\operatorname{Re}\bigl(\omega_{n,+}-\omega_{n,-}\bigr).
\label{eq:def_real_splitting}
\end{equation}
Thus a bath can split a double pole in two different ways: it can separate the
branches vertically through $\Delta\gamma_n$, or horizontally through
$\Delta\Omega_n$.
\\\\
\textbf{Envelope versus pointwise limit}:
In BTZ--BTZ systems, the large-$n$ behavior can contain oscillatory factors from
sine or gamma function ratios. Therefore a sequence may fail to converge pointwise. We define the envelope by removing oscillatory factors and keeping the prefactor of the large-$n$ trend:
\begin{equation}
|\Delta\gamma_n|_{\rm env}
\sim
h\,n^{\Delta_S+\Delta_B-2}.
\label{eq:envelope_scaling}
\end{equation}
This distinction is important near resonances, where the modes can
oscillate while the envelope stays simple.

\section{Quasinormal Modes for the Open System Analytically}
\label{sec:QNM_AdS_schld_BTZ}
The following analytical results hold for small coupling (small-$h$) perturbations (e.g. $h \approx 0.5$ ). Previously we have emphasized that the methodology we take is independent of what we call the system and what we call the bath. In the following sections we swap the roles of the system and the bath and we match the results \emph{analytically} to what was observed before \cite{Karch:2025hof} numerically.
\\\\
Below, we outline five scenarios that each require a somewhat distinct mathematical approach.
\begin{enumerate}
    \item \textbf{Empty AdS - Bath: Non-Rotating BTZ} $\rightarrow$ Simple Pole
    \item \textbf{Non-Rotating BTZ - Bath: Empty AdS} $\rightarrow$ Double pole treatment
    \item \textbf{Non-Rotating BTZ - Bath: Non-Rotating BTZ} $\rightarrow$ Resonant Frequency treatment
    \item \textbf{Rotating BTZ - Bath: Empty AdS}
    \item \textbf{General BTZ - Bath: General BTZ} $\rightarrow$ Putting everything together.
\end{enumerate}
The general rule is: the \textbf{system} creates most of the complexity of the calculation because the residue(s) come from the system's Green's function poles. For each scenario: the analytical formula of spectral density, damping width, real shift, and the resulting plots matching \cite{Karch:2025hof} are given.

\subsection{System 1.1: AdS System and Schwarzschild BTZ Bath (m=0, J=0)}
This section outlines the general analytical steps used to derive the poles of the system after coupling to an environment.
\\\\
\textbf{Step 1: Write down expressions for $\alpha$,  $\beta$, $G^R_{AdS}$ and poles $\Omega_n$}
\\\\
\begin{equation}
    \alpha_{AdS}(\omega)=\frac{\Gamma(1-\Delta)\Gamma(l+1)}{\Gamma(\frac{l-\omega-\Delta+2}{2})\Gamma(\frac{l+\omega-\Delta+2}{2})}
\end{equation}
\begin{equation}
    \beta_{AdS}(\omega)=\frac{\Gamma(\Delta-1)\Gamma(l+1)}{\Gamma(\frac{l-\omega+\Delta}{2})\Gamma(\frac{l+\omega+\Delta}{2})}
\end{equation}
\begin{equation}
    G_{AdS}^R=(2\Delta-d)\frac{\beta_{AdS}}{\alpha_{AdS}}=(2\Delta-2)\frac{\Gamma(\Delta-1)\Gamma(\frac{l-\omega-\Delta+2}{2})\Gamma(\frac{l+\omega-\Delta+2}{2})}{\Gamma(1-\Delta)\Gamma(\frac{l-\omega+\Delta}{2})\Gamma(\frac{l+\omega+\Delta}{2})}
\end{equation}
The standard reflecting boundary conditions of AdS requires the nonnormalizable mode $\alpha = 0$. This condition determines the normal modes of empty AdS:
\begin{equation}
    \Omega_{n,\pm}=\pm(l+\Delta+2n)
\end{equation}
We define the magnitude of the original modes as:
\begin{equation}
    \Omega_n=\abs{\Omega_{n,\pm}}=l+\Del+2n
\end{equation}
\textbf{Step 2: Use the Laurent Expansion for the Green's function}
\begin{equation}
    G_{AdS}^R\simeq\frac{R_{n,\pm}}{\omega-\Omega_{n,\pm}}
\end{equation}
By definition $R_{n,\pm}=\lim_{\omega \to \Omega_{n,\pm}} (\omega-\Omega_{n,\pm})G^R_{AdS}(\omega)$. For empty AdS, the two sets of poles are $\Omega_\pm = \pm(2-\Delta+2n)$, we apply the limit separately at each pole: $R_{n,+} = \lim_{\omega \to \Omega_n} (\omega-\Omega_n)G^R_{AdS}(\omega), R_{n,-} = \lim_{\omega \to -\Omega_n} (\omega+\Omega_n)G^R_{AdS}(\omega)$ where $\Omega_n = 2-\Delta+2n$ is the magnitude of the original modes for empty AdS. For simiplicity, let's denote the arguments of Gamma functions as:
\begin{equation}
    a_1(\omega)=\frac{1}{2}(l-\omega-\Delta+2) \quad a_2(\omega)=\frac{1}{2}(l+\omega-\Delta+2)
\end{equation}
\begin{equation}
    b_1(\omega)=\frac{1}{2}(l-\omega+\Delta) \quad b_2(\omega)=\frac{1}{2}(l+\omega+\Delta)
\end{equation}
The relation between them
\begin{equation}
b_1(\omega)=a_1(\omega)-1+\Delta \quad b_2(\omega)=a_2(\omega)-1+\Delta
\end{equation}
Now, the retarded green's function is
\begin{equation}
    G_{AdS}^R(\omega)=(2\Delta-2)\frac{\Gamma(\Delta-1)\Gamma(a_1(\omega))\Gamma(a_2(\omega))}{\Gamma(1-\Delta)\Gamma(b_1(\omega))\Gamma(b_2(\omega))}
\end{equation}
\textbf{Step 3: Obtain Epsilon and choose a branch}
Under the small-h approximation, the new poles are close to the old ones
\begin{equation}
    \omega=\Omega_n+\delta\omega
\end{equation} 
We define $\epsilon$ as
\begin{equation}
    \epsilon\equiv a(\omega)-a(\Omega_n)=-\frac{1}{2}(\omega-\Omega_n)
\end{equation}
Here we choose either the $\pm$ branch for the original poles for the next steps to obtain half of the perturbed modes. Repeat the same steps for the other branch. We proceed with the $+$ branch for now.
\\\\
\textbf{Step 4: Use Gamma function expansion around the poles, and Taylor expand everything else}
\\\\
The pole expansion is
\begin{equation}
    \Gamma(a_1(\omega)+\epsilon)=\Gamma(-n+\epsilon)\simeq\frac{(-1)^n}{n!\epsilon}[1+\psi(n+1)\epsilon+O(\epsilon^2)]\simeq\frac{(-1)^n}{n!\epsilon}
\end{equation}
Taylor expanding every other factor since they don't have poles at the $-n$ integers. 
\begin{equation}
    \Gamma(a_2(\Omega_n)+\epsilon)=\Gamma(a_2(\Omega_n))[1+\psi(a_2(\Omega_n)+O(\epsilon^2)]\simeq\Gamma(a_2(\Omega_n))
\end{equation}
\begin{equation}
    \Gamma(b_1(\Omega_n)+\epsilon)=\Gamma(b_1(\Omega_n))[1+\psi(b_1(\Omega_n)+O(\epsilon^2)]\simeq\Gamma(b_1(\Omega_n))
\end{equation}
\begin{equation}
    \Gamma(b_2(\Omega_n)+\epsilon)=\Gamma(b_2(\Omega_n))[1+\psi(b_2(\Omega_n)+O(\epsilon^2)]\simeq\Gamma(b_2(\Omega_n))
\end{equation}
\textbf{Step 5: Write Final Form of Green's function under small h shift, and read off residue}
\begin{align}
    G_{AdS}^R(\omega_{n,\mathrm{branch1}})&=(2\Delta-2)\frac{\Gamma(\Delta-1)\Gamma(a_2(\Omega_n))}{\Gamma(1-\Delta)\Gamma(b_1(\Omega_n))\Gamma(b_2(\Omega_n)))}\frac{(-1)^n}{n!\epsilon}\\&=\frac{(2\Delta-2)\frac{\Gamma(\Delta-1)\Gamma(a_2(\Omega_n))}{\Gamma(1-\Delta)\Gamma(b_1(\Omega_n))\Gamma(b_2(\Omega_n))))}\frac{-2(-1)^n}{n!}}{\omega-\Omega_n}
\end{align}
Therefore for both branches, we have
\begin{equation} R_{n,\pm}=\mp(2\Delta-2)\frac{\Gamma(\Delta-1)\Gamma(a_2(\Omega_n))}{\Gamma(1-\Delta)\Gamma(b_1(\Omega_n))\Gamma(b_2(\Omega_n)))}\frac{(-1)^n}{2n!} 
\end{equation}
Many simplifications can be done, but this form is already good for using it computationally. No blow ups at integers!
\\\\ 
\textbf{Step 6: Obtain pole shifting expression}
Now, we use the general Open System expression we obtained in the Markovian approximation in the first section: 
\begin{equation}
G^R_{\rm system,open}=\frac{1}{(G_{\rm system}^R)^{-1}-h^2G^R_{\rm Bath}}.
\end{equation}
The poles of the open system:
\begin{equation}
1-h^2G^R_{system}G^R_{Bath}=0
\end{equation}
In our case:
\begin{equation}
(G^R_{AdS})^{-1}-h^2G^R_{BTZ}=0
\end{equation}
\begin{equation}
G^R_{AdS}\simeq\frac{R_{n,\pm}}{\omega_{n,\pm}-\Omega_{n,\pm}}
\end{equation}
So
\begin{equation}
\frac{\omega_{n,\pm}-\Omega_{n,\pm}}{R_{n,\pm}}-h^2G^R_B(\Omega_{n,\pm})=0
\end{equation}
\begin{equation} \omega_{n,\pm}=\Omega_{n,\pm}+h^2R_{n,\pm}G^R_B(\Omega_{n,\pm})+O(h^4)  
\end{equation}
This equation works for systems \emph{with simple poles}, the same equation can be applied for different branches of $\Omega_n$. For clarity, the perturbation on each branch is
\begin{align}
\omega_{n,+}&=\Omega_n -h^2 Z_n G_B^R(E_n)+\mathcal{O}(h^4)\\ 
\omega_{n,-}&=-\Omega_n +h^2 Z_n G_B^R(-E_n)+\mathcal{O}(h^4)
\end{align}
where $Z_n=\abs{R_{n,\pm}}$

\subsubsection{Dissipation Width and Real Shift}
The empty-AdS system uses the alternate convention
\begin{equation}
\Delta_S=2-\Del, \qquad \Om_n=2-\Del+2n.
\end{equation}
The system residue for the positive branch is
\begin{equation}
 R_n=-Z_n,
 \qquad
 Z_n=4(\Del-1)^2\left[\frac{(2-\Del)_n}{n!}\right]^2.
 \label{eq:case1-Z}
\end{equation}
The details of the simplification of the residue are in Appendix \ref{app:residue-simplification}. The bath is non-rotating BTZ with mass \(M_B=M\) and dimension \(\Del\).  Its spectral density is \eqref{eq:BTZ-rho-nonrot}.  From the simple pole formula,
\begin{align}
\omega_{n,+} = \Omega_n-h^2Z_nG_B^R(\Omega_n)+O(h^4),\quad \omega_{n,-} = -\Omega_n+h^2Z_nG_B^R(-\Omega_n)+O(h^4)
\end{align}
Writing  $\omega_{n,\pm}=\Omega_{n,\pm}+\delta\Omega_{n,\pm}-i \Gamma_{n,\pm}/2$

\begin{equation}
\delta\Om_{n,\pm}=h^2R_{n,\pm}\operatorname{Re}G_B^R(\Om_{n,\pm}),
 \label{eq:case1-real}
\end{equation}
\begin{equation}
\Gamma_{n,\pm}= -2 h^2 R_{n,\pm} \Im G^R_B(\Omega_{n,\pm}) = -2 h^2 R_{n,\pm}\rho_B(\Om_{n,\pm}).
\label{eq:case1-width}
\end{equation}
Substituting the analytic density gives
\begin{equation}
\frac{\Gamma_n}{h^2}=
 -4(\Del-1)^2\left[\frac{(2-\Del)_n}{n!}\right]^2
 \nu(\Del)\frac{\sin(\pi\Del)}{\pi^2}
 \sinh\!\left(\frac{\pi(2-\Del+2n)}{\sqrt M}\right)\left|\Gamma\!\left(\frac{\Del}{2}+\frac{\ii(2-\Del+2n)}{2\sqrt M}\right)\right|^4.
 \label{eq:case1-exact-Gamma}
\end{equation}
From equation \eqref{eq:G_BTZ}, we can see that $G_B^R(-\Omega_n) = [G_B^R(\Omega_n)]^*$. Therefore $\Re G_B^R(-\Omega_n) = \Re G_B^R(\Omega_n)$ and $\Im G_B^R(-\Omega_n) = -\Im G_B^R(\Omega_n)$. This tells us immediately that \eqref{eq:case1-exact-Gamma} stays the same for the $\pm$ towers, that is for the negative tower of the modes, we have $\Gamma_{n,-} = -2h^2R_{n,-} \Im G_B^R(-\Omega_n) = -2h^2(+Z_n)(-\Im G_B^R(\Omega_n))$ so $\Gamma_{n,-} = \Gamma_{n,+}$. Both towers have the same damping as expected.
The real shift is also fully analytic. For the positive branch, let
\begin{equation}
 y_n=\frac{2-\Del+2n}{2\sqrt M} = {\frac{\Omega_n}{2\sqrt M}}.
\end{equation}
Then using \eqref{eq:BTZ-Re-nonrot}, $\delta\Om_{n,+} = -h^2 Z_n Re G_B^R(E_n)$ and $\delta\Om_{n,-} = +h^2 Z_n Re G_B^R(E_n)$. Therefore $\delta\Om_{n,-}=-\delta\Om_{n,+} $
\begin{equation}
\frac{\delta\Om_{n,+}}{h^2}=Z_n\nu(\Del)\frac{|\Gamma(\Del/2+\ii y_n)|^4}{\pi^2}\left[\sin^2\frac{\pi\Del}{2}\cosh^2(\pi y_n)-\cos^2\frac{\pi\Del}{2}\sinh^2(\pi y_n)\right].
\label{eq:case1-exact-real}
\end{equation}
\subsubsection{High-n Limit of Dissipation Width and Real Width}
The high-$n$ limit of the dissipation and real shift quantities are presented here, and a detailed derivation is also provided.
\begin{equation}
\lim_{n\to\infty}\frac{\Gamma_n}{h^2}=\frac{16}{\pi}(\Delta-1)^2\sin^2(\pi\Delta)M^{1-\Delta}.
\label{eq:case1plateau}
\end{equation}
The large-$n$ real shift for the positive branch of the modes $\Omega_{n,+}$ follows from the ratio
\begin{equation}
\frac{\operatorname{Re}G_B^R(\omega)}{\operatorname{Im}G_B^R(\omega)}\longrightarrow -\cot(\pi\Delta),
 \qquad \omega\to+
\infty.
\end{equation}
This formula comes from the form of the Green's function of the bath. Here, we provide a quick derivation. At zero momentum, write
\begin{equation}
G_B^R(\omega)
=
\mathcal N_\Delta
\left[
\frac{\Gamma\!\left(\frac{\Delta}{2}-ix\right)}
{\Gamma\!\left(1-\frac{\Delta}{2}-ix\right)}
\right]^2,
\qquad
x\equiv
\frac{\omega}{2\sqrt M},
\qquad
\mathcal N_\Delta\in\mathbb R .
\end{equation}
Using the large-\(|z|\) identity
\begin{equation}
\frac{\Gamma(z+a)}{\Gamma(z+b)}
\sim z^{a-b},
\end{equation}
with \(z=-ix\), \(a=\Delta/2\), and
\(b=1-\Delta/2\), gives
\begin{equation}
G_B^R(\omega)
\sim
\mathcal N_\Delta(-ix)^{2\Delta-2}
=
\mathcal N_\Delta x^{2\Delta-2}
e^{-i\pi(\Delta-1)},
\qquad \omega\to+\infty .
\end{equation}
Therefore,
\begin{align}
\Re G_B^R(\omega)
&\sim
\mathcal N_\Delta x^{2\Delta-2}
\cos\!\bigl[\pi(\Delta-1)\bigr],\\
\Im G_B^R(\omega)
&\sim
-\mathcal N_\Delta x^{2\Delta-2}
\sin\!\bigl[\pi(\Delta-1)\bigr],
\end{align}
and hence
\begin{equation}
\frac{\Re G_B^R(\omega)}
{\Im G_B^R(\omega)}
\longrightarrow
-\cot\!\bigl[\pi(\Delta-1)\bigr]
=
-\cot(\pi\Delta),
\qquad \omega\to+\infty .
\end{equation}
Then
\begin{equation}
\lim_{n\to\infty}\frac{\delta\Omega_n}{h^2}
 =\frac12\cot(\pi\Delta)
 \lim_{n\to\infty}\frac{\Gamma_n}{h^2}
 =\frac{8}{\pi}(\Delta-1)^2\sin(\pi\Delta)\cos(\pi\Delta)M^{1-
\Delta}.
 \label{eq:case1realPlateau}
\end{equation}

\subsubsection{High-n/UV limit: full derivation}
In the $n\rightarrow\infty$ limit,
\begin{equation}
 \frac{(2-\Del)_n}{n!}
 =\frac{\Gamma(n+2-\Del)}{\Gamma(2-\Del)\Gamma(n+1)}
 =\frac{n^{1-\Del}}{\Gamma(2-\Del)}\left[1+O(n^{-1})\right].
\end{equation}
Therefore
\begin{equation}
 Z_n=\frac{4(\Del-1)^2}{\Gamma(2-\Del)^2}n^{2-2\Del}
 \left[1+O(n^{-1})\right].
 \label{eq:case1-Z-asymp}
\end{equation}
Second, from \eqref{eq:BTZ-rho-asymp-nonrot},
\begin{equation}
 \rho_B(\Om_n)=-2\nu(\Del)\sin(\pi\Del)
 \left(\frac{\Om_n}{2\sqrt M}\right)^{2\Del-2}
 \left[1+O(n^{-1})\right].
\end{equation}
Since \(\Om_n=2n+2-\Del\),
\begin{equation}
 \left(\frac{\Om_n}{2\sqrt M}\right)^{2\Del-2}
 =\left(\frac{n}{\sqrt M}\right)^{2\Del-2}\left[1+O(n^{-1})\right].
\end{equation}
Multiplying with \eqref{eq:case1-Z-asymp},
\begin{align}
 \lim_{n\to\infty}\frac{\Gamma_n}{h^2}
 &=\frac{4(\Del-1)^2}{\Gamma(2-\Del)^2}
 \left[-2\nu(\Del)\sin(\pi\Del)M^{1-\Del}\right]\nonumber\\
 &=-\frac{8(\Del-1)^2\nu(\Del)\sin(\pi\Del)}{\Gamma(2-\Del)^2}M^{1-\Del}.
\end{align}
Now substitute
\begin{equation}
 \nu(\Del)=2(\Del-1)\frac{\Gamma(2-\Del)}{\Gamma(\Del)},
\end{equation}
which gives
\begin{equation}
 \lim_{n\to\infty}\frac{\Gamma_n}{h^2}
 =-\frac{16(\Del-1)^3\sin(\pi\Del)}{\Gamma(\Del)\Gamma(2-\Del)}M^{1-\Del}.
\end{equation}
Finally,
\begin{equation}
 \Gamma(\Del)\Gamma(2-\Del)=(1-\Del)\frac{\pi}{\sin(\pi\Del)}.
\end{equation}
Thus
\begin{equation}
\lim_{n\to\infty}\frac{\Gamma_{n,\pm}}{h^2}
 =\frac{16}{\pi}(\Del-1)^2\sin^2(\pi\Del)M^{1-\Del}.
 \label{eq:case1-plateau}
\end{equation}
This is the high-overtone damping plateau. From eq \eqref{eq:case1-width} and as illustrated in Figure \ref{fig:my_plot1}, the damping width, $\frac{\Gamma_n}{h^2}$, is independent of n.
\\\\
For the shift in the real part of the modes, we use
\begin{equation} 
\begin{cases}
\frac{\operatorname{Re}G_B^R(\omega)}{\operatorname{Im}G_B^R(\omega)}\longrightarrow -\cot(\pi\Del),
\qquad \omega\to+\infty\\
{\frac{\operatorname{Re}G_B^R(\omega)}{\operatorname{Im}G_B^R(\omega)}\longrightarrow +\cot(\pi\Del),
\qquad \omega\to-\infty}
\end{cases}
\end{equation}
Since \(\Gamma_n=2h^2Z_n\operatorname{Im}G_B^R(\Om_n)\) and \(\delta\Om_n=-h^2Z_n\operatorname{Re}G_B^R(\Om_n)\),
\begin{equation}
\frac{\delta\Om_n}{h^2}\longrightarrow-\frac12\left(\frac{\Gamma_n}{h^2}\right)\frac{\operatorname{Re}G_B^R}{\operatorname{Im}G_B^R}=\frac12\cot(\pi\Del)\lim_{n\to\infty}\frac{\Gamma_n}{h^2}.
\end{equation}
Therefore
\begin{align}
\lim_{n\to\infty}\frac{\delta\Om_{n,+}}{h^2}&=+ \frac{8}{\pi}(\Del-1)^2\sin(\pi\Del)\cos(\pi\Del)M^{1-\Del}.\\
\lim_{n\to\infty}\frac{\delta\Om_{n,-}}{h^2}&=- \frac{8}{\pi}(\Del-1)^2\sin(\pi\Del)\cos(\pi\Del)M^{1-\Del}.
\label{eq:case1-real-plateau}
\end{align}

\subsubsection{Dissipation Width and Real Shift Plots}
\begin{figure}[H]
    \centering
    \includegraphics[width=.7\textwidth]{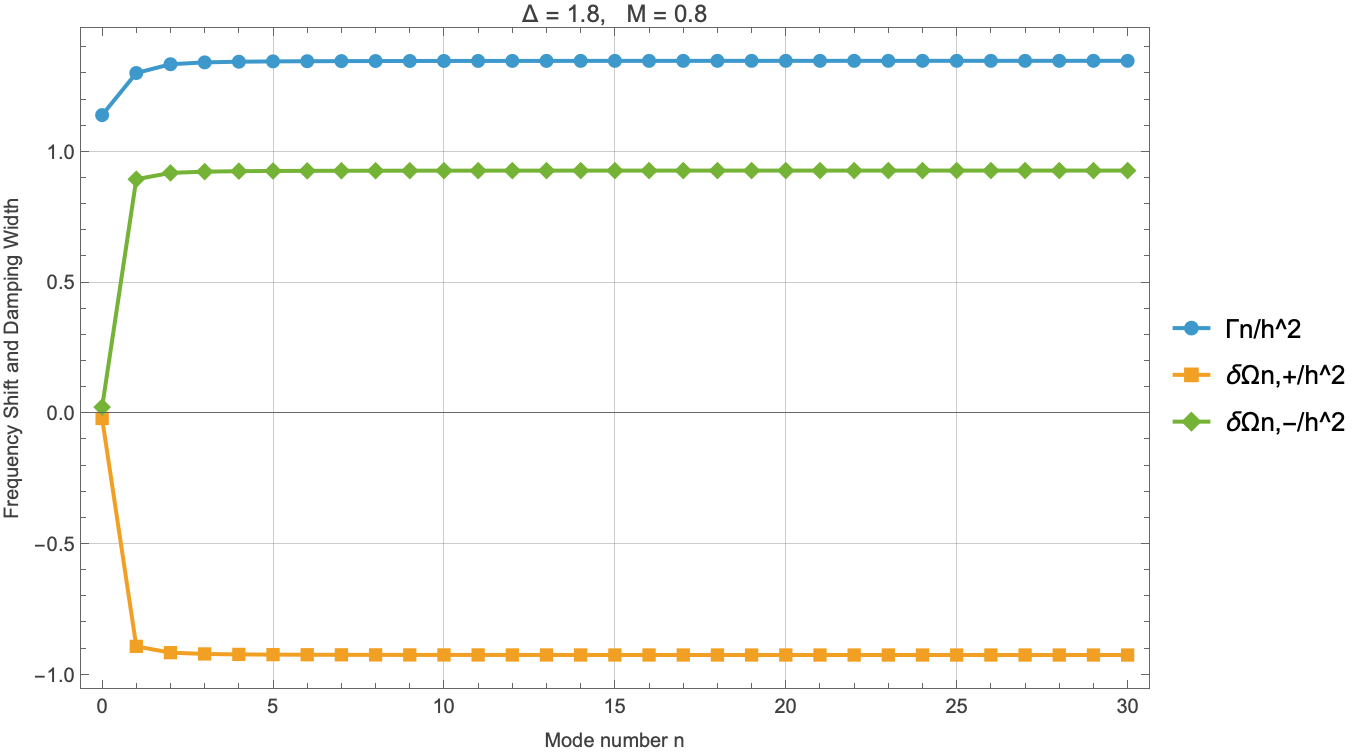}
    \caption{
    Real frequency shift ($\delta\Omega_n/h^2$) and damping width ($\Gamma_n/h^2$) as functions of mode number (n). Both quantities rapidly approach approximately constant values at large $n$.}
    \label{fig:my_plot1}
\end{figure}
\begin{figure}[H]
    \centering
    \includegraphics[width=.7\textwidth]{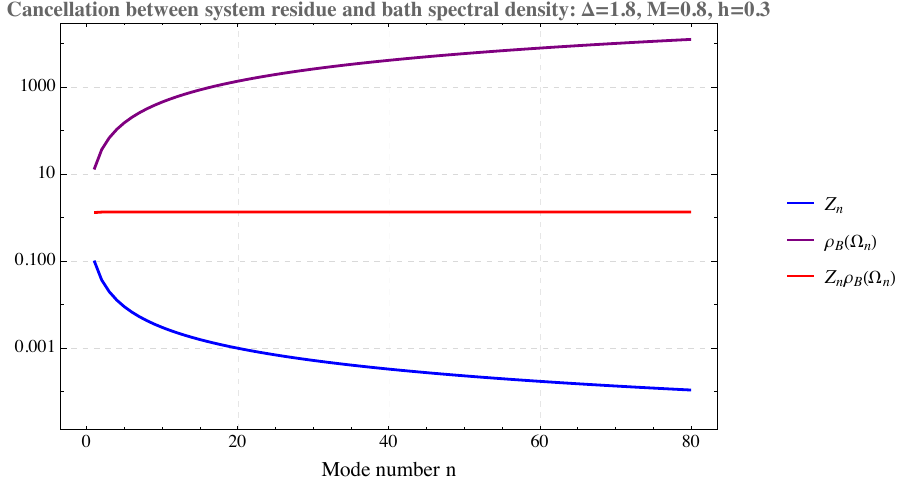}
    \caption{
    Cancellation mechanism behind the high-overtone damping plateau. The residue factor ($Z_n$) decreases with mode number, while the bath spectral density $(\rho_B(\Omega_n)$) increases. Their product ($Z_n\rho_B(\Omega_n$)) approaches a constant, explaining why ($\Gamma_n/h^2$) becomes independent of (n) at large overtone number. Note that we are plotting here the magnitude of $Z_n \rho_B(\Omega)$.}  \label{fig:my_plot2}
\end{figure}
\begin{figure}[H]
\centering
\includegraphics[width=0.6\textwidth]
{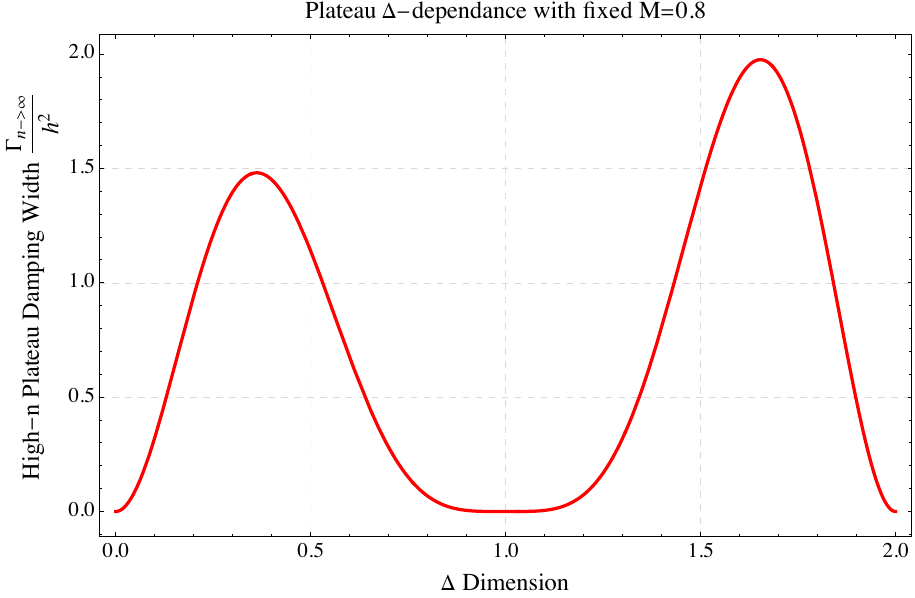}
\caption{Dependence of the high-overtone damping plateau on the operator dimension ($\Delta$) at fixed ($M=0.8$). The plateau is non-monotonic in ($\Delta$), vanishing near special values and developing regions of stronger dissipation.}
\label{fig:my_plot3}
\end{figure}
\begin{figure}[H]
    \centering
    \includegraphics[width=0.8\textwidth]{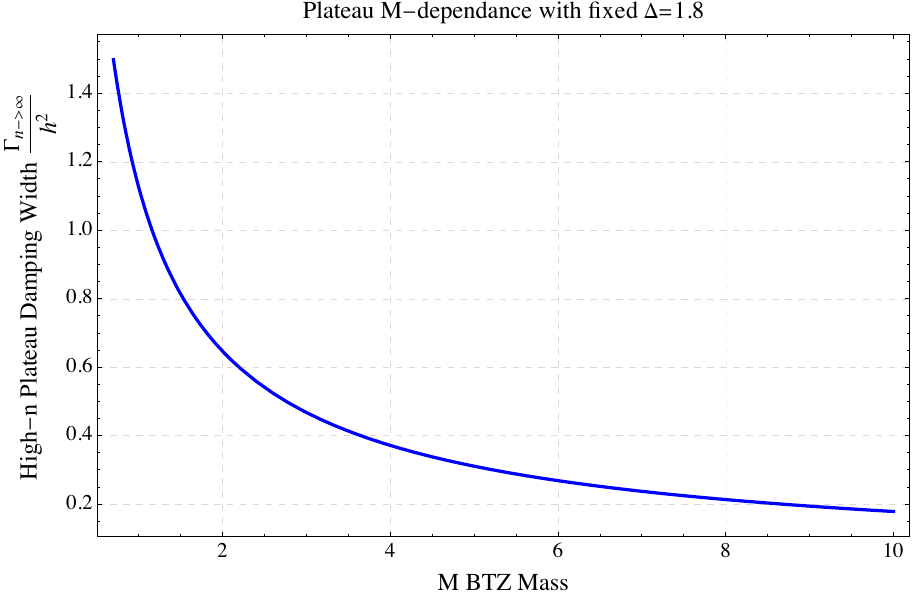}
    \caption{Dependence of the high-overtone damping plateau on the BTZ mass ($M$) at fixed ($\Delta=1.8$). The plateau decreases as ($M$) increases, matching the analytic scaling ($\Gamma_{n\to\infty}/h^2 \propto M^{1-\Delta}$).}
    \label{fig:my_plot4}
\end{figure}

Due to the marginal deformation, we have partial cancellation betwen $Z_n$ and $\rho_B(\Omega_n)$ that results in a constant (See Figure \ref{fig:my_plot2}). At high $n$ limit $\delta \Omega/h^2$ and $\delta \Gamma/h^2$ form a plateau as is illustrated in Figure \ref{fig:my_plot1} that depends on $\Delta$ and $M$. See Figure \ref{fig:my_plot3} and Figure \ref{fig:my_plot4} respectively.

\subsubsection{Quasinormal Modes for AdS tracing out BTZ}
The plots below are obtained analytically using the perturbative expressions derived above and show good qualitative agreement with the numerical results of \cite{Karch:2025hof}. This provides a first benchmark of our analytic results. Despite retaining only the leading $\mathcal{O}(h^2)$ correction to the pole locations, the analytic expressions capture the qualitative behavior of the quasinormal modes even for couplings up to $h\sim0.8$, although quantitative deviations become increasingly important at larger h. 
\begin{figure}[H]
\centering
\includegraphics[width=0.7\linewidth]{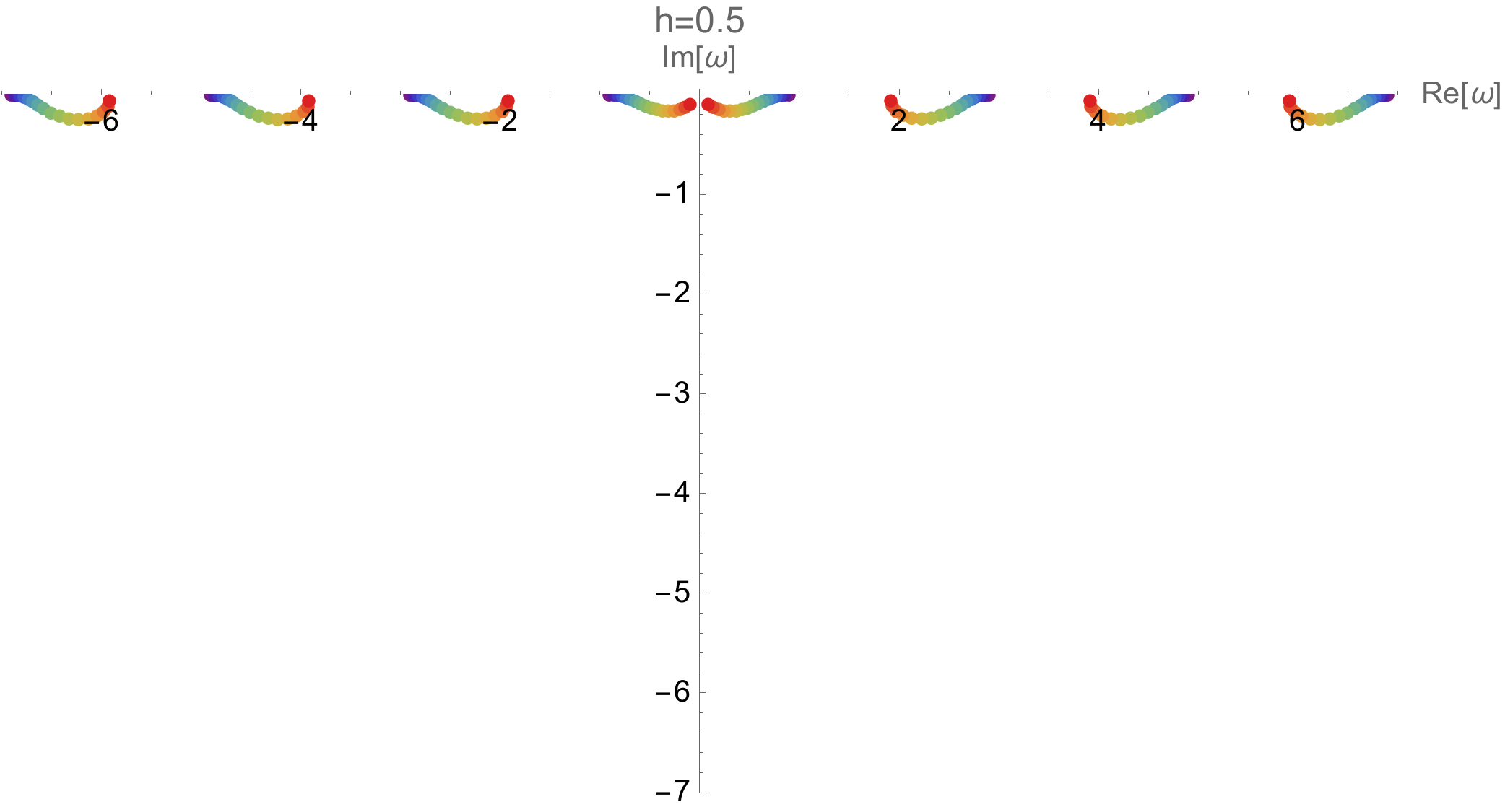}
\caption{AdS tracing out BTZ modes for $h=.5$ and $M=.8$. The colors indicate different values for $\Delta$ varying from $1.1$ to $1.9$ in increments of $.05$. The plot reproduces the same qualitative behavior as the numerical results in \cite{Karch:2025hof}. Over a substantial part of the $\Delta$ range, the discrepancy between the numerical and analytic mode locations is consistent with an $\mathcal{O}(h^4)$ correction, as expected from truncating the analytic pole position at $\mathcal{O}(h^2)$. Near the upper end of the range, $\Delta \simeq 1.8$ and $1.9$, $h=.5$ is not small enough for perturbation theory to agree with the numerics at order $h^4$.}
\label{fig:placeholder}
\end{figure}
\begin{figure}[H]
\centering
\includegraphics[width=0.5\linewidth]{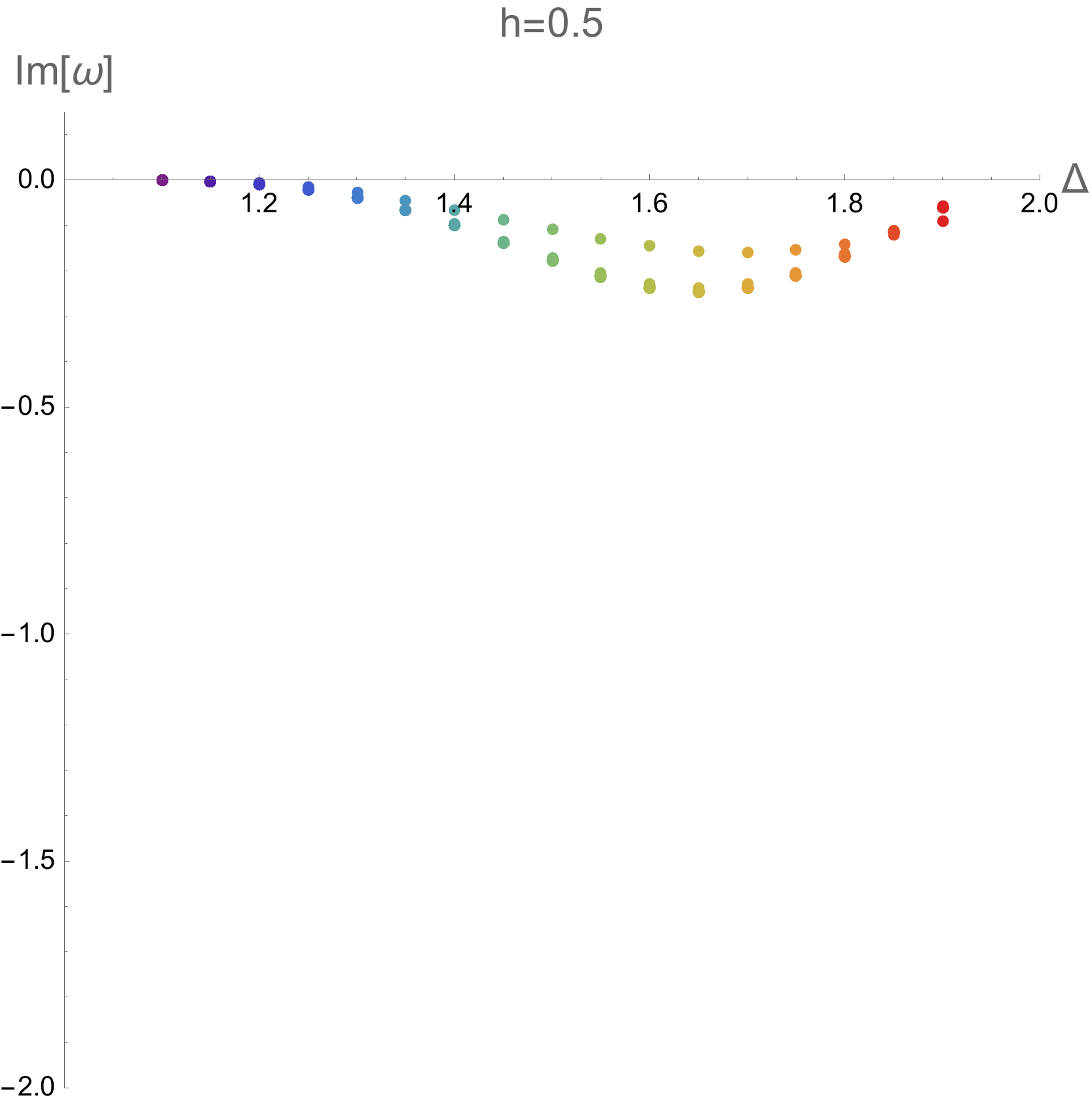}
\caption{Dissipation of AdS due to coupling to a BTZ after tracing out the BTZ degrees of freedom.}
\label{fig:placeholder}
\end{figure}

\subsection{Schwarzschild BTZ System - AdS as the bath}

At zero angular momentum $J=0$ and radial quantum number $m=0$ the two BTZ poles sit on top of each other, so the object that replaces the usual simple
residue are the first two Laurent coefficients/residues of a double pole. Now we follow virtually the same steps except we now have to go to the next order in $\epsilon$ when expanding the Gamma functions around the poles.
\\\\
\textbf{Gamma Function Pole Expansion}
\\\\
We start from the retarded BTZ Green's function
\begin{equation}
G^R_{\mathrm{BTZ}}(\omega)
=
-(2\Delta-2)\frac{\Gamma(2-\Delta)}{\Gamma(\Delta)}
\left[
\frac{
\Gamma\!\left(\frac{\Delta}{2}-\frac{i\omega}{2\sqrt M}\right)}
{
\Gamma\!\left(1-\frac{\Delta}{2}-\frac{i\omega}{2\sqrt M}\right)}
\right]^2 = (2\Delta-2)\frac{\beta_{\mathrm{BTZ}}(\omega)}{\alpha_{\mathrm{BTZ}}(\omega)}
\label{eq:G_BTZ}
\end{equation}
It is useful to introduce the following notation: 
\begin{equation}
a(\omega)\equiv \frac{\Delta}{2}-\frac{i\omega}{2\sqrt M},
\qquad
b(\omega)\equiv 1-\frac{\Delta}{2}-\frac{i\omega}{2\sqrt M}.
\end{equation}
These two arguments are related:
\begin{equation}
b(\omega)=1-\Delta+a(\omega),
\end{equation}
the denominator gamma function is evaluated at \(1-\Delta+a(\omega)\).
\\\\
The BTZ poles occur when the numerator gamma function has a pole:
\begin{equation}
a(\Omega_n)=-n,
\qquad n=0,1,2,\ldots .
\end{equation}
This gives
\begin{equation}
\Omega_n=-i\sqrt M(2n+\Delta).
\end{equation}
Near this frequency write the shifted pole as
\begin{equation}
\omega=\Omega_n+\delta\omega,
\qquad
\epsilon\equiv a(\omega)-a(\Omega_n) =-\frac{i}{2\sqrt M}(\omega-\Omega_n).
\end{equation}
Then
\begin{equation}
a(\omega)=-n+\epsilon,
\qquad
b(\omega)=1-\Delta-n+\epsilon.
\end{equation}
Define
\begin{equation}
b_n\equiv 1-\Delta-n.
\end{equation}
From the expression for $G^R_{\mathrm{BTZ}}$, the singularity comes from \(\Gamma(-n+\epsilon)\).  The standard expansion
of the gamma function around a negative integer is
\begin{equation}
\Gamma(-n+\epsilon)=\frac{(-1)^n}{n!\,\epsilon}\left[1+\left(H_n-\gamma_{\mathrm E}\right)\epsilon+O(\epsilon^2)\right],
\end{equation}
where \(H_n\) is the \(n\)-th harmonic number and \(\gamma_{\mathrm E}\) is the
Euler--Mascheroni constant.  Since
\begin{equation}
H_n-\gamma_{\mathrm E}=\psi(n+1),
\end{equation}
this can also be written as
\begin{equation}
\Gamma(-n+\epsilon)
=
\frac{(-1)^n}{n!\,\epsilon}
\left[
1+\psi(n+1)\epsilon+O(\epsilon^2)
\right].
\end{equation}
Now, look at the denominator of \ref{eq:G_BTZ}. It is regular for non-integer \(\Delta\), so we expand it in the
ordinary Taylor series
\begin{equation}
\Gamma(b_n+\epsilon)
=
\Gamma(b_n)
\left[
1+\psi(b_n)\epsilon+O(\epsilon^2)
\right].
\end{equation}
Equivalently,
\begin{equation}
\frac{1}{\Gamma(b_n+\epsilon)}
=
\frac{1}{\Gamma(b_n)}
\left[
1-\psi(b_n)\epsilon+O(\epsilon^2)
\right].
\end{equation}
Therefore
\begin{align}
\frac{\Gamma(-n+\epsilon)}{\Gamma(b_n+\epsilon)}
&=
\frac{(-1)^n}{n!\,\Gamma(b_n)}
\frac{1}{\epsilon}
\left[
1+\psi(n+1)\epsilon+O(\epsilon^2)
\right]
\left[
1-\psi(b_n)\epsilon+O(\epsilon^2)
\right]
\nonumber\\
&=
\frac{(-1)^n}{n!\,\Gamma(b_n)}
\left[
\frac{1}{\epsilon}
+
\psi(n+1)-\psi(b_n)
+O(\epsilon)
\right].
\end{align}
We define
\begin{equation}
L_n\equiv \psi(n+1)-\psi(1-\Delta-n).
\end{equation}
With this definition,
\begin{equation}
\frac{\Gamma(-n+\epsilon)}{\Gamma(1-\Delta-n+\epsilon)}
=
\frac{(-1)^n}{n!\,\Gamma(1-\Delta-n)}
\left[
\frac{1}{\epsilon}+L_n+O(\epsilon)
\right].
\end{equation}
Since the BTZ Green's function contains the square of the gamma function ratio, one obtains
\begin{equation}
\left[
\frac{\Gamma(-n+\epsilon)}{\Gamma(1-\Delta-n+\epsilon)}
\right]^2
=
\frac{1}{(n!)^2\Gamma(1-\Delta-n)^2}
\left[
\frac{1}{\epsilon^2}
+
\frac{2L_n}{\epsilon}
+
O(1)
\right].
\end{equation}
The factor \((-1)^n\) drops out because the expression is squared.
\\\\
Converting the expansion variable \(\epsilon\) back to the physical frequency
variable where
\begin{equation}
\epsilon=-\frac{i}{2\sqrt M}(\omega-\Omega_n),
\end{equation}
we have
\begin{equation}
\frac{1}{\epsilon}
=
\frac{2i\sqrt M}{\omega-\Omega_n},
\qquad
\frac{1}{\epsilon^2}
=
-\frac{4M}{(\omega-\Omega_n)^2}.
\end{equation}
Substituting into the Green's function gives
\begin{align}
G^R_{R,\mathrm{BTZ}}(\omega)
&=
-(2\Delta-2)\frac{\Gamma(2-\Delta)}{\Gamma(\Delta)}
\frac{1}{(n!)^2\Gamma(1-\Delta-n)^2}
\left[
-\frac{4M}{(\omega-\Omega_n)^2}
+
\frac{4i\sqrt M\,L_n}{\omega-\Omega_n}
+
O(1)
\right]
\nonumber\\
&=
\frac{A_n}{(\omega-\Omega_n)^2}
+
\frac{B_n}{\omega-\Omega_n}
+
O(1),
\end{align}
where
\begin{equation}
A_n
=
4M(2\Delta-2)\frac{\Gamma(2-\Delta)}{\Gamma(\Delta)}
\frac{1}{(n!)^2\Gamma(1-\Delta-n)^2}
\end{equation}
and
\begin{equation}
B_n
=
-4i\sqrt M(2\Delta-2)\frac{\Gamma(2-\Delta)}{\Gamma(\Delta)}
\frac{L_n}{(n!)^2\Gamma(1-\Delta-n)^2}
\end{equation}
These are the unsimplified Laurent coefficients/ Residues.
\\\\
To simplify these expressions, we use the reflection identity
\begin{equation}
\Gamma(z)\Gamma(1-z)=\frac{\pi}{\sin(\pi z)}.
\end{equation}
First,
\begin{equation}
\Gamma(1-\Delta-n)
=
\frac{\pi}{\sin[\pi(\Delta+n)]\,\Gamma(\Delta+n)}
=
\frac{\pi}{(-1)^n\sin(\pi\Delta)\,\Gamma(\Delta+n)}.
\end{equation}
Therefore
\begin{equation}
\frac{1}{\Gamma(1-\Delta-n)^2}
=
\frac{\sin^2(\pi\Delta)\,\Gamma(\Delta+n)^2}{\pi^2}.
\end{equation}
Second,
\begin{equation}
\Gamma(2-\Delta)
=
(1-\Delta)\Gamma(1-\Delta),
\end{equation}
and again using reflection,
\begin{equation}
\Gamma(1-\Delta)
=
\frac{\pi}{\sin(\pi\Delta)\Gamma(\Delta)}.
\end{equation}
Hence
\begin{equation}
\frac{\Gamma(2-\Delta)}{\Gamma(\Delta)}
=
\frac{(1-\Delta)\pi}{\sin(\pi\Delta)\Gamma(\Delta)^2}.
\end{equation}
Putting these together,
\begin{align}
(2\Delta-2)\frac{\Gamma(2-\Delta)}{\Gamma(\Delta)}
\frac{1}{\Gamma(1-\Delta-n)^2}
&=
2(\Delta-1)
\frac{(1-\Delta)\pi}{\sin(\pi\Delta)\Gamma(\Delta)^2}
\frac{\sin^2(\pi\Delta)\Gamma(\Delta+n)^2}{\pi^2}
\nonumber\\
&=
-2(\Delta-1)^2
\frac{\sin(\pi\Delta)}{\pi}
\left[
\frac{\Gamma(\Delta+n)}{\Gamma(\Delta)}
\right]^2 .
\end{align}
Using the Pochhammer symbol
\begin{equation}
(\Delta)_n\equiv \frac{\Gamma(\Delta+n)}{\Gamma(\Delta)},
\end{equation}
we finally get
\begin{equation}
A_n
=
-\frac{8M(\Delta-1)^2\sin(\pi\Delta)}{\pi}
\left[
\frac{(\Delta)_n}{n!}
\right]^2
\end{equation}
and
\begin{equation}
B_n
=
\frac{8i\sqrt M(\Delta-1)^2\sin(\pi\Delta)}{\pi}
\left[
\frac{(\Delta)_n}{n!}
\right]^2
L_n
\end{equation}
Notice that the double pole came from only the fact that the gamma function pole is squared. 
\\\\
\textbf{Corresponding Shifted Pole Equation}
\\\\
The open-system pole equation is
\begin{equation}
\left(G^R_{\mathrm{BTZ}}(\omega)\right)^{-1}
-
h^2G^R_{\mathrm{AdS}}(\omega)
=
0.
\end{equation}
Near the double pole,
\begin{align}
G^R_{\rm BTZ}(\omega)
&\simeq \frac{A_n}{x^2}+\frac{B_n}{x},
\qquad x=\omega-\Omega_n .
\end{align}
Factor out the leading double pole:
\begin{align}
G^R_{\rm BTZ}(\omega)
&=
\frac{A_n}{x^2}
\left(
1+\frac{B_n}{A_n}x
\right).
\end{align}
Therefore
\begin{align}
\left(G^R_{\rm BTZ}(\omega)\right)^{-1}
&=
\frac{x^2}{A_n}
\left(
1+\frac{B_n}{A_n}x
\right)^{-1}.
\end{align}
Using the geometric expansion
\begin{align}
(1+y)^{-1}
=
1-y+y^2+\mathcal O(y^3),
\qquad
y=\frac{B_n}{A_n}x,
\end{align}
we get
\begin{align}
\left(G^R_{\rm BTZ}(\omega)\right)^{-1}
&=
\frac{x^2}{A_n}
\left[
1-\frac{B_n}{A_n}x
+\mathcal O(x^2)
\right]
\\
&=
\frac{x^2}{A_n}
-
\frac{B_n x^3}{A_n^2}
+
\mathcal O(x^4)
\end{align}
Hence
\begin{equation}
\left(G^R_{\mathrm{BTZ}}(\omega)\right)^{-1}
=
\frac{x^2}{A_n}
-
\frac{B_nx^3}{A_n^2}
+
O(x^4).
\end{equation}
At leading order the pole equation becomes
\begin{equation}
\frac{x^2}{A_n}
-
h^2G^R_{\mathrm{AdS}}(\Omega_n)
=
0,
\end{equation}
so
\begin{equation}
x
=
\pm h
\sqrt{
A_nG^R_{\mathrm{AdS}}(\Omega_n)
}
+
O(h^2).
\end{equation}
Therefore
\begin{equation}
\omega_{n,\pm}
=
\Omega_n
\pm h
\sqrt{
A_nG^R_{\mathrm{AdS}}(\Omega_n)
}
+
O(h^2).
\end{equation}

To obtain the next correction to the shifted pole, we must expand the pole equation
one order beyond the leading square-root splitting.  Let
\begin{align}
x \equiv \omega-\Omega_n ,
\end{align}
so that near the double pole of the BTZ Green's function,
\begin{align}
G^R_{\rm BTZ}(\omega)
=
\frac{A_n}{x^2}
+
\frac{B_n}{x}
+
O(1).
\end{align}
Inverting this Laurent expansion gives
\begin{align}
\left(G^R_{\rm BTZ}(\omega)\right)^{-1}
=
\frac{x^2}{A_n}
-
\frac{B_n x^3}{A_n^2}
+
O(x^4).
\end{align}
The open system pole equation is therefore
\begin{align}
\left(G^R_{\rm BTZ}(\omega)\right)^{-1}
-
h^2G^R_{\rm AdS}(\omega)
=
0,
\end{align}
or, using the expansion above,
\begin{align}
\frac{x^2}{A_n}
-
\frac{B_n x^3}{A_n^2}
-
h^2G^R_{\rm AdS}(\omega)
+
O(x^4)
=
0.
\end{align}
Since the leading equation gives $\omega \sim O(h)$, we write the next order shifted pole as the ansatz
\begin{align}
\omega_{n,\pm}
=
\Omega_n
+
h s_{n,\pm}
+
h^2 t_n
+
O(h^3)
\end{align}
or equivalently
\begin{align}
x
=
h s_{n,\pm}
+
h^2 t_n
+
O(h^3).
\end{align}
Now expand each term in the pole equation.
First,
\begin{align}
x^2
&=
\left(hs_{n,\pm}+h^2t_n\right)^2
=
h^2s_{n,\pm}^2
+
2h^3s_{n,\pm}t_n
+
O(h^4),
\\
x^3
&=
\left(hs_{n,\pm}+h^2t_n\right)^3
=
h^3s_{n,\pm}^3
+
O(h^4).
\end{align}
Next, because the AdS bath Green's function is regular at $\omega=\Omega_n$, we Taylor expand it:
\begin{equation}
G^R_{\mathrm{AdS}}(\omega)
\simeq
G^R_{\mathrm{AdS}}(\Omega_n)
+
(\omega-\Omega)\,\frac{\partial G^R_{\mathrm{AdS}}}{\partial\omega}(\Omega_n)
+
\cdots,
\end{equation}
\begin{align}
G^R_{\rm AdS}(\omega)
&=
G_0
+
xG_1
+
O(x^2),
\end{align}
where we defined
\begin{align}
G_0\equiv G^R_{\rm AdS}(\Omega_n),
\qquad
G_1\equiv \partial_\omega G^R_{\rm AdS}(\Omega_n).
\end{align}
Since this term is multiplied by $h^2$, we only need the first two terms:
\begin{align}
h^2G^R_{\rm AdS}(\omega)
=
h^2G_0
+
h^2xG_1
+
O(h^2x^2)
=
h^2G_0
+
h^3s_{n,\pm}G_1
+
O(h^4).
\end{align}
Substituting everything back into the pole equation gives
\begin{align}
0
&=
\frac{1}{A_n}
\left(
h^2s_{n,\pm}^2
+
2h^3s_{n,\pm}t_n
\right)
-
\frac{B_n}{A_n^2}
\left(
h^3s_{n,\pm}^3
\right)
-
\left(
h^2G_0
+
h^3s_{n,\pm}G_1
\right)
+
O(h^4)
\\
&=
h^2
\left(
\frac{s_{n,\pm}^2}{A_n}
-
G_0
\right)
+
h^3
\left(
\frac{2s_{n,\pm}t_n}{A_n}
-
\frac{B_n s_{n,\pm}^3}{A_n^2}
-
s_{n,\pm}G_1
\right)
+
O(h^4).
\end{align}
Now the $O(h^3)$ equation fixes $t_n$:
\begin{align}
\frac{2s_{n,\pm}t_n}{A_n}
-
\frac{B_n s_{n,\pm}^3}{A_n^2}
-
s_{n,\pm}G_1
=
0.
\end{align}
Dividing by $s_{n,\pm}$ gives
\begin{align}
\frac{2t_n}{A_n}
-
\frac{B_n s_{n,\pm}^2}{A_n^2}
-
G_1
=
0.
\end{align}
Using the leading-order result $s_{n,\pm}^2=A_nG_0$, we find
\begin{align}
\frac{2t_n}{A_n}
-
\frac{B_n}{A_n}G_0
-
G_1
=
0.
\end{align}
Therefore
\begin{align}
t_n
=
\frac{A_n}{2}G_1
+
\frac{B_n}{2}G_0.
\end{align}
Restoring the definitions of $G_0$ and $G_1$,
\begin{align}
t_n
=
\frac{A_n}{2}\,
\partial_\omega G^R_{\rm AdS}(\Omega_n)
+
\frac{B_n}{2}\,
G^R_{\rm AdS}(\Omega_n).
\end{align}
Thus the pole position through second order in $h$ is
\begin{align}
\omega_{n,\pm}
=
\Omega_n
\pm
h
\sqrt{
A_nG^R_{\rm AdS}(\Omega_n)
}
+
h^2
\left[
\frac{A_n}{2}\,
\partial_\omega G^R_{\rm AdS}(\Omega_n)
+
\frac{B_n}{2}\,
G^R_{\rm AdS}(\Omega_n)
\right]
+
O(h^3).
\end{align}

\subsubsection{Dissipation Quantities}
For making the notation much cleaner, define $g_n\equiv G_{AdS}^R(\Omega_n)$. The branch gaps/Imaginary part of the modes are
\begin{equation}
\gamma_{n,\pm}=\sqrt{M_S}(2n+\Del_S)
 -\operatorname{Im}\left[
 \pm h\sqrt{A_{n}g_n}
 +\frac{h^2}{2}(A_{n}g'_n+B_{n}g_n)
 \right]+O(h^3).
 \label{eq:case2-gamma-exact}
\end{equation}
The real parts are
\begin{equation}
\operatorname{Re}\omega_{n,\pm}=
 \operatorname{Re}\left[
 \pm h\sqrt{A_{n}g_n}
 +\frac{h^2}{2}(A_{n}g'_n+B_{n}g_n)
 \right]+O(h^3).
 \label{eq:case2-real-exact}
\end{equation}
The central damping shift for the BTZ(system)-AdS(bath), and the damping splitting:
\begin{equation}
 \delta\gamma_{{\rm cent},n}= -\frac{h^2}{2}\operatorname{Im}(A_{n}g'_n+B_{n}g_n)+O(h^3)
 \label{eq:case2-cent-exact}
\end{equation}
\begin{equation}
\Delta\gamma_n=-2h\operatorname{Im}\sqrt{A_{S,n}g_n}+O(h^3)
\label{eq:case2-split-exact}
\end{equation}

\begin{equation}
\Gamma_{\pm,n}= \pm
h
\sqrt{
A_ng_n
}
+
h^2
\left[
\frac{A_n}{2}\,
 g_n'
+
\frac{B_n}{2}\,
g_n
\right].
\end{equation}

\subsubsection{UV/High-n Dissipation}
Let
\begin{equation}
\lambda_n=\sqrt{M_S}(2n+\Del)
\end{equation}
The leading Laurent coefficient is
\begin{equation}
A_{S,n}
=
-\frac{8M_S(\Delta_S-1)^2\sin(\pi\Delta_S)}{\pi}
\left[\frac{(\Delta_S)_n}{n!}\right]^2 .
\end{equation}
Using
\begin{align}
\frac{(\Delta_S)_n}{n!}
&=
\frac{\Gamma(n+\Delta_S)}
{\Gamma(\Delta_S)\Gamma(n+1)}
\nonumber\\
&=
\frac{n^{\Delta_S-1}}{\Gamma(\Delta_S)}
\left[
1+\frac{\Delta_S(\Delta_S-1)}{2n}
+O(n^{-2})
\right],
\end{align}
we obtain
\begin{equation}
A_{S,n}
=
\mathcal A_{\Delta_S}M_Sn^{2\Delta_S-2}
\left[
1+\frac{\Delta_S(\Delta_S-1)}{n}
+O(n^{-2})
\right],
\end{equation}
where
\begin{equation}
\mathcal A_{\Delta_S}
\equiv
-\frac{
8(\Delta_S-1)^2\sin(\pi\Delta_S)
}{
\pi\,\Gamma(\Delta_S)^2
}.
\end{equation}
In particular,
\(A_{S,n}=\mathcal A_{\Delta_S}M_S
n^{2\Delta_S-2}[1+O(n^{-1})]\),

\begin{equation}
g_n=\calE_{\Delta_B}\left(\frac{\lambda_n}{2}\right)^{2\Delta_B-2}\left[1+O(n^{-2})\right]=\calE_{\Delta_B}M_S^{\Delta_B-1}n^{2\Delta_B-2}\left[1+O(n^{-1})\right].
\end{equation}
Therefore
\begin{equation}
A_{S,n}g_n=\calA_{\Del_S}\calE_{\Delta_B}M_S^{\Delta_B}n^{2(\Del_S+\Delta_B)-4}\left[1+O(n^{-1})\right].
\label{eq:case2-Ag-asymp}
\end{equation}
The leading splitting is
\begin{equation}
\sqrt{A_{S,n}g_n}= \left(\calA_{\Del_S}\calE_{\Delta_B}M_S^{\Delta_B}\right)^{1/2}n^{\Del_S+\Delta_B-2}\left[1+O(n^{-1})\right],
\label{eq:case2-sqrt-asymp}
\end{equation}
with the square-root branch chosen consistently.
Thus the leading high-n branch dissipation shift is
\begin{equation}
\delta\gamma_{n,\pm}=\mp h\operatorname{Im}\left[\left(\calA_{\Del_S}\calE_{\Delta_B}M_S^{\Delta_B}\right)^{1/2}n^{\Del_S+\Delta_B-2}\right]+\cdots.
 \label{eq:case2-dgam-asymp-general}
\end{equation}
There is a true leading plateau only if
\begin{equation}
\Del_S+\Delta_B=2.
\end{equation}
Set
\begin{equation}
 \Delta_B=2-\Del_S\equiv 2-\Del,
 \qquad M_S=M.
\end{equation}
Then
\begin{equation}
 \calE_{2-\Del}=2(\Del-1)\frac{\Gamma(\Del)}{\Gamma(2-\Del)}.
\end{equation}
Using
\begin{equation}
 \calA_{\Del}=-\frac{8(\Del-1)^2\sin(\pi\Del)}{\pi\Gamma(\Del)^2},
\end{equation}
and
\begin{equation}
 \Gamma(\Del)\Gamma(2-\Del)=(1-\Del)\frac{\pi}{\sin(\pi\Del)},
\end{equation}
one obtains
\begin{equation}
 \calA_{\Del}\calE_{2-\Del}=\frac{16}{\pi^2}(\Del-1)^2\sin^2(\pi\Del).
\end{equation}
Therefore
\begin{equation}
\lim_{n\to\infty}A_{S,n}g_n
 =\frac{16}{\pi^2}(\Del-1)^2\sin^2(\pi\Del)M^{2-\Del}.
 \label{eq:case2-Ag-plateau}
\end{equation}
The leading order-\(h\) splitting coefficient is real and positive for \(1<\Del<2\):
\begin{equation}
\lim_{n\to\infty}\sqrt{A_{S,n}g_n}
 =\frac{4}{\pi}(\Del-1)|\sin(\pi\Del)|M^{1-\Del/2}.
 \label{eq:case2-splitting-plateau}
\end{equation}
Consequently the order-\(h\) effect is a real-frequency splitting and not a damping splitting:
\begin{equation}
\lim_{n\to\infty}\Delta\gamma_n^{(h)}=0, \qquad \lim_{n\to\infty}\frac{\operatorname{Re}(\omega_{n,+}-\omega_{n,-})}{h}=\frac{8}{\pi}(\Del-1)|\sin(\pi\Del)|M^{1-\Del/2}.
 \label{eq:case2-real-split-plateau}
\end{equation}
The damping induced by the empty-AdS bath first appears in the central order-\(h^2\) term.  From the definition of the the green's function for AdS,
\begin{equation}
 \frac{g'_n}{g_n}=\frac{2\ii(1-\Del)}{\sqrt M(2n+\Del)}+O(n^{-3}),
\end{equation}
so \(A_{S,n}g'_n\to0\) when \(A_{S,n}g_n\) has a plateau.  Similarly,
\begin{equation}
 B_{S,n}g_n=-\frac{\ii}{\sqrt M}L_n^{(\Del)}A_{S,n}g_n.
\end{equation}
Since \(L_n^{(\Del)}\to-\pi\cot(\pi\Del)\),
\begin{equation}
 B_{S,n}g_n\to \ii\frac{\pi\cot(\pi\Del)}{\sqrt M}
 \left[\frac{16}{\pi^2}(\Del-1)^2\sin^2(\pi\Del)M^{2-\Del}\right].
\end{equation}
Thus
\begin{align}
 \lim_{n\to\infty}\frac{\delta\gamma_{{\rm cent},n}}{h^2}
 &=-\frac12\operatorname{Im}\lim_{n\to\infty}(A_{S,n}g'_n+B_{S,n}g_n)\nonumber\\
 &=-\frac{8}{\pi}(\Del-1)^2\sin^2(\pi\Del)\cot(\pi\Del)M^{3/2-\Del}.
\end{align}
Equivalently,
\begin{equation}
\lim_{n\to\infty}\frac{\delta\gamma_{{\rm cent},n}}{h^2}=-\frac{8}{\pi}(\Del-1)^2\sin(\pi\Del)\cos(\pi\Del)M^{3/2-\Del}.
 \label{eq:case2-cent-plateau}
\end{equation}
This is the clean double-pole analogue of the high-overtone damping plateau in System 2.

\paragraph{High-$n$/UV asymptotics of dissipation width}
\begin{equation}
\omega_{n,\pm}-\Omega_{S,n}
=
\pm h\sqrt{A_{S,n}g_n}
+\frac{h^2}{2}
\left(A_{S,n}g_n'+B_{S,n}g_n\right)
+O(h^3).
\end{equation}
Introduce
\begin{equation}
d=\Delta_S+\Delta_B,
\qquad
\mathcal P_{\Delta_S,\Delta_B}
\equiv
\mathcal A_{\Delta_S}\mathcal E_{\Delta_B}
M_S^{\Delta_B}.
\end{equation}
Using
\begin{align}
A_{S,n}g_n
&=
\mathcal P_{\Delta_S,\Delta_B}\,
n^{2d-4}
\left[
1+\frac{\Delta_S(d-2)}{n}
+O(n^{-2})
\right],\\
\frac{g_n'}{g_n}
&=
\frac{2i(\Delta_B-1)}{\lambda_n}
+O(n^{-3}),\\
\frac{B_{S,n}}{A_{S,n}}
&=
-\frac{i}{\sqrt{M_S}}L_{S,n},
\qquad
L_{S,n}
=
-\pi\cot(\pi\Delta_S)
+\frac{1-\Delta_S}{n}
+O(n^{-2}),
\end{align}
one finds
\begin{equation}
\begin{aligned}
\omega_{\pm,n}-\Omega_n
={}&
\pm h\,
\mathcal P_{\Delta_S,\Delta_B}^{1/2}
n^{d-2}
\left[
1+\frac{\Delta_S(d-2)}{2n}
+O(n^{-2})
\right]
\\
&+
\frac{i h^2\mathcal P_{\Delta_S,\Delta_B}}
{2\sqrt{M_S}}\,
n^{2d-4}
\left[
1+\frac{\Delta_S(d-2)}{n}
+O(n^{-2})
\right]
\\
&\hspace{1.2cm}\times
\left[
\pi\cot(\pi\Delta_S)
+\frac{d-2}{n}
+O(n^{-2})
\right]
+O(h^3),
\end{aligned}
\end{equation} 
When \(\mathcal P_{\Delta_S,\Delta_B}>0\), the leading
order-\(h\) splitting is real, and hence
\begin{equation}
\Gamma_{UV,\pm}
=
-\frac{\pi h^2\mathcal P_{\Delta_S,\Delta_B}}
{2\sqrt{M_S}}
\cot(\pi\Delta_S)\,
n^{2(\Delta_S+\Delta_B)-4}
\left[1+O(n^{-1})\right]
+O(h^3).
\end{equation}

\subsubsection{Quasinormal Modes for BTZ tracing out AdS Bath}
Using the analytical expressions, the quasinormal modes and dissipation can be plotted directly.
\begin{figure}[H]
    \centering
    \includegraphics[width=0.8\linewidth]{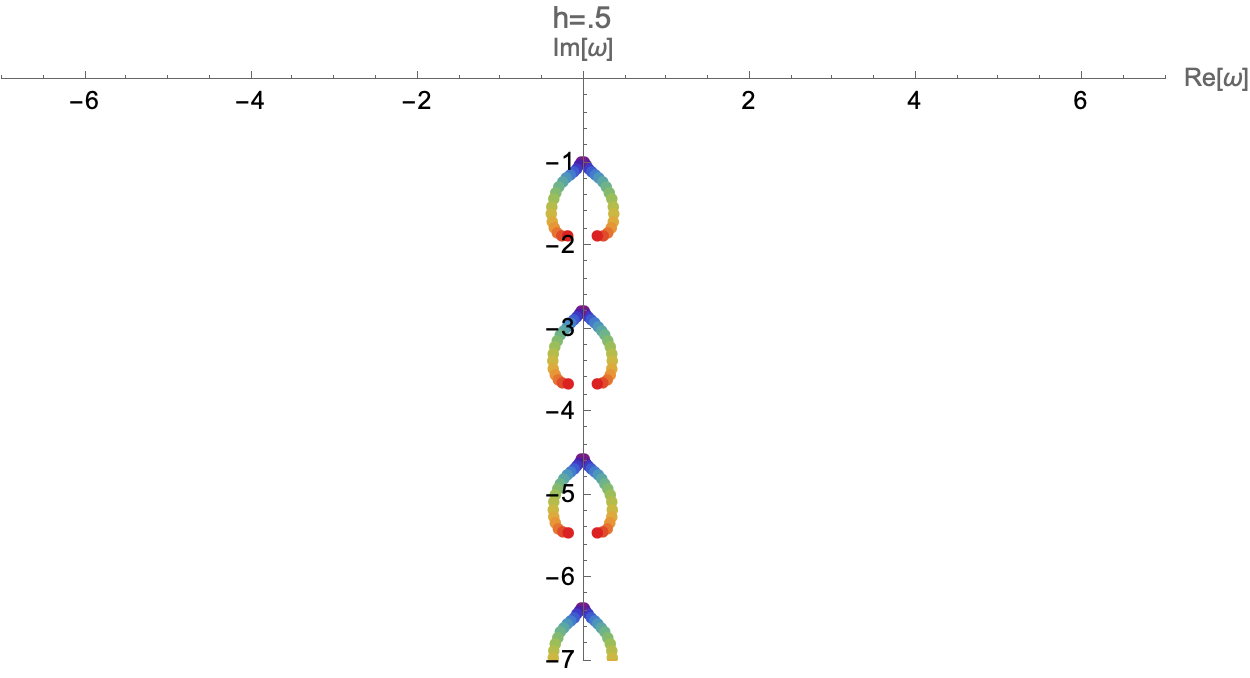}
    \caption{BTZ after tracing out AdS gains a real part}
    \label{fig:placeholder}
\end{figure}

\begin{figure}[H]
    \centering
    \includegraphics[width=0.8\linewidth]{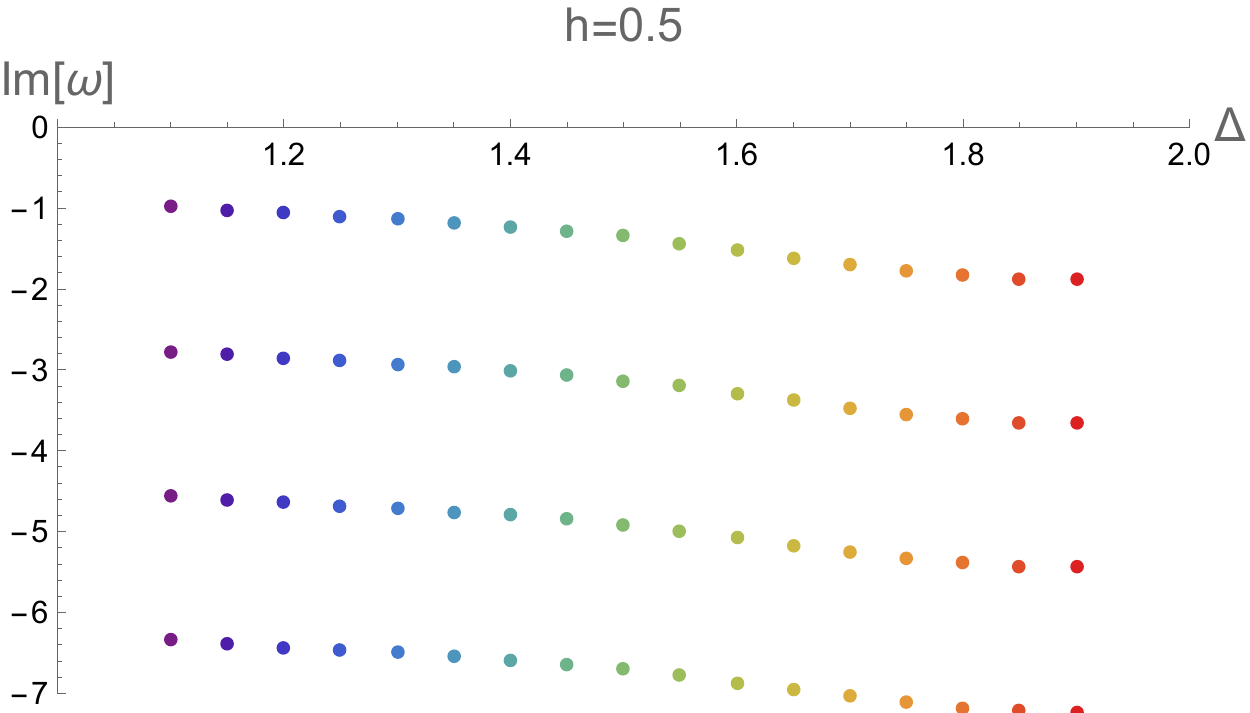}
    \caption{Branch + of Dissipation, the other branch looks the same since they are symmetric}
    \label{fig:placeholder}
\end{figure}
Combining the previous AdS-BTZ and BTZ-AdS case we fully reproduce the expected plot to good accuracy.
\begin{figure}[H]
    \centering
    \includegraphics[width=0.8\linewidth]{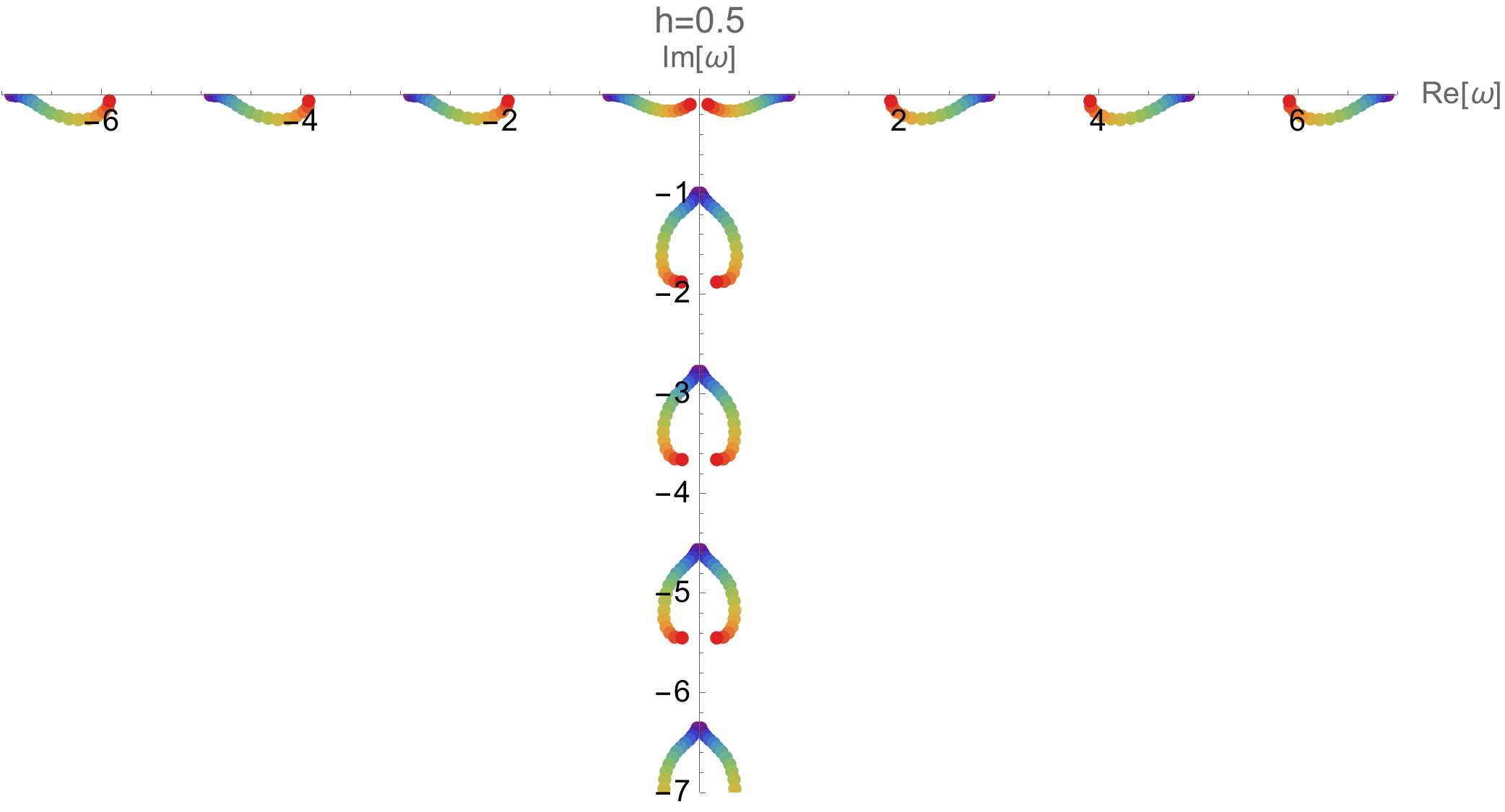}
    \caption{Combined Case 1.1 and 1.2 reproduce figure in \cite{Karch:2025hof}}
    \label{fig:placeholder}
\end{figure}

\begin{figure}[H]
    \centering
    \includegraphics[width=0.6\linewidth]{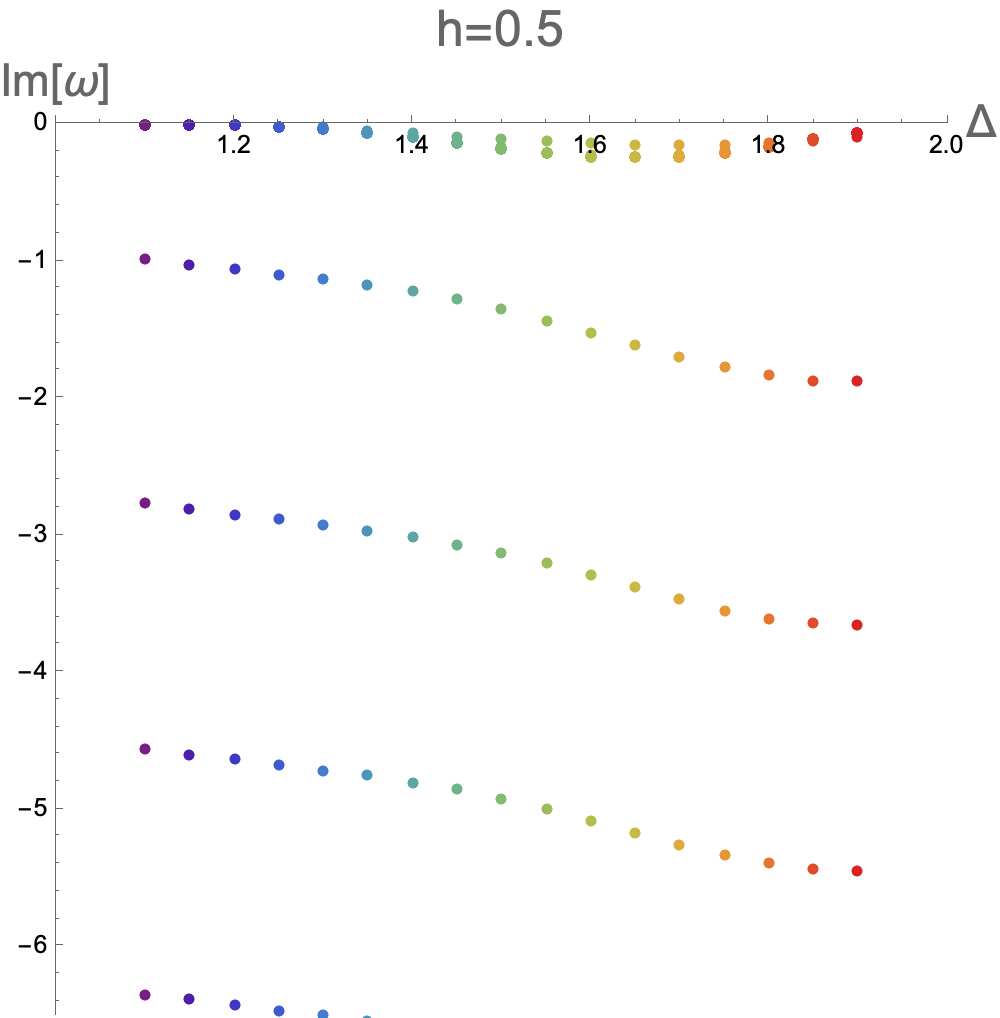}
    \caption{Dissipation combining both cases using the analytic formulae for $\omega_n$ in Table \ref{tab:perturbative-poles}.}
    \label{fig:placeholder}
\end{figure}

\subsection{System 3: non-rotating BTZ system and non-rotating BTZ bath}

\subsubsection{Non-resonant expression}

The system pole and Laurent data are the same as in System 2.  The bath has mass \(M_B\) and dimension \(\Del_B\).  Let
\begin{equation}
 r=\sqrt{M_S/M_B}.
\end{equation}
At the system pole,
\begin{equation}
 a_{B,n}=\frac{\Del_B}{2}-\frac{r}{2}(2n+\Del_S),
 \qquad
 b_{B,n}=1-\frac{\Del_B}{2}-\frac{r}{2}(2n+\Del_S).
\end{equation}
If \(a_{B,n}\notin\{0,-1,-2,\ldots\}\), the bath is analytic and
\begin{equation}
g_n=G_B^R(\Om_{S,n})=-\nu(\Del_B)
 \left[\frac{\Gamma(a_{B,n})}{\Gamma(b_{B,n})}\right]^2,
 \label{eq:case3-g-exact}
\end{equation}
\begin{equation}
g'_n=-\frac{\ii}{\sqrt{M_B}}\left[\psi(a_{B,n})-\psi(b_{B,n})\right]g_n.
 \label{eq:case3-gprime-exact}
\end{equation}
The poles are
\begin{equation}
\omega_{n,\pm}=\Om_{S,n}
 \pm h\sqrt{A_{S,n}g_n}
 +\frac{h^2}{2}(A_{S,n}g'_n+B_{S,n}g_n)+O(h^3).
 \label{eq:case3-nonres-omega}
\end{equation}
The branch gaps are
\begin{equation}
\gamma_{n,\pm}=\sqrt{M_S}(2n+\Del_S)-\operatorname{Im}\left[\pm h\sqrt{A_{S,n}g_n}+\frac{h^2}{2}(A_{S,n}g'_n+B_{S,n}g_n)\right]+O(h^3).
 \label{eq:case3-nonres-gamma}
\end{equation}
The real gaps are the real parts of \eqref{eq:case3-nonres-omega}.
The bath spectral density on the real axis is
\begin{equation}
\rho_B(\omega)=-\nu(\Del_B)\frac{\sin(\pi\Del_B)}{\pi^2}\sinh\!\left(\frac{\pi\omega}{\sqrt{M_B}}\right)\left|\Gamma\!\left(\frac{\Del_B}{2}+\frac{\ii\omega}{2\sqrt{M_B}}\right)\right|^4.
\end{equation}
However, the damping of a BTZ system pole is determined by the analytic value \(g_n\) at the complex system pole.

\subsubsection{Non-resonant high-$n$ plateau}

The subtlety in BTZ--BTZ is that \(a_{B,n}\) and \(b_{B,n}\) run along the negative real axis.  Use the reflection formula:
\begin{equation}
 \Gamma(z)=\frac{\pi}{\sin(\pi z)\Gamma(1-z)}.
\end{equation}
Then
\begin{equation}
 \frac{\Gamma(a_{B,n})}{\Gamma(b_{B,n})}
 =\frac{\sin(\pi b_{B,n})}{\sin(\pi a_{B,n})}
 \frac{\Gamma(1-b_{B,n})}{\Gamma(1-a_{B,n})}.
\end{equation}
Since
\begin{equation}
 (1-b_{B,n})-(1-a_{B,n})=a_{B,n}-b_{B,n}=\Del_B-1,
\end{equation}
we get, away from the zeros of the sine denominator,
\begin{equation}
 \frac{\Gamma(1-b_{B,n})}{\Gamma(1-a_{B,n})}
 =(-a_{B,n})^{\Del_B-1}\left[1+O(n^{-1})\right].
\end{equation}
Define 
\begin{equation}
\mathcal S_n^{(B)}=\frac{\sin(\pi b_{B,n})}{\sin(\pi a_{B,n})}.
 \label{eq:SnB}
\end{equation}
Then
\begin{equation}
g_n=-\nu(\Del_B)\left(\mathcal S_n^{(B)}\right)^2(rn)^{2\Del_B-2}\left[1+O(n^{-1})\right].
 \label{eq:case3-g-asymp}
\end{equation}
Because \(\mathcal S_n^{(B)}\) generally oscillates with \(n\) and diverges near resonances, a pointwise high-\(n\) limit need not exist.  What always exists away from resonant subsequences is the analytic high-\(n\) asymptotic envelope.

Using \(A_{S,n}=\calA_{\Del_S}M_Sn^{2\Del_S-2}[1+O(n^{-1})]\),
\begin{equation}
A_{S,n}g_n=-\calA_{\Del_S}\nu(\Del_B)M_S r^{2\Del_B-2}\left(\mathcal S_n^{(B)}\right)^2n^{2(\Del_S+\Del_B)-4}\left[1+O(n^{-1})\right].
 \label{eq:case3-Ag-asymp}
\end{equation}
For \(1<\Del_B<2\), \(g_n<0\) away from resonances when \(\mathcal S_n^{(B)}\) is real, so \(A_{S,n}g_n<0\).  Then the leading splitting is primarily along the imaginary direction:
\begin{equation}
 \sqrt{A_{S,n}g_n}
 =\ii\sqrt{\calA_{\Del_S}\nu(\Del_B)M_S}
 r^{\Del_B-1}|\mathcal S_n^{(B)}|
 n^{\Del_S+\Del_B-2}\left[1+O(n^{-1})\right],
\end{equation}
up to branch choice.  Therefore the leading damping splitting has envelope
\begin{equation}
|\Delta\gamma_n|\sim
 2h\sqrt{\calA_{\Del_S}\nu(\Del_B)M_S}
 r^{\Del_B-1}|\mathcal S_n^{(B)}|
 n^{\Del_S+\Del_B-2}.
 \label{eq:case3-gap-envelope}
\end{equation}
A leading plateau can occur only at the level of the envelope if
\begin{equation}
 \Del_S+\Del_B=2.
\end{equation}
Even then, a true pointwise limit exists only if \(\mathcal S_n^{(B)}\) has a limit or one studies an averaged/envelope quantity.

\subsubsection{Resonance}
A bath double pole collides with the system pole when
\begin{equation}
\sqrt{M_S}(2n+\Del_S)=\sqrt{M_B}(2m+\Del_B).
 \label{eq:case3-resonance}
\end{equation}
Let
\begin{equation}
 \Om=\Om_{S,n}=\Om_{B,m}.
\end{equation}
The bath Laurent data are
\begin{equation}
 A_{B,m}=-\frac{8M_B(\Del_B-1)^2\sin(\pi\Del_B)}{\pi}
 \left[\frac{(\Del_B)_m}{m!}\right]^2,
\end{equation}
\begin{equation}
 B_{B,m}=\frac{8\ii\sqrt{M_B}(\Del_B-1)^2\sin(\pi\Del_B)}{\pi}
 \left[\frac{(\Del_B)_m}{m!}\right]^2
 \left[\psi(m+1)-\psi(m+\Del_B)-\pi\cot(\pi\Del_B)\right].
\end{equation}
The four branches are
\begin{equation}
\omega_{n,m,j}=\Om+h^{1/2}q_j
 +h\frac{A_{S,n}B_{B,m}+B_{S,n}A_{B,m}}{4q_j^2}
 +O(h^{3/2}),
 \qquad q_j^4=A_{S,n}A_{B,m}.
 \label{eq:case3-res-omega}
\end{equation}
The branch gaps are
\begin{equation}
\gamma_{n,m,j}= -\operatorname{Im}\Om
 -h^{1/2}\operatorname{Im}q_j
 -h\operatorname{Im}\left(\frac{A_{S,n}B_{B,m}+B_{S,n}A_{B,m}}{4q_j^2}\right)+O(h^{3/2}).
 \label{eq:case3-res-gamma}
\end{equation}
The real gaps are obtained by replacing \(-\operatorname{Im}\) with \(\operatorname{Re}\) in the two correction terms.
\\\\
More, explicitly, the dimensions at which the resonances occur are 
\begin{equation}
    \Delta=2\frac{r(n+1)-m}{1+r}
\end{equation}
with 
\begin{equation}
    r=\sqrt{M_S/M_B}
\end{equation}
For equal masses, the resonances occur at 
\begin{equation}
    \Delta=n-m+1
\end{equation}
Therefore, for equal masses, every integer produces a resonance. 

\subsubsection{Plots comparison to \cite{Karch:2025hof}} 
 Here we run $\Delta$ from 1 to 2 divided in 17 intervals starting from 1.1 to 1.9.

\begin{figure}[H]
    \centering
    \includegraphics[width=0.8\linewidth]{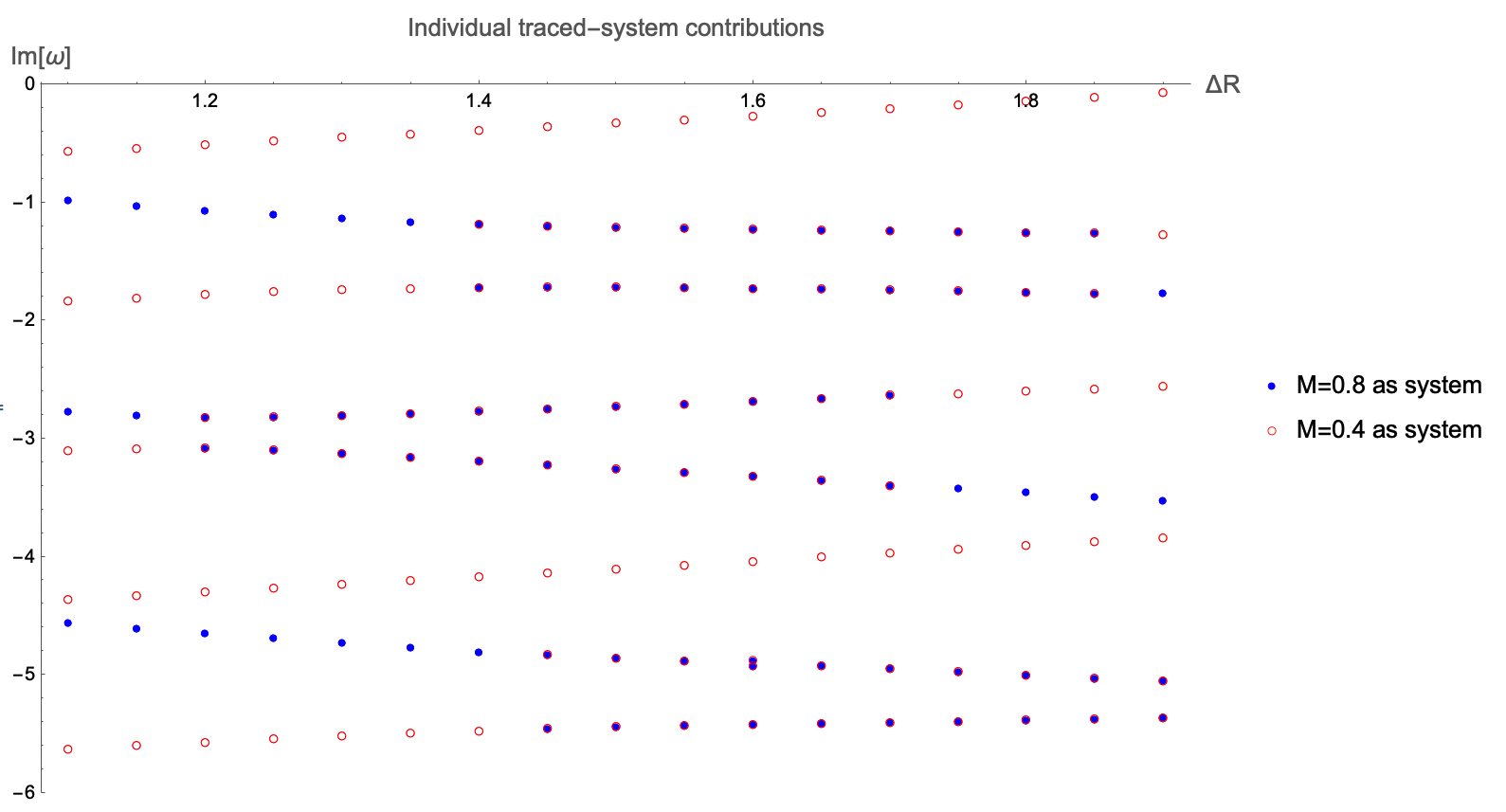}
    \caption{Two coupled BTZs of different masses. Red dots represents the case of taking the system to be the BTZ of M=.4 and tracing out the other while blue represents the scenario of taking the BTZ of M=.8 as the system and tracing out the other.}
    \label{fig:placeholder}
\end{figure}

\begin{figure}[H]
    \centering
    \includegraphics[width=0.8\linewidth]{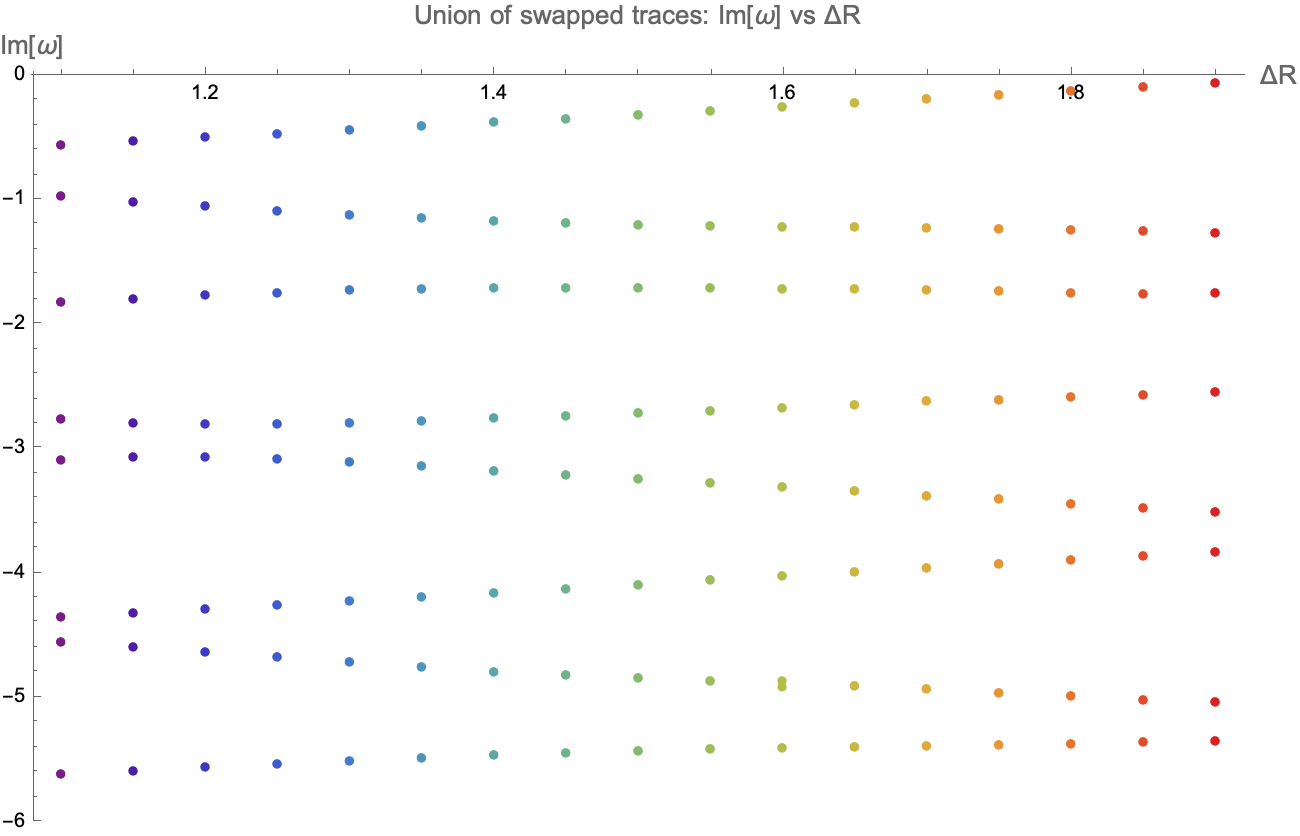}
    \caption{Full BTZ-BTZ dissipation. Reproduces figure 9 in \cite{Karch:2025hof}}
    \label{fig:placeholder}
\end{figure}
\begin{figure}[H]
    \centering
    \begin{minipage}[c]{0.45\linewidth}
        \centering
        \includegraphics[width=\linewidth]{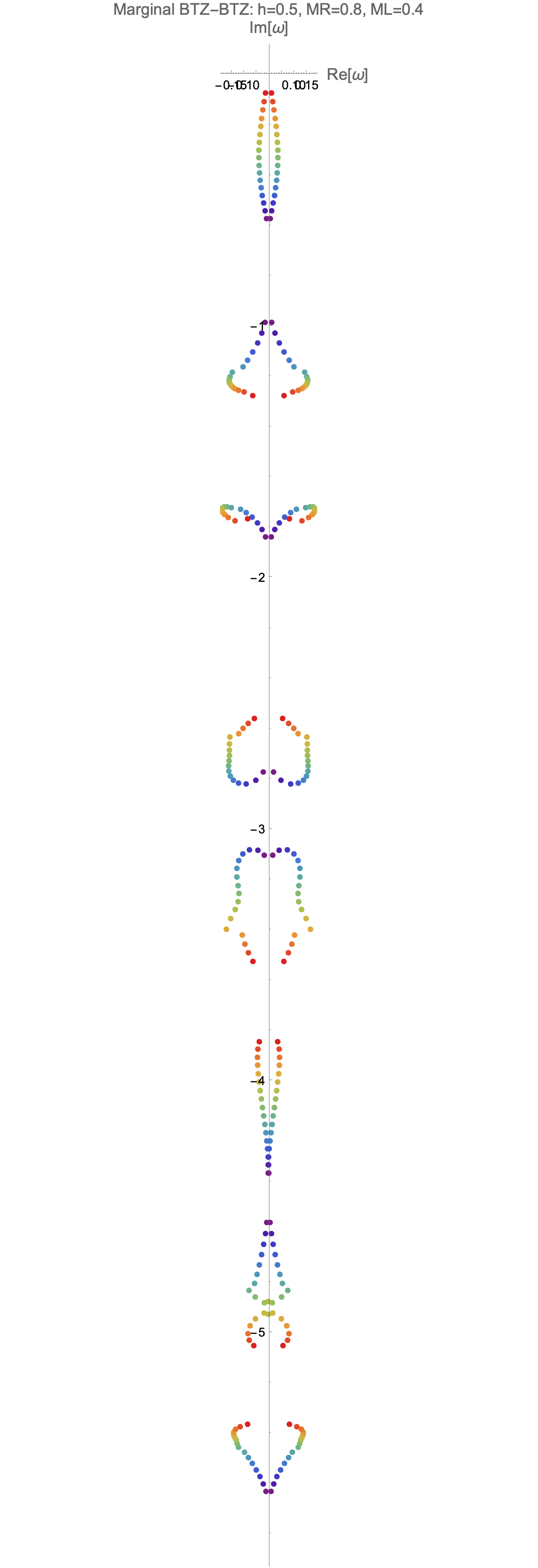}
    \end{minipage}
    \hfill
    \begin{minipage}[c]{0.23\linewidth}
        \centering
        \includegraphics[width=\linewidth]{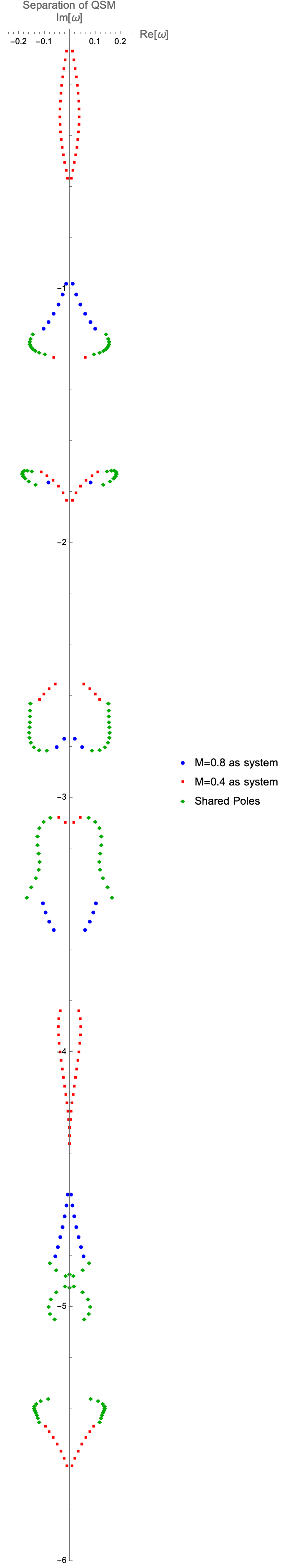}
    \end{minipage}
    \caption{BTZ-BTZ Quasinormal Modes. Left: The analytics reproducing Figure 10 in
    \cite{Karch:2025hof}, where the real part is more sensitive to
    perturbations. Right: individual BTZ--BTZ modes and their hybridization
    due to resonances.}
    \label{fig:btz_btz_modes}
\end{figure}
\subsection{System 4: Spinning BTZ system and AdS as the bath}

\subsubsection{Generic simple chiral pole}

The system pole is
\begin{equation}
 \Om_{S,n,\sigma}=-\sigma m-\ii\kap_{S\sigma}(2n+\Del_S),
 \qquad \kap_{S\sigma}=r_{S+}+\sigma r_{S-}.
\end{equation}
Define
\begin{equation}
    A_{S,\sigma}=\frac{\Delta}{2}-\frac{i(\omega+\sigma m)}{2\kappa_\sigma} \qquad B_{S,\sigma}=1-\frac{\Delta}{2}-\frac{i(\omega+\sigma m)}{2\kappa_\sigma}
\end{equation}
The AdS Green's function is
\begin{equation}
    G_{\AdS}^R(\omega)=(2\Delta-2)\frac{\Gamma(1-\Delta)}{\Gamma(\Delta-1)}\frac{\Gamma(\frac{\Delta-\omega}{2})\Gamma(\frac{\Delta+\omega}{2})}{\Gamma(\frac{2-\Delta-\omega}{2})\Gamma(\frac{2-\Delta+\omega}{2})}
\end{equation}
Define
\begin{equation}
 C_{S,n,\sigma}=\frac{2\ii\kap_{S\sigma}(-1)^n}{n!\Gamma(1-\Del_S-n)}.
\end{equation}
Define the ratio of Gamma functions as
\begin{equation}
 F_{S,-\sigma}(\Om_{S,n,\sigma})=
 \frac{\Gamma(A_{S,-\sigma}(\Om_{S,n,\sigma},m))}
      {\Gamma(B_{S,-\sigma}(\Om_{S,n,\sigma},m))}.
\end{equation}
The residue is
\begin{equation}
 R_{S,n,\sigma}=-\nu(\Del_S)C_{S,n,\sigma}F_{S,-\sigma}(\Om_{S,n,\sigma}).
 \label{eq:case4-R}
\end{equation}
The empty-AdS bath value is
\begin{equation}
 g_{n,\sigma}=G_{\text{AdS}}^R(\Om_{S,n,\sigma};\Delta_B).
\end{equation}
Since the system pole is simple and the bath is analytic at the complex pole,
\begin{equation}
\omega_{n,\sigma}=\Om_{S,n,\sigma}+h^2R_{S,n,\sigma}g_{n,\sigma}+O(h^4).
 \label{eq:case4-omega}
\end{equation}
The damping gap and real part are
\begin{equation}
 \gamma_{n,\sigma}=\kap_{S\sigma}(2n+\Del_S)
 -h^2\operatorname{Im}\left[R_{S,n,\sigma}g_{n,\sigma}\right]+O(h^4),
 \label{eq:case4-gamma}
\end{equation}
\begin{equation}
 \operatorname{Re}\omega_{n,\sigma}=-\sigma m
 +h^2\operatorname{Re}\left[R_{S,n,\sigma}g_{n,\sigma}\right]+O(h^4).
 \label{eq:case4-real}
\end{equation}
\subsubsection{High-overtone asymptotics}

At \(\Om_{S,n,\sigma}\), the opposite chiral argument is
\begin{equation}
 A_{S,-\sigma}=\frac{\Del_S}{2}
 -\frac{\kap_{S\sigma}}{2\kap_{S,-\sigma}}(2n+\Del_S)
 +\ii\frac{\sigma m}{\kap_{S,-\sigma}},
\end{equation}
\begin{equation}
 B_{S,-\sigma}=1-\frac{\Del_S}{2}
 -\frac{\kap_{S\sigma}}{2\kap_{S,-\sigma}}(2n+\Del_S)
 +\ii\frac{\sigma m}{\kap_{S,-\sigma}}.
\end{equation}
Using reflection as in System 3,
\begin{equation}
 \frac{\Gamma(A_{S,-\sigma})}{\Gamma(B_{S,-\sigma})}
 =\frac{\sin(\pi B_{S,-\sigma})}{\sin(\pi A_{S,-\sigma})}
 \left[\frac{\kap_{S\sigma}}{2\kap_{S,-\sigma}}(2n+\Del_S)\right]^{\Del_S-1}
 \left[1+O(n^{-1})\right].
\end{equation}
Also
\begin{equation}
 C_{S,n,\sigma}=\frac{2\ii\kap_{S\sigma}(\Del_S)_n}{n!\Gamma(1-\Del_S)}
 =2\ii\kap_{S\sigma}\frac{\sin(\pi\Del_S)}{\pi}
 n^{\Del_S-1}\left[1+O(n^{-1})\right].
\end{equation}
Therefore in the high-$n$ limit,
\begin{equation}
 R_{S,n,\sigma}=\mathcal R_{\sigma,n}^{\rm osc}\,
 n^{2\Del_S-2}\left[1+O(n^{-1})\right],
 \label{eq:case4-R-asymp}
\end{equation}
where the oscillatory prefactor is
\begin{equation}
\mathcal R_{\sigma,n}^{\rm osc}=
 -\nu(\Del_S)
 \left[2\ii\kap_{S\sigma}\frac{\sin(\pi\Del_S)}{\pi}\right]
 \left(\frac{\kap_{S\sigma}}{\kap_{S,-\sigma}}\right)^{\Del_S-1}
 \frac{\sin(\pi B_{S,-\sigma})}{\sin(\pi A_{S,-\sigma})}.
 \label{eq:case4-Rosc}
\end{equation}
The empty-AdS bath analytic value at the same pole obeys
\begin{equation}
 g_{n,\sigma}=\calE_{\Delta_B}\left(\frac{\kap_{S\sigma}(2n+\Del_S)}{2}\right)^{2\Delta_B-2}
 \left[1+O(n^{-1})\right].
\end{equation}
Thus
\begin{equation}
R_{S,n,\sigma}g_{n,\sigma}
 =\mathcal R_{\sigma,n}^{\rm osc}\,
 \calE_{\Delta_B}\kap_{S\sigma}^{2\Delta_B-2}
 n^{2(\Del_S+\Delta_B)-4}\left[1+O(n^{-1})\right].
 \label{eq:case4-Rg-asymp}
\end{equation}
The high-overtone gap shift is
\begin{equation}
\delta\gamma_{n,\sigma}
 =-h^2\operatorname{Im}\left[\mathcal R_{\sigma,n}^{\rm osc}\,
 \calE_{\Delta_B}\kap_{S\sigma}^{2\Delta_B-2}
 n^{2(\Del_S+\Delta_B)-4}\right]+\cdots.
 \label{eq:case4-highn-gamma}
\end{equation}
A true nonzero plateau requires
\begin{equation}
 \Del_S+\Delta_B=2
\end{equation}
and a non-oscillatory limit of the sine ratio in \eqref{eq:case4-Rosc}.  Otherwise the high-\(n\) answer is the analytic envelope \eqref{eq:case4-highn-gamma}, not a single number.

\subsubsection{Coincident chiral poles}
If the two chiral system poles coincide, the system pole is double.  This happens when
\begin{equation}
 m=0,
 \qquad
 \kap_{S+}(2n_++\Del_S)=\kap_{S-}(2n_-+\Del_S).
\end{equation}
Use the double-pole Laurent data
\begin{equation}
 G_S^R(\Om+x)=\frac{A_{S,2}}{x^2}+\frac{A_{S,1}}{x}+O(1).
\end{equation}
Then with
\begin{equation}
 g=G_{\text{AdS}}^R(\Om;\Delta_B),
 \qquad g'=\partial_\omega G_{\text{AdS}}^R(\Om;\Delta_B),
\end{equation}
\begin{equation}
 \omega_\pm=\Om\pm h\sqrt{A_{S,2}g}
 +\frac{h^2}{2}(A_{S,2}g'+A_{S,1}g)+O(h^3),
\end{equation}
and
\begin{equation}
 \gamma_\pm=-\operatorname{Im}\Om
 \mp h\operatorname{Im}\sqrt{A_{S,2}g}
 -\frac{h^2}{2}\operatorname{Im}(A_{S,2}g'+A_{S,1}g)+O(h^3).
\end{equation}
This contains the non-rotating \(m=J=0\) BTZ system as a special case.

\subsubsection{Spinning BTZ-AdS Plots}
\begin{figure}[H]
    \centering
    \begin{minipage}[t]{0.49\linewidth}
        \centering
        \includegraphics[width=\linewidth]{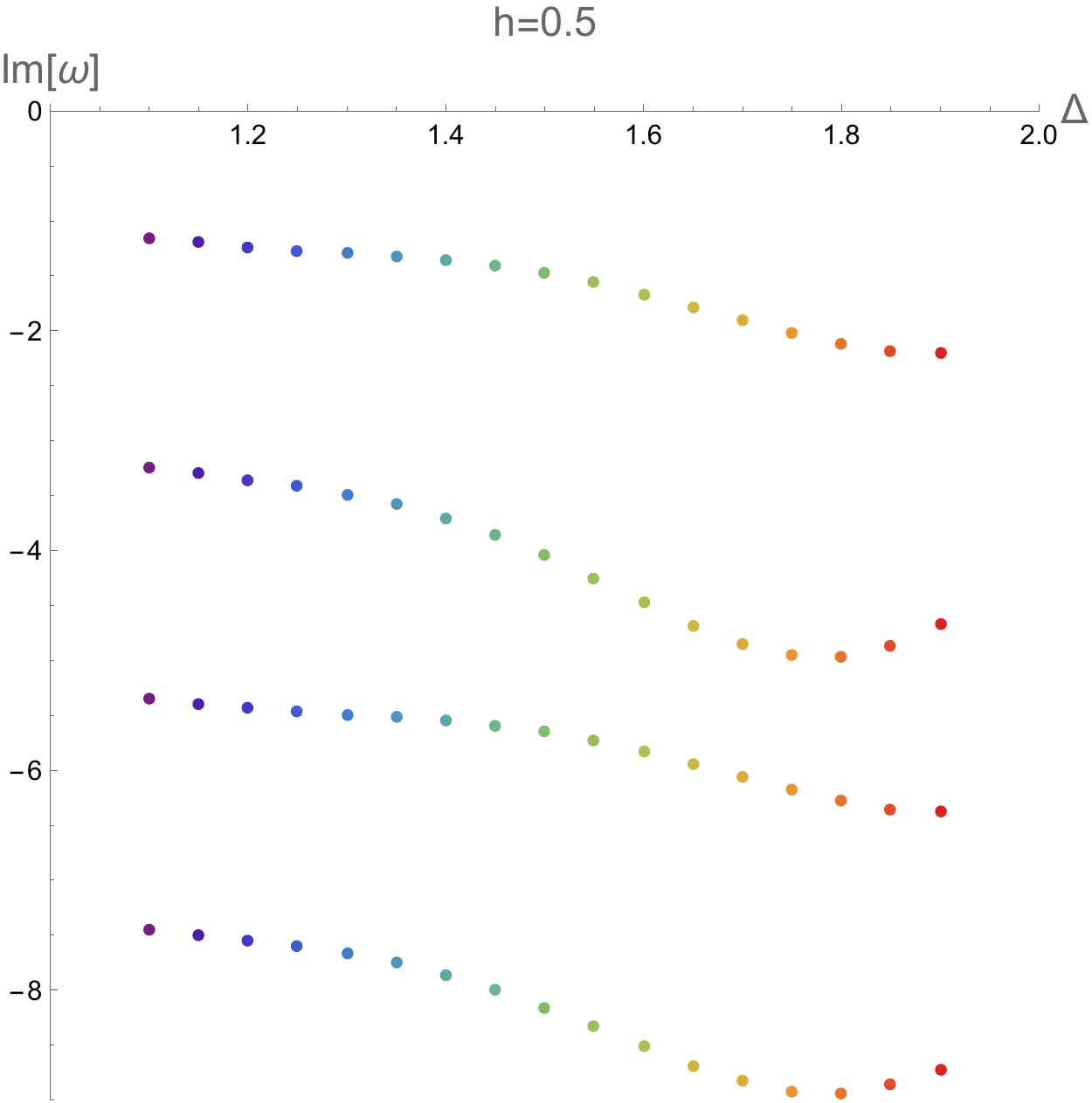}
        \par\smallskip
        {\small (a) Positive chirality}
    \end{minipage}\hfill
    \begin{minipage}[t]{0.49\linewidth}
        \centering
        \includegraphics[width=\linewidth]{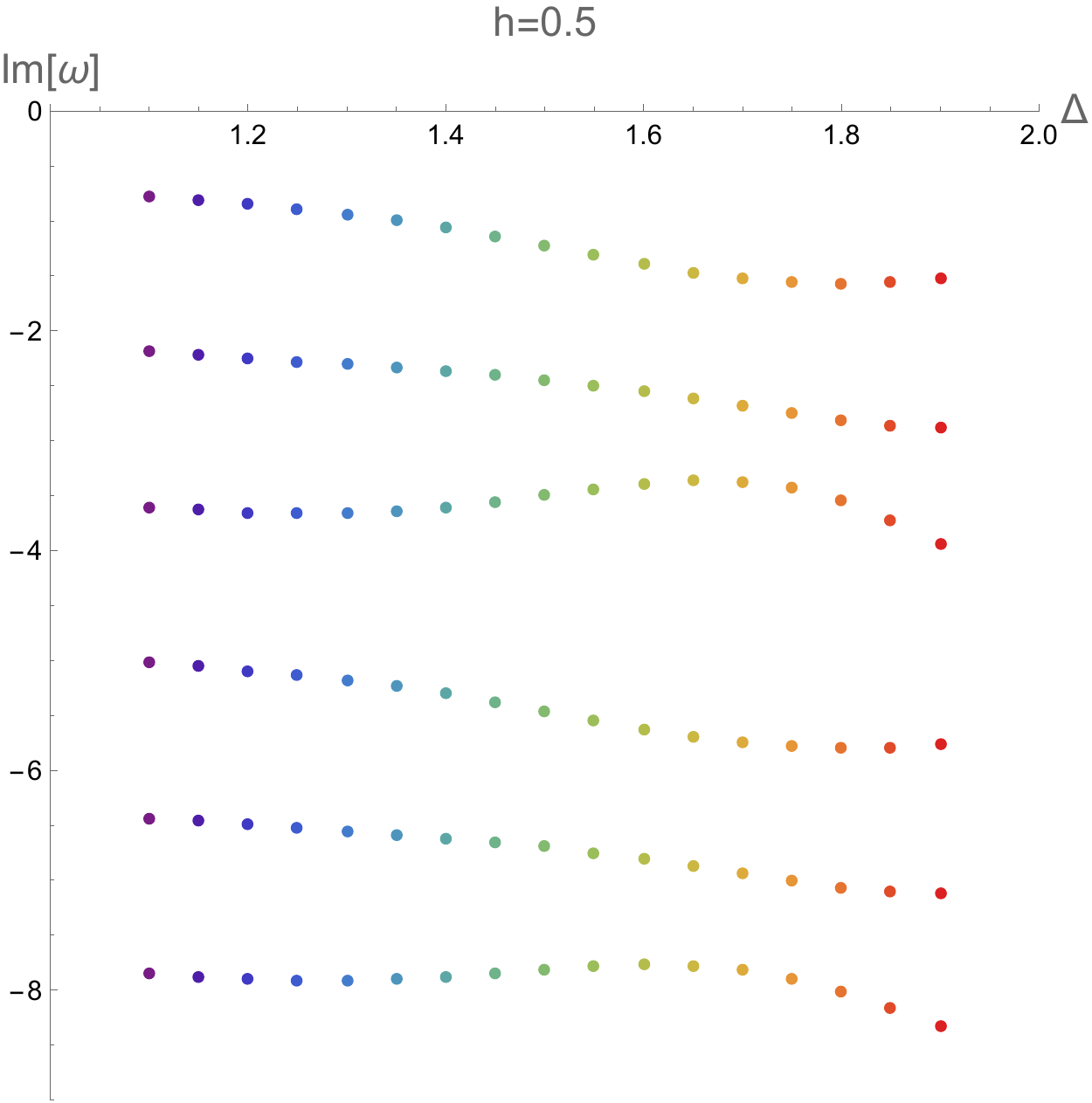}
        \par\smallskip
        {\small (b) Negative chirality}
    \end{minipage}

    \caption{Imaginary parts of the spinning BTZ modes versus $\Delta_S$
    at $h=0.5$, $M=0.8$, $J=0.3$, and $m=0$, with
    $\Delta_B=2-\Delta_S$. The pronounced bends in the second and fourth
    positive-chirality bands ($n_+=1,3$) correspond to those in the third
    and sixth negative-chirality bands ($n_-=2,5$). These paired modes
    approach the chiral coincidence condition
    $\kappa_{S+}(2n_++\Delta_S)
    =\kappa_{S-}(2n_-+\Delta_S)$.
    Their decreasing pole separation enhances the opposite-chirality
    factor in the residue $R_{S,n,\sigma}$.}
    \label{fig:spinning-btz-delta}
\end{figure}

\begin{figure}[H]
    \centering
    \begin{minipage}[t]{0.49\linewidth}
        \centering
        \includegraphics[width=\linewidth]{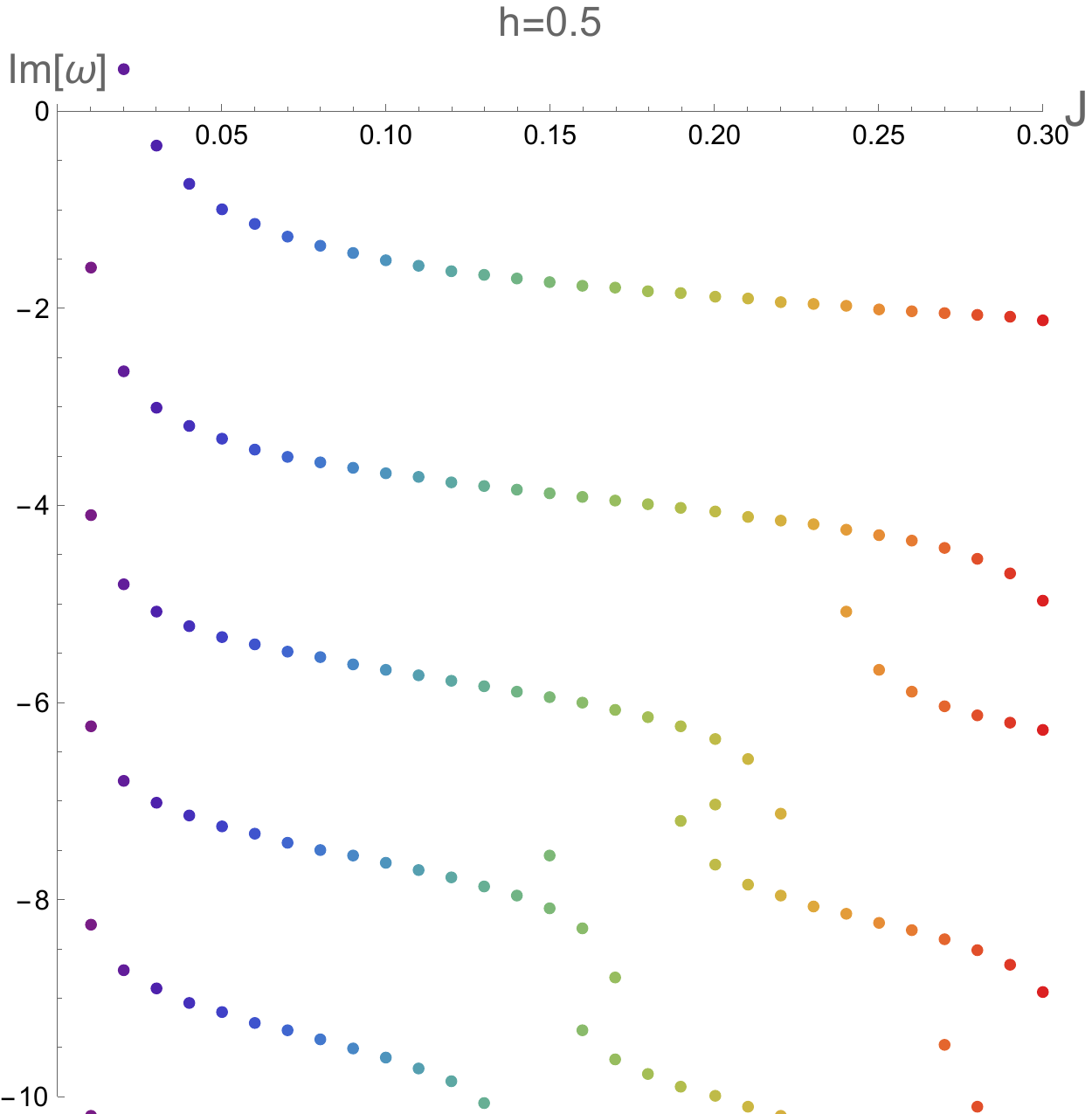}
        \par\smallskip
        {\small (a) Positive chirality}
    \end{minipage}\hfill
    \begin{minipage}[t]{0.49\linewidth}
        \centering
        \includegraphics[width=\linewidth]{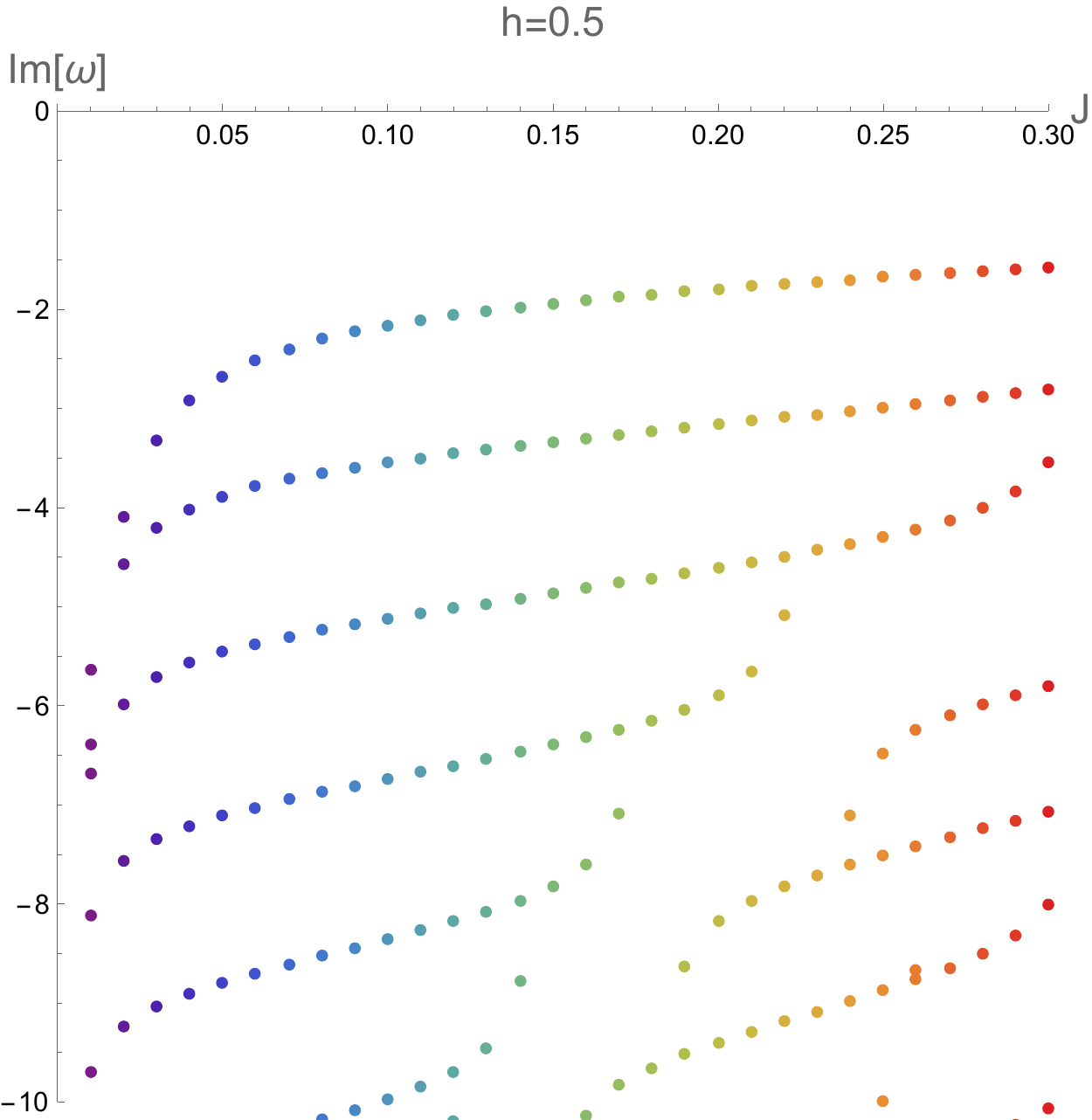}
        \par\smallskip
        {\small (b) Negative chirality}
    \end{minipage}

    \caption{Imaginary parts of the spinning BTZ modes versus $J$ at $h=0.5$, $M=0.8$, $m=0$, and fixed $\Delta_S=1.8$($\Delta_B=0.2$). The positive-chirality poles move downward as $\kappa_{S+}=\sqrt{M+J}$ increases, whereas the negative-chirality poles move upward as $\kappa_{S-}=\sqrt{M-J}$ decreases. The sharp higher-overtone excursions arise when poles from opposite chiral sectors approach coincidence, enhancing the residue correction.}
    \label{fig:spinning-btz-J}
\end{figure}

\subsection{General BTZ system and general BTZ bath}
For each side \(a=S,B\), define
\begin{equation}
 \kap_{a\sigma}=r_{a+}+\sigma r_{a-},
\end{equation}
and the chiral pole set at frequency \(\Om\):
\begin{equation}
 \calC_a(\Om)=\left\{(\sigma,n):\Om=-\sigma m-\ii\kap_{a\sigma}(2n+\Del_a),\quad n\in\mathbb Z_{\ge0},\quad \sigma=\pm\right\}.
\end{equation}
The pole order is
\begin{equation}
 p_a(\Om)=|\calC_a(\Om)|\in\{0,1,2\}.
\end{equation}
At a system pole,
\begin{equation}
 P=p_S(\Om),
 \qquad Q=p_B(\Om),
 \qquad N=P+Q.
\end{equation}
Construct Laurent data
\begin{equation}
 G_S^R(\Om+x)=\frac{S_0+S_1x+\cdots}{x^P},
 \qquad
 G_B^R(\Om+x)=\frac{B_0+B_1x+\cdots}{x^Q}.
\end{equation}
Then all local branches are
\begin{equation}
\omega_j=\Om+h^{2/N}q_j
 +h^{4/N}\frac{S_0B_1+S_1B_0}{Nq_j^{N-2}}+O(h^{6/N}),
 \qquad q_j^N=S_0B_0.
 \label{eq:case5-general-omega}
\end{equation}
The damping and real-gap expressions are
\begin{equation}
\gamma_j=-\operatorname{Im}\Om
 -h^{2/N}\operatorname{Im}q_j
 -h^{4/N}\operatorname{Im}\left(\frac{S_0B_1+S_1B_0}{Nq_j^{N-2}}\right)+\cdots,
 \label{eq:case5-general-gamma}
\end{equation}
\begin{equation}
 \operatorname{Re}\omega_j=\operatorname{Re}\Om
 +h^{2/N}\operatorname{Re}q_j
 +h^{4/N}\operatorname{Re}\left(\frac{S_0B_1+S_1B_0}{Nq_j^{N-2}}\right)+\cdots.
\end{equation}
These two equations work for every BTZ-BTZ case: spinning-spinning, spinning-non-spinning, equal masses, unequal masses, resonances, and chiral degeneracy.

\subsubsection{How to build the Residues}

For a singular chiral sector, write
\begin{equation}
 F_{a\sigma}=\frac{\Gamma(A_{a\sigma})}{\Gamma(B_{a\sigma})}.
\end{equation}
If \((\sigma,n)\in\calC_a(\Om)\), then
\begin{equation}
 F_{a\sigma}(\Om+x)=\frac{C_{a,n,\sigma}}{x}+D_{a,n,\sigma}+O(x),
\end{equation}
where
\begin{equation}
 C_{a,n,\sigma}=\frac{2\ii\kap_{a\sigma}(-1)^n}{n!\Gamma(1-\Del_a-n)},
\end{equation}
\begin{equation}
 D_{a,n,\sigma}=\frac{(-1)^n}{n!\Gamma(1-\Del_a-n)}
 \left[\psi(n+1)-\psi(1-\Del_a-n)\right].
\end{equation}
If only one sector is singular, the residue is
\begin{equation}
 S_{a,0}=R_{a,n,\sigma}=-\nu(\Del_a)C_{a,n,\sigma}F_{a,-\sigma}(\Om).
\end{equation}
If both sectors are singular, the double-pole data are
\begin{equation}
S_{a,0}=A_{a,2}=-\nu(\Del_a)C_{a,+}C_{a,-},
\end{equation}
\begin{equation}
 S_{a,1}=A_{a,1}=-\nu(\Del_a)(C_{a,+}D_{a,-}+C_{a,-}D_{a,+}).
\end{equation}
If the side is analytic at \(\Om\), then
\begin{equation}
 S_{a,0}=G_a^R(\Om),
 \qquad S_{a,1}=\partial_\omega G_a^R(\Om).
\end{equation}
The derivative of a BTZ Green function is
\begin{equation}
 \frac{\partial_\omega G_{\BTZ}^R}{G_{\BTZ}^R}
 =-\ii\sum_{\sigma=\pm}\frac{1}{2\kap_\sigma}
 \left[\psi(A_\sigma)-\psi(B_\sigma)\right].
\end{equation}

\subsubsection{High-$n$/UV general scaling}

For a BTZ pole family with overtone \(n\), both the simple-pole residue and the double-pole principal coefficient scale, up to known oscillatory sine ratios in generic spinning cases, as
\begin{equation}
 S_0\sim n^{2\Del_S-2}.
\end{equation}
If the bath is empty AdS and analytic at the system pole, then
\begin{equation}
 B_0\sim n^{2\Delta_B-2}.
\end{equation}
If the bath is BTZ and analytic at the system pole, then away from resonances
\begin{equation}
 B_0\sim \mathcal S_n^2 n^{2\Del_B-2},
\end{equation}
where \(\mathcal S_n\) is an explicitly known sine ratio such as \eqref{eq:SnB}.  If the bath has a colliding BTZ pole, its Laurent principal coefficient also scales as
\begin{equation}
 B_0\sim n_B^{2\Del_B-2}.
\end{equation}
Therefore the leading branch shift scales as
\begin{equation}
 x_j^{(0)}\sim h^{2/N}n^{\frac{2(\Delta_S+\Delta_B)-4}{N}},
 \label{eq:master-scaling}
\end{equation}
 For resonant BTZ bath poles, \(n_B\) must be related to \(n\) by the resonance condition.

This scaling immediately explains the different high-$n$ behaviors:
\begin{itemize}[leftmargin=2em]
\item Empty AdS system + BTZ bath: \(N=1\) in the local formula, but the residue and spectral density combine with complementary powers to give the plateau \eqref{eq:case1-plateau}.
\item Non-rotating BTZ system + empty AdS bath: \(N=2\).  The leading split is order \(h\), and it has a plateau if \(\Del_S+\Delta_B=2\).
\item Non-resonant BTZ--BTZ: \(N=2\).  The envelope is a plateau if \(\Del_S+\Del_B=2\), but a pointwise limit can be spoiled by the sine ratio.
\item Double--double BTZ resonance: \(N=4\).  For identical BTZs the leading split scales as \(h^{1/2}n^{\Del-1}\), so there is no plateau.
\item Generic spinning BTZ + empty AdS: \(N=1\) for simple chiral poles; the gap shift scales as \(h^2n^{2(\Del_S+\Delta_B)-4}\) up to oscillatory factors.
\end{itemize}

\section{Non-Perturbative Optical Depth Law for AdS-BTZ System}
\label{sec:optical_depth}
In this final section, we proceed to derive an expression that relates transmission-dissipation of the AdS system and the BTZ bath, and allows the system to be reinterpreted as an optical depth law equation in the UV limit. We first show how the Dissipation-Transmission connection appears in the weak-coupling regime using the expressions we derived previously, and then we will proceed to derive the non-perturbative version of this expression without relying on perturbation theory.

\subsection{Weak-Coupling Relation of Dissipation -  Transmission}
For the AdS-BTZ system, we found the following expressions
\begin{equation}
\Gamma_{UV}=\lim_{n\to\infty}\frac{\Gamma_n}{h^2}
  =\frac{16}{\pi}(\Delta-1)^2\sin^2(\pi\Delta)M^{1-\Delta}.
  \label{eq:width-plateau-system1}
\end{equation}
The dissipation plateau is just half of this since $\Gamma/2=\gamma$
\begin{equation}
  \lim_{n\to\infty}\frac{\gamma_n}{h^2}
  =\frac{8}{\pi}(\Delta-1)^2\sin^2(\pi\Delta)M^{1-\Delta}.
  \label{eq:gamma-plateau-system1}
\end{equation}
For convenience later, let's define here the quantity \textit{Relaxation Transmission Coefficient} as
\begin{equation}
    \chi^{UV}_{relax}\equiv\pi M^{\Delta-1}\lim_{n\rightarrow\infty}\frac{\Gamma_n}{h^2}=\pi M^{\Delta-1}\Gamma_{UV}=16(\Delta-1)^2\sin^2(\pi\Delta)
\end{equation}
In \cite{Karch:2025hof} the \textit{energy transmission coefficient} for a scalar crossing the transparent boundary is
\begin{equation}
  |t|^2=\frac{4h^2(d-2\Delta_L)^2\sin^2[\pi(d/2-\Delta_L)]}{\left[1+h^2(d-2\Delta_L)^2\right]^2}.
  \label{eq:KY-trans-general}
\end{equation}
In our \(d=2\) marginal setup, the empty-AdS side uses
\begin{equation}
  \Delta_L=2-\Delta,
\end{equation}
so
\begin{equation}
  d-2\Delta_L=2-2(2-\Delta)=2(\Delta-1),
\end{equation}
and
\begin{equation}
  \sin[\pi(d/2-\Delta_L)]=\sin[\pi(1-(2-\Delta))]
  =\sin[\pi(\Delta-1)].
\end{equation}
Therefore
\begin{equation}
|t|^2=\frac{16h^2(\Delta-1)^2\sin^2[\pi(\Delta-1)]}
  {\left[1+4h^2(\Delta-1)^2\right]^2}.
  \label{eq:trans-d2}
\end{equation}
Expanding at small \(h\),
\begin{equation}
  |t|^2=h^2\,16(\Delta-1)^2\sin^2[\pi(\Delta-1)]+O(h^4).
\end{equation}
Because \(\sin^2[\pi(\Delta-1)]=\sin^2(\pi\Delta)\), 
\begin{equation}
  |t|^2=h^2\,16(\Delta-1)^2\sin^2(\pi\Delta)+O(h^4).
\end{equation}
Compare with our expression we found above
\begin{equation}
 \chi_{\rm relax}^{UV}=\lim_{h\to0}\frac{|t|^2}{h^2}.
  \label{eq:chi-trans-relation}
\end{equation}
Equivalently,
\begin{equation}
  \frac{\Gamma_{UV}}{h^2}=\frac{1}{\pi}M^{1-\Delta}\lim_{h\to0}\frac{|t|^2}{h^2}+O(h^4).
\end{equation}
The high-n/UV Dissipation plateau is exactly the weak-coupling Transmission coefficient obtained from bulk energy flux multiplied by a mass factor. The lesson is \textbf{weak coupling transmission is related to high-frequency dissipation.} For completeness, another interesting relation is
\begin{equation}
   \chi_{\rm relax}^{UV}=\frac{\dd}{\dd(h^2)}|t|^2\bigg|_{h=0},
\end{equation}

\subsection{The General Weak-coupling Dissipation Plateau}
 Before tackling the  problem at any coupling, it is useful to derive the Dissipation Width plateau for the AdS-General BTZ system at weak coupling.
\\\\
For a simple real system pole with residue \(R_n=-Z_n\), the open pole for the positive branch is 
\begin{equation}
        \omega_n=\Om_n-h^2 Z_n G^R_B(\Om_n)+O(h^4).
\end{equation}
Writing
\begin{equation}       \omega_n=\Om_n+\delta\Om_n-{i\over2}\Gamma_n,
\end{equation}
one gets the Fermi's golden rule expressions
\begin{equation}
        \delta\Om_n=-h^2Z_n\Re G^R_B(\Om_n),
        \qquad
        \Gamma_n=h^2Z_n\rho_B(\Om_n).
        \label{eq:weakGamma}
\end{equation}
Now, we do simplifications in the residue and green's function. Use Stirling's formula in the form
\begin{equation}
        {\Gamma(n+a)\over\Gamma(n+b)}\sim n^{a-b},
        \qquad n\to\infty.
\end{equation}
From \eqref{eq:case1-Z},
\begin{equation}
        Z_n
        =4(\Delta-1)^2\left[{\Gamma(n+2-\Delta)
        \over \Gamma(2-\Delta)\Gamma(n+1)}\right]^2
        \sim {4(\Delta-1)^2\over\Gamma(2-\Delta)^2}n^{2-2\Delta}.
        \label{eq:Zasymp}
\end{equation}
For large real \(\omega\) at fixed \(m\), use
\begin{equation}
        |\Gamma(a+iy)|^2\sim 2\pi e^{-\pi y}y^{2a-1},
        \qquad y\to+\infty.
\end{equation}
Since \(y_\pm\to+\infty\),
\begin{align}
        \rho_{\BTZ}(\omega,m)
        &\simeq -\nu_\Delta {\sin(\pi\Delta)\over\pi^2}
        \left({1\over2}e^{\pi(y_++y_-)}\right)
        (2\pi)^2e^{-\pi(y_++y_-)}(y_+y_-)^{\Delta-1}
        \\
        &=-2\nu_\Delta\sin(\pi\Delta)(y_+y_-)^{\Delta-1}.
\end{align}
Using
\begin{equation}
        y_+y_-={\omega^2-m^2\over4\kap_+\kap_-},
\end{equation}
and fixed \(m\),
\begin{equation}
        \rho_{\BTZ}(\Om_n,m)
        \simeq -2\nu_\Delta\sin(\pi\Delta)
        {n^{2\Delta-2}\over K^{\Delta-1}}.
        \label{eq:rhoasymp}
\end{equation}
Therefore
\begin{align}
        \lim_{n\to\infty}{\Gamma_n\over h^2}
        &= {4(\Delta-1)^2\over\Gamma(2-\Delta)^2}
        \left[-2\nu_\Delta\sin(\pi\Delta)K^{1-\Delta}\right]
        \\
        &= -{8(\Delta-1)^2\nu_\Delta\sin(\pi\Delta)\over\Gamma(2-\Delta)^2}K^{1-\Delta}.
\end{align}
Substituting
\begin{equation}
        \nu_\Delta=2(\Delta-1){\Gamma(2-\Delta)\over\Gamma(\Delta)}
\end{equation}
gives
\begin{equation}
        \lim_{n\to\infty}{\Gamma_n\over h^2}
        =-{16(\Delta-1)^3\sin(\pi\Delta)
        \over\Gamma(\Delta)\Gamma(2-\Delta)}K^{1-\Delta}.
\end{equation}
Finally,
\begin{equation}
        \Gamma(\Delta)\Gamma(2-\Delta)
        =(1-\Delta){\pi\over\sin(\pi\Delta)},
\end{equation}
so
\begin{equation}
        \;
        \lim_{n\to\infty}{\Gamma_n\over h^2}
        ={16\over\pi}(\Delta-1)^2\sin^2(\pi\Delta)K^{1-\Delta}\;  .
        \label{eq:weakspinplateau}
\end{equation}
For \(J=0\), \(K=M\), and this reduces to the previous non-rotating plateau. The same BTZ asymptotics gives the relation between real part and imaginary we showed in previous notes
\begin{equation}
        {\Re G^R_{\BTZ}(\omega,m)\over\Im G^R_{\BTZ}(\omega,m)}
        \longrightarrow -\cot(\pi\Delta),
        \qquad \omega\to+\infty.
\end{equation}
Since \(\rho=2\Im G\), the weak-coupling real shift obeys
\begin{equation}
        \lim_{n\to\infty}{\delta\Om_n\over h^2}
        ={1\over2}\cot(\pi\Delta)
        \lim_{n\to\infty}{\Gamma_n\over h^2}.
\end{equation}
Therefore
\begin{equation}
        \;
        \lim_{n\to\infty}{\delta\Om_n\over h^2}
        ={8\over\pi}(\Delta-1)^2\sin(\pi\Delta)\cos(\pi\Delta)K^{1-\Delta}\; 
        \label{eq:weakrealplateau}
\end{equation}

\subsection{The Optical Depth Theorem for AdS-BTZ}

Previously, in the perturbative weak-coupling calculation, we expanded each Green's function in the open pole equation before taking the UV limit. For the non-perturbative calculation, we take first the UV limit of the open pole equation, and solve an algebraic equation that encodes the information about the complex shift on the original poles.  The result is finite in \(h\) and reproduces the perturbative result in the weak coupling regime.  
\\\\
We start with a general form for the shift in the AdS poles, with $s$ being an complex number, and our second main formula will be the open pole master equation: 
\begin{equation}
        \omega=\Om_n+s=2-\Delta+2n+s,
\end{equation}
\begin{equation}
    1=h^2G^R_BG^R_S
\end{equation}
Substitute \(\delta=2-\Delta\) into the AdS Green's function:
\begin{equation}
        G^R_{\AdS}
        =C_E
        {\Gamma\!\bigl({2-\Delta-\omega\over2}\bigr)
         \Gamma\!\bigl({2-\Delta+\omega\over2}\bigr)
        \over
         \Gamma\!\bigl({\Delta-\omega\over2}\bigr)
         \Gamma\!\bigl({\Delta+\omega\over2}\bigr)} ,
        \label{eq:GEstart}
\end{equation}
where
\begin{equation}
        C_E=(2(2-\Delta)-2){\Gamma(\Delta-1)\over\Gamma(1-\Delta)}
        =-2(\Delta-1){\Gamma(\Delta-1)\over\Gamma(1-\Delta)} .
\end{equation}
For \(\omega=2-\Delta+2n+s\), the four arguments are
\begin{align}
        {2-\Delta-\omega\over2}&=-n-{s\over2},
        \\
        {2-\Delta+\omega\over2}&=n+2-\Delta+{s\over2},
        \\
        {\Delta-\omega\over2}&=\Delta-1-n-{s\over2},
        \\
        {\Delta+\omega\over2}&=n+1+{s\over2}.
\end{align}
Therefore
\begin{equation}
        G^R_{\rm AdS}=C_E
        {\Gamma(-n-s/2)\Gamma(n+2-\Delta+s/2)
        \over
        \Gamma(\Delta-1-n-s/2)\Gamma(n+1+s/2)}.
        \label{eq:GEfourargs}
\end{equation}
Use
\begin{equation}
        \Gamma(z)={\pi\over\sin(\pi z)\Gamma(1-z)}.
\end{equation}
First,
\begin{align}
        \Gamma(-n-s/2)
        &={\pi\over \sin[-\pi n-\pi s/2] \Gamma(n+1+s/2)}
        \\
        &={(-1)^{n+1}\pi\over \sin(\pi s/2)\Gamma(n+1+s/2)}.
\end{align}
Second, with
\begin{equation}
        A\equiv \pi(\Delta-1),
\end{equation}
we have
\begin{align}
        \Gamma(\Delta-1-n-s/2)
        &={\pi\over \sin[A-\pi n-\pi s/2]\Gamma(n+2-\Delta+s/2)}
        \\
        &={(-1)^n\pi\over \sin[A-\pi s/2]\Gamma(n+2-\Delta+s/2)}.
\end{align}
Thus
\begin{equation}
        {\Gamma(-n-s/2)\over\Gamma(\Delta-1-n-s/2)}
        =-{\sin[A-\pi s/2]\over\sin(\pi s/2)}
        {\Gamma(n+2-\Delta+s/2)\over\Gamma(n+1+s/2)}.
\end{equation}
Substitution into \eqref{eq:GEfourargs} gives the exact rearrangement
\begin{equation}
        G^R_{\AdS}(\Om_n+s;2-\Delta)
        =-C_E {\sin[A-\pi s/2]\over\sin(\pi s/2)}
        \left[{\Gamma(n+2-\Delta+s/2)\over\Gamma(n+1+s/2)}\right]^2.
        \label{eq:GEuniformExact}
\end{equation}
Now use the gamma-ratio asymptotic uniformly for fixed \(s\) and large-n:
\begin{equation}
        {\Gamma(n+2-\Delta+s/2)\over\Gamma(n+1+s/2)}
        =n^{1-\Delta}\left[1+O(n^{-1})\right].
\end{equation}
Hence
\begin{equation}
        \;
        G^R_{\AdS}(\Om_n+s;2-\Delta)
        =2(\Delta-1){\Gamma(\Delta-1)\over\Gamma(1-\Delta)}
        n^{2-2\Delta}{\sin[A-\pi s/2]\over\sin(\pi s/2)}
        \left[1+O(n^{-1})\right]\; .
        \label{eq:GEuniform}
\end{equation}
This formula is a great improvement over the simple residue approximation.  The simple-residue approximation is recovered by taking \(s\to0\), where \(\sin(\pi s/2)\sim\pi s/2\).  We keep the full s shift.
\\\\
Next, we need \(G^R_{\BTZ}(\Om_n+s,m)\) at large positive real part.  From \eqref{eq:BTZ-G}
\begin{equation}
        G^R_{\BTZ}=-\nu_\Delta\prod_{\sigma=\pm}{\Gamma(\Delta/2-iy_\sigma)
        \over \Gamma(1-\Delta/2-iy_\sigma)}.
\end{equation}
Let
\begin{equation}
        z_\sigma=-iy_\sigma.
\end{equation}
and 
\begin{equation}
    K^2=M^2-J^2
\end{equation}
For fixed \(m\), fixed \(s\), and \(n\to\infty\), one has \(|z_\sigma|\to\infty\).  The ratio asymptotic gives
\begin{equation}
        {\Gamma(z_\sigma+\Delta/2)\over\Gamma(z_\sigma+1-\Delta/2)}
        =z_\sigma^{\Delta-1}\left[1+O(n^{-1})\right].
\end{equation}
Thus
\begin{equation}
        G^R_{\BTZ}(\omega,m)
        =-\nu_\Delta\prod_{\sigma=\pm}(-iy_\sigma)^{\Delta-1}
        \left[1+O(n^{-1})\right].
\end{equation}
For positive large \(\omega\), we get
\begin{equation}
        (-iy_\sigma)^{\Delta-1}=e^{-i\pi(\Delta-1)/2}y_\sigma^{\Delta-1}
        \left[1+O(n^{-1})\right].
\end{equation}
Therefore
\begin{equation}
        G^R_{\BTZ}(\omega,m)
        =-\nu_\Delta e^{-i\pi(\Delta-1)}(y_+y_-)^{\Delta-1}
        \left[1+O(n^{-1})\right].
\end{equation}
Since \(-e^{-i\pi(\Delta-1)}=e^{-i\pi\Delta}\), this is
\begin{equation}
        G^R_{\BTZ}(\omega,m)
        =\nu_\Delta e^{-i\pi\Delta}(y_+y_-)^{\Delta-1}
        \left[1+O(n^{-1})\right].
        \label{eq:GBTZasymp1}
\end{equation}
At fixed \(m\),
\begin{equation}
        y_+y_-={\omega^2-m^2\over4K}
        ={(2n)^2\over4K}\left[1+O(n^{-1})\right]
        ={n^2\over K}\left[1+O(n^{-1})\right].
\end{equation}
Therefore
\begin{equation}
        \;
        G^R_{\BTZ}(\Om_n+s,m)
        =\nu_\Delta e^{-i\pi\Delta}{n^{2\Delta-2}\over K^{\Delta-1}}
        \left[1+O(n^{-1})\right]\;  .
        \label{eq:GBTZuniform}
\end{equation}
Multiply \eqref{eq:GEuniform} and \eqref{eq:GBTZuniform}.  The powers of \(n\) cancel. \textbf{This signals that marginal coupling is highly relevant in our interpretation. A non-marginal version might be interesting!}
\begin{equation}
        n^{2-2\Delta}n^{2\Delta-2}=1.
\end{equation}
The constant multiplying all the expression is
\begin{align}
        C_{\rm prod}
        &=2(\Delta-1){\Gamma(\Delta-1)\over\Gamma(1-\Delta)}
        \nu_\Delta e^{-i\pi\Delta}K^{1-\Delta}
        \\
        &=2(\Delta-1){\Gamma(\Delta-1)\over\Gamma(1-\Delta)}
        \left[2(\Delta-1){\Gamma(2-\Delta)\over\Gamma(\Delta)}\right]
        e^{-i\pi\Delta}K^{1-\Delta}.
\end{align}
Use
\begin{equation}
        \Gamma(2-\Delta)=(1-\Delta)\Gamma(1-\Delta),
        \qquad
        \Gamma(\Delta)=(\Delta-1)\Gamma(\Delta-1).
\end{equation}
Then
\begin{equation}
        C_{\rm prod}
        =-4(\Delta-1)^2 e^{-i\pi\Delta}K^{1-\Delta}.
\end{equation}
Since
\begin{equation}
        -e^{-i\pi\Delta}=e^{-i\pi(\Delta-1)}=e^{-iA},
\end{equation}
one may write
\begin{equation}
        G^R_{\AdS}(\Om_n+s)G^R_{\BTZ}(\Om_n+s,m)
        =4(\Delta-1)^2K^{1-\Delta}e^{-iA}
        {\sin(A-\pi s/2)\over\sin(\pi s/2)}
        \left[1+O(n^{-1})\right].
\end{equation}
Define
\begin{equation}
        \lambda^2\equiv 4h^2(\Delta-1)^2K^{1-\Delta}.
        \label{eq:lambdaDefinition}
\end{equation}
The full pole equation becomes, to leading order in \(n\),
\begin{equation}
        \;
        \lambda^2 e^{-iA}{\sin(A-\pi s/2)\over\sin(\pi s/2)}=1\;  .
        \label{eq:trigPole}
\end{equation}
This is the AdS-BTZ high-$n$ pole equation.  All detailed gamma functions and all plateau powers disappeared.  The only surviving quantities are \(\Delta\), the effective dimensionless coupling \(\lambda\), and the BTZ scale \(K\), which is $K=\sqrt{M^2-J^2}$.

In solving this equation, let
\begin{equation}
        \theta={\pi s\over2},
        \qquad
        u=e^{i\theta}.
\end{equation}
Equation \eqref{eq:trigPole} is
\begin{equation}
        \sin\theta=\lambda^2 e^{-iA}\sin(A-\theta).
        \label{eq:sinEq}
\end{equation}
Now
\begin{equation}
        \sin\theta={u-u^{-1}\over2i},
\end{equation}
and
\begin{equation}
        \sin(A-\theta)={e^{iA}u^{-1}-e^{-iA}u\over2i}.
\end{equation}
Substituting into \eqref{eq:sinEq},
\begin{equation}
        u-u^{-1}=\lambda^2 e^{-iA}\left(e^{iA}u^{-1}-e^{-iA}u\right)
        =\lambda^2\left(u^{-1}-e^{-2iA}u\right).
\end{equation}
Multiplying by \(u\),
\begin{equation}
        u^2-1=\lambda^2\left(1-e^{-2iA}u^2\right).
\end{equation}
Thus
\begin{equation}
        u^2\left(1+\lambda^2e^{-2iA}\right)=1+\lambda^2.
\end{equation}
Since \(u^2=e^{2i\theta}=e^{i\pi s}\),
\begin{equation}
        \;
        e^{i\pi s_{\rm UV}}
        ={1+\lambda^2\over 1+\lambda^2e^{-2iA}}\;,
        \qquad A=\pi(\Delta-1).
\end{equation}
To extract the damping width and real shift, define
\begin{equation}
        R\equiv {1+\lambda^2\over 1+\lambda^2e^{-2iA}}.
\end{equation}
Let
\begin{equation}
        s_{\rm UV}=\delta_{\rm UV}-{i\over2}\Gamma_{\rm UV}.
\end{equation}
Then
\begin{equation}
        e^{i\pi s_{\rm UV}}=e^{i\pi\delta_{\rm UV}+\pi\Gamma_{\rm UV}/2}.
\end{equation}
Hence
\begin{equation}
        {\pi\Gamma_{\rm UV}\over2}=\log|R|,
        \qquad
        \pi\delta_{\rm UV}=\arg R.
\end{equation}
Because \(1+\lambda^2>0\),
\begin{equation}
        |R|^2={(1+\lambda^2)^2\over |1+\lambda^2e^{-2iA}|^2}
        ={(1+\lambda^2)^2\over1+\lambda^4+2\lambda^2\cos(2A)}.
\end{equation}
Therefore
\begin{equation}
        \Gamma_{\rm UV}={1\over\pi}
        \log{(1+\lambda^2)^2\over1+\lambda^4+2\lambda^2\cos(2A)}.
        \label{eq:GammaRaw}
\end{equation}
Also,
\begin{equation}
        1+\lambda^4+2\lambda^2\cos(2A)
        =(1+\lambda^2)^2-4\lambda^2\sin^2 A.
\end{equation}
Define
\begin{equation}
        \mathcal{T}_{\rm UV}={4\lambda^2\sin^2A\over(1+\lambda^2)^2}
\end{equation}
Then
\begin{equation}
        {1+\lambda^4+2\lambda^2\cos(2A)\over(1+\lambda^2)^2}
        =1-\cal{T}_{\rm UV},
\end{equation}
and \eqref{eq:GammaRaw} becomes
\begin{equation}
        \;
        \pi\Gamma_{\rm UV}=-\log(1-\cal{T}_{\rm UV})\;  .
\end{equation}
For the real shift,
\begin{equation}
        \delta_{\rm UV}=-{1\over\pi}\arg\left(1+\lambda^2e^{-2iA}\right).
\end{equation}
Since
\begin{equation}
        1+\lambda^2e^{-2iA}
        =1+\lambda^2\cos(2A)-i\lambda^2\sin(2A),
\end{equation}
we get the real shift
\begin{equation}
        \;
        \delta_{\rm UV}={1\over\pi}
        \operatorname{atan2}\!\bigl( \lambda^2\sin(2A),\,1+\lambda^2\cos(2A)\bigr)\;  .
\end{equation}
where, 
\[
\operatorname{atan2}(y,x)\equiv \rm arg(x+i y)\in(-\pi,\pi]
=
\begin{cases}
\arctan\!\left(\dfrac{y}{x}\right), & x>0,\\[6pt]
\arctan\!\left(\dfrac{y}{x}\right)+\pi, & x<0,\; y\ge 0,\\[6pt]
\arctan\!\left(\dfrac{y}{x}\right)-\pi, & x<0,\; y<0,\\[6pt]
+\dfrac{\pi}{2}, & x=0,\; y>0,\\[6pt]
-\dfrac{\pi}{2}, & x=0,\; y<0.
\end{cases}
\]

\subsection{Interpretation and Open Holographic AdS-BTZ Optical Theorem}

\begin{figure}[H]
  \centering
  \includegraphics[width=1\textwidth]{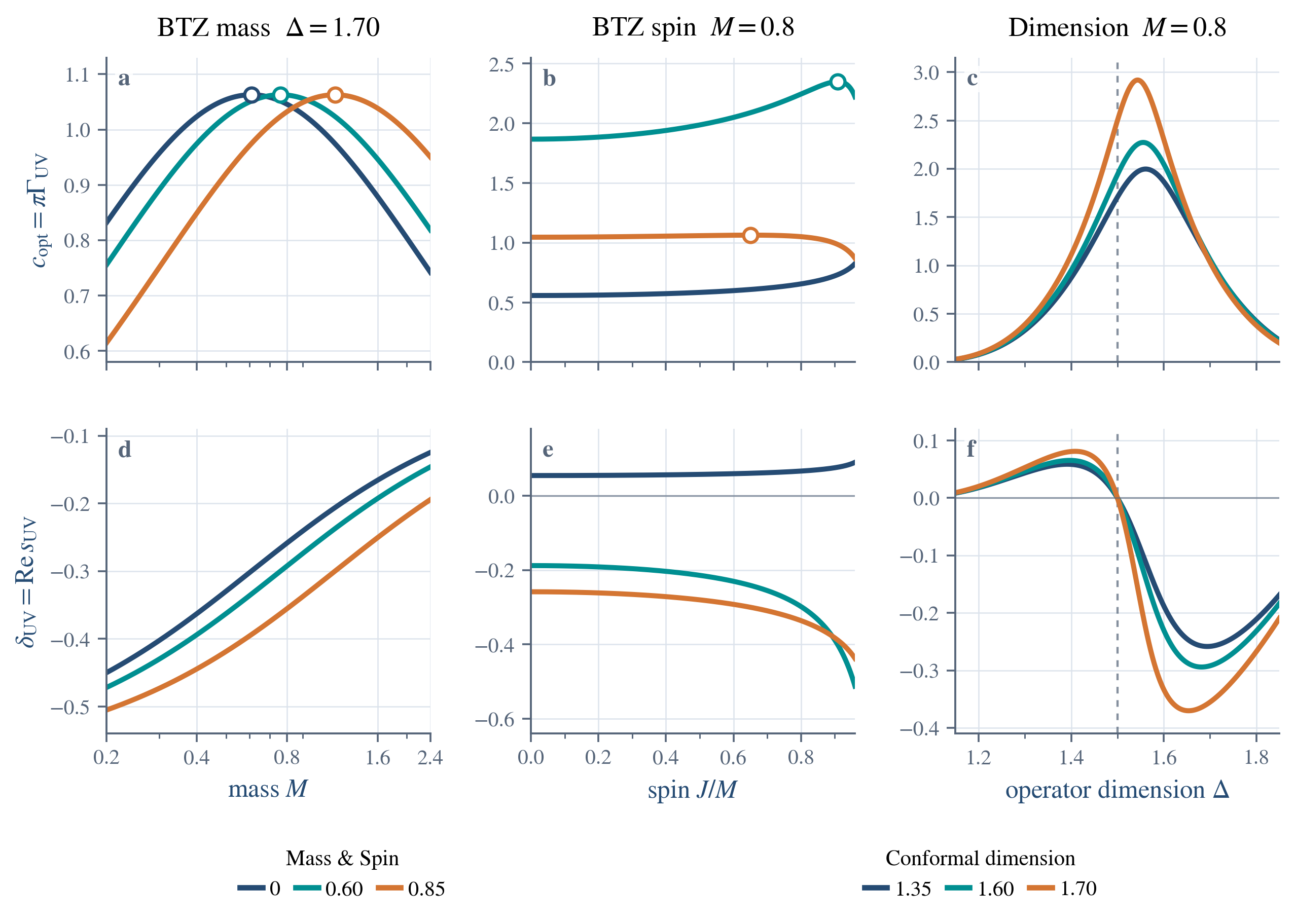}
  \caption{Combined parameter atlas for the Sec.~6 ultraviolet optical-depth
  law, with $h=0.6$. The upper panels give $c_{\mathrm{opt}}=\pi\Gamma_{\mathrm{UV}}=-\ln(1-T_{\mathrm{UV}})$; the lower panels give the principal real shift $\delta_{\mathrm{UV}}=\mathrm{Re}\,s_{\mathrm{UV}}$. Columns vary: $M$ at $\Delta=1.70$, $J/M$ at $M=0.8$, and $\Delta$ at $M=0.8$, respectively. Blue, teal, and orange denote $J/M=0,0.60,0.85$ in the left and right columns and $\Delta=1.35,1.60,1.70$ in the middle column. Open circles locate $\lambda=1$ on fixed-$\Delta$ curves; depth is invariant under $\lambda\leftrightarrow1/\lambda$ while the phase is not. The dashed line marks $\Delta=3/2$.}
  \label{fig:optical-depth-parameter-atlas}
\end{figure}

The central formula is
\begin{equation}
        \pi\Gamma_{\rm UV}=-\log(1-\Tcal_{\rm UV}).
        \label{eq:opticaldepth}
\end{equation}
The quantity
\begin{equation}
        \;
        \mathfrak c_{\rm opt}\equiv \pi\Gamma_{\rm UV}\;
\end{equation}
is the finite-coupling relaxation observable for AdS-BTZ systems.  It has the same definition as  \textbf{optical depth} in optics associated with a single boundary encounter.
\\\\
At weak coupling,
\begin{equation}
        \mathfrak c_{\rm opt}=\Tcal_{\rm UV}+{1\over2}\Tcal_{\rm UV}^2+{1\over3}\Tcal_{\rm UV}^3+\cdots .
\end{equation}
Thus the old result is
\begin{equation}
        \mathfrak c_{\rm opt}\approx \Tcal_{\rm UV}.
\end{equation}
\\\\
\textbf{Open Holographic AdS-BTZ Optical Depth Theorem}:
\\
Consider an empty global \(\AdS_3\) system in alternate quantization of dimension \(2-\Delta\), with \(1<\Delta<2\), coupled by transparent B.C to a spinning BTZ bath in standard quantization of dimension \(\Delta\).  Let \(m\) be fixed and let the poles of AdS be
\begin{equation}
        \Om_n=2-\Delta+2n,
\end{equation}
Then the poles of the empty AdS after the bath has been traced out are
\begin{equation}
        \omega_n=\Om_n+s_{\rm UV}
\end{equation}
where \(s_{\rm UV}\) is determined by
\begin{equation}
        \;
        e^{i\pi s_{\rm UV}}
        ={1+\lambda^2\over 1+\lambda^2 e^{-2\pi i(\Delta-1)}}\;,
        \qquad
        \;
        \lambda^2={4h^2(\Delta-1)^2\over K^{\Delta-1}}\;,
        \label{eq:mainShift}
\end{equation}
with \(K=\kap_+\kap_-=r_+^2-r_-^2\).
\\\\
Writing
\begin{equation}
        s_{\rm UV}=\delta_{\rm UV}-{i\over2}\Gamma_{\rm UV},
\end{equation}
one obtains
\begin{equation}
        \;
        \Gamma_{\rm UV}=-{1\over\pi}\log(1-\Tcal_{\rm UV})\;,
        \qquad
        \;
        \Tcal_{\rm UV}={4\lambda^2\sin^2\pi(\Delta-1)\over(1+\lambda^2)^2}\;,
        \label{eq:GammaOptical}
\end{equation}
and
\begin{equation}
        \;
        \delta_{\rm UV}={1\over\pi}
        \operatorname{atan2}\!\left(
        \lambda^2\sin 2\pi(\Delta-1),\,1+\lambda^2\cos 2\pi(\Delta-1)
        \right)\; .
        \label{eq:deltaUV}
\end{equation}
For an analysis of the dependance of the Dissipation Width and Real Shift on $\Delta, M, J$, see Figure \ref{fig:optical-depth-parameter-atlas}

\subsection{Scattering Bounce Time intuitive interpretation}

\begin{figure}[H]
    \centering
    \includegraphics[width=0.8\linewidth]{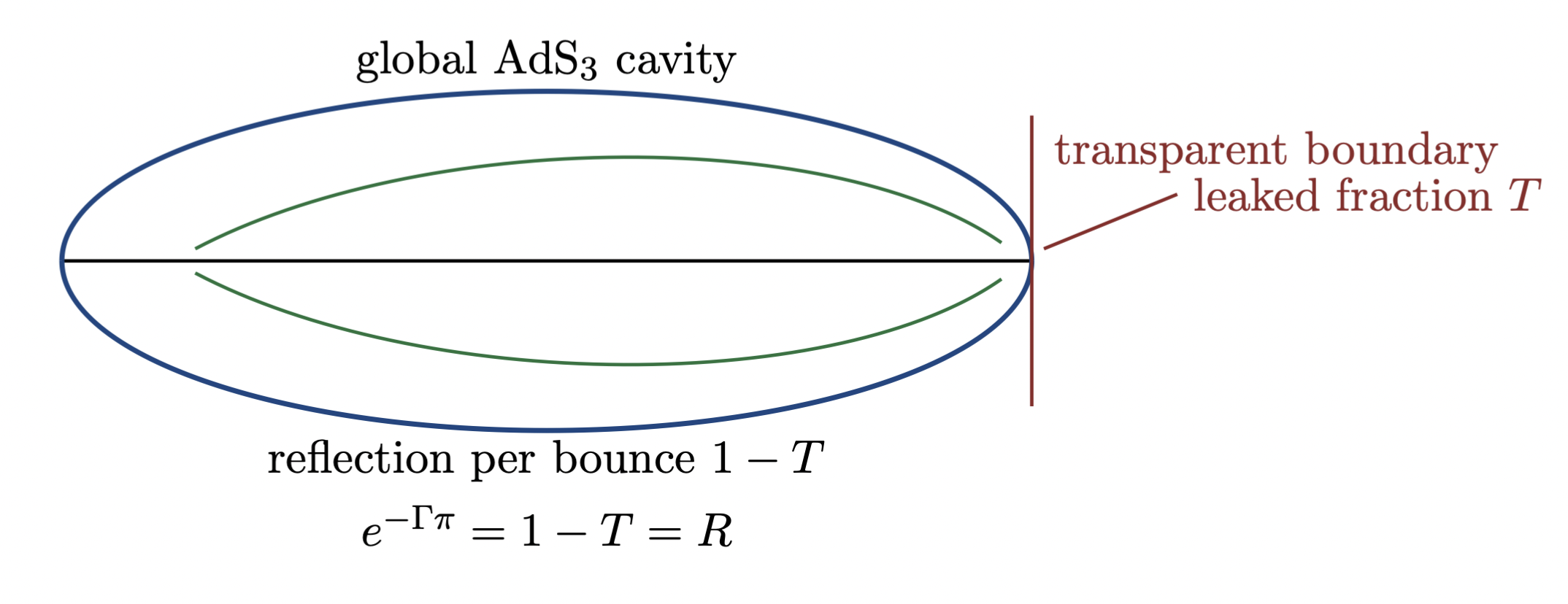}
    \caption{High-frequency optical-depth AdS-BTZ picture.  The $\AdS$ wave returns to the boundary once per time $\pi$, leaking some energy on each bounce. A one-bounce transmission probability $T$ becomes an exponential dissipation.}
    \label{fig:placeholder}
\end{figure}

Empty global \(\AdS_3\) normal-mode frequencies are separated by
\begin{equation}
        \Om_{n+1}-\Om_n=2.
\end{equation}
Therefore the natural return time of the wave to the boundary is
\begin{equation}
        t_{\rm bounce}=2\pi/\rm spacing=\pi.
\end{equation}
If the reflection probability per bounce is \(1-\Tcal_{\rm UV}\), then
\begin{equation}
        e^{-\Gamma_{\rm UV}t_{\rm bounce}}=1-\Tcal_{\rm UV},
\end{equation}
which gives \eqref{eq:opticaldepth}.  This explains how we get the logarithm.

\subsection{Sanity Checks and consequences}
\subsubsection{Check 1: weak coupling reproduces the residue plateau}

For \(\lambda\ll1\),
\begin{equation}
        \Tcal_{\rm UV}=4\lambda^2\sin^2 A+O(\lambda^4).
\end{equation}
Thus
\begin{equation}
        \Gamma_{\rm UV}={4\lambda^2\sin^2 A\over\pi}+O(\lambda^4).
\end{equation}
Using
\begin{equation}
        \lambda^2=4h^2(\Delta-1)^2K^{1-\Delta},
        \qquad
        \sin^2A=\sin^2\pi(\Delta-1)=\sin^2\pi\Delta,
\end{equation}
one gets
\begin{equation}
        {\Gamma_{\rm UV}\over h^2}
        ={16\over\pi}(\Delta-1)^2\sin^2(\pi\Delta)K^{1-\Delta}+O(h^2),
\end{equation}
which is exactly \eqref{eq:weakspinplateau}.
\\\\
For the real shift,
\begin{equation}
        \delta_{\rm UV}={\lambda^2\sin(2A)\over\pi}+O(\lambda^4)
        ={8h^2\over\pi}(\Delta-1)^2\sin(\pi\Delta)\cos(\pi\Delta)K^{1-\Delta}+O(h^4),
\end{equation}
which matches \eqref{eq:weakrealplateau}.
 
\subsubsection{Check 2: non-rotating BTZ}

For \(J=0\),
\begin{equation}
        K=M.
\end{equation}
The theorem becomes
\begin{equation}
        \lambda^2={4h^2(\Delta-1)^2\over M^{\Delta-1}}.
\end{equation}
The weak-coupling limit is
\begin{equation}
        \lim_{n\to\infty}{\Gamma_n\over h^2}
        ={16\over\pi}(\Delta-1)^2\sin^2(\pi\Delta)M^{1-\Delta},
\end{equation}
which is precisely the non-rotating plateau.

\subsubsection{Check 3: positivity and stability}

For real \(\lambda\) and real \(A\),
\begin{equation}
        0\le \Tcal_{\rm UV}\le1.
\end{equation}
Indeed,
\begin{equation}
        (1+\lambda^2)^2-4\lambda^2\sin^2A
        =(1-\lambda^2)^2+4\lambda^2\cos^2A\ge0.
\end{equation}
Hence
\begin{equation}
        \Gamma_{\rm UV}\ge0.
\end{equation}
The only way to reach \(\Tcal_{\rm UV}=1\) is
\begin{equation}
        \lambda=1,
        \qquad
        \cos A=0.
\end{equation}
That is,
\begin{equation}
        \lambda=1,
        \qquad
        \Delta={3\over2}.
\end{equation}
At this perfectly transmitting point, \(\Gamma_{\rm UV}\to\infty\). This point corresponds to a conformally coupled scalar. 

\subsubsection{Check 4: strong/weak duality}

The transmission coefficient obeys
\begin{equation}
        \Tcal_{\rm UV}(\lambda)=\Tcal_{\rm UV}(1/\lambda).
\end{equation}
Therefore
\begin{equation}
        \Gamma_{\rm UV}(\lambda)=\Gamma_{\rm UV}(1/\lambda).
\end{equation}
The real shift is not invariant. Since
\begin{equation}
        \delta_{\rm UV}(\lambda)=-{1\over\pi}\arg(1+
        \lambda^2e^{-2iA}),
\end{equation}
we have
\begin{align}
        \delta_{\rm UV}(1/\lambda)
        &=-{1\over\pi}\arg(1+
        \lambda^{-2}e^{-2iA})
        \\
        &=-{1\over\pi}\arg(\lambda^2+e^{-2iA})
        \\
        &=-{1\over\pi}\arg\left[e^{-2iA}(1+
        \lambda^2e^{2iA})\right].
\end{align}
On the continuous branch connected to weak coupling,
\begin{equation}
        \delta_{\rm UV}(1/\lambda)=2(\Delta-1)-\delta_{\rm UV}(\lambda)
        \quad \text{mod }2.
\end{equation}
Thus strong/weak duality leaves the damping invariant but \textbf{maps the real part to the other quantization branch.}
\subsubsection{Check 5: near-extremal spinning BTZ}

For fixed \(h\) and \(\Delta>1\),
\begin{equation}
        K=r_+^2-r_-^2\to0
\end{equation}
near extremality, so
\begin{equation}
        \lambda^2={4h^2(\Delta-1)^2\over K^{\Delta-1}}\to\infty.
\end{equation}
By strong/weak duality,
\begin{equation}
        \Tcal_{\rm UV}\sim {4\sin^2A\over\lambda^2}\to0,
        \qquad
        \Gamma_{\rm UV}\to0.
\end{equation}
Thus the high-n dissipation is suppressed as the BTZ bath approaches extremality. The exactly extremal case is a singular limit and should be treated separately.

\subsection{Optical relaxation depth}
We define the quantity Optical relaxation depth $\mathfrak c_{\rm opt}$, which depends only on $\Delta, h, K$, the only bath details involving $K$,
\begin{equation}
        \;
        \mathfrak c_{\rm opt}\equiv 
        \pi\Gamma_{\rm UV}
        =-\log(1-\Tcal_{\rm UV})\;,
\end{equation}
 This object has several advantages over \(\Tcal\).

\begin{enumerate}
\item It is additive under independent losses:
\begin{equation}
        (1-\Tcal_{12})=(1-\Tcal_1)(1-\Tcal_2)
        \quad\Rightarrow\quad
        \mathfrak c_{12}=\mathfrak c_1+\mathfrak c_2.
\end{equation}
\item It reduces to the weak-coupling transmissivity:
\begin{equation}
        \mathfrak c_{\rm opt}=\Tcal_{\rm UV}+O(\Tcal_{\rm UV}^2).
\end{equation}
Thus previous weak-coupling computations were linearization of \(\mathfrak c_{\rm opt}\).

\item It makes perfect transmission singular:
\begin{equation}
        \Tcal_{\rm UV}=1
        \quad\Rightarrow\quad
        \mathfrak c_{\rm opt}=\infty.
\end{equation}
This is physically correct for trapped waves: if every boundary encounter transmits all energy, there is no trapped quasinormal modes.

\item It is exactly strong/weak invariant:
\begin{equation}
        \mathfrak c_{\rm opt}(\lambda)=\mathfrak c_{\rm opt}(1/\lambda).
\end{equation}
\end{enumerate}

\subsection{Strong-Weak Duality and the Optical Limit}
\label{sec:optical-figures}

The coupling used in the quasinormal modes analysis of ~\cite{Karch:2025hof}
is $g=(2\Delta_L-d)h$.  For our $d=2$ assignment
$\Delta_L=2-\Delta$, this means $g=-2h(\Delta-1)$;
in particular, the relevant positive coupling in the plots is
$|g|=2|h|(\Delta-1)$.
\\\\
Writing $A=\pi(\Delta-1)$ and $s_{\rm UV}=\delta_{\rm UV}-i\Gamma_{\rm UV}/2$,
the coupling equations take the compact form
\begin{align}
 T_{\rm UV}&=\sin^2 A\,\operatorname{sech}^2(\log|g|),\\
 c_{\rm opt}&\equiv\pi\Gamma_{\rm UV}=-\log(1-T_{\rm UV}).
 \label{eq:optical-duality-display}
\end{align}
The associated phase is 
\begin{equation}
 \delta_{\rm UV}=\frac1\pi\operatorname{atan2}\!\bigl(
 g^2\sin(2A),\\[-2pt]
 1+g^2\cos(2A)\bigr).
\label{eq:optical-phase-display}
\end{equation}
The first line is even under $\log|g|\mapsto-\log|g|$.
The real shift instead obeys
\begin{equation}
 \delta_{\rm UV}(|g|)+\delta_{\rm UV}(1/|g|)
 =2(\Delta-1)\pmod{2}.
 \label{eq:phase-complement}
\end{equation}
Thus the width is unchanged by reciprocal coupling, whereas the real
frequency moves toward the other AdS quantization tower: in the weak limit
$s_{\rm UV}\to0$, while in the strong limit
$s_{\rm UV}\to2(\Delta-1)\pmod 2$.  The modulo-two freedom corresponds
to the spacing between neighboring AdS modes.  The microscopic duality of
~\cite{Karch:2025hof} also exchanges the two quantizations and sends
$g\mapsto-1/g$; the reciprocal pairing shown here holds for the
UV damping at fixed $\Delta$.

\begin{figure}[H]
 \centering
 \includegraphics[width=0.7\linewidth]{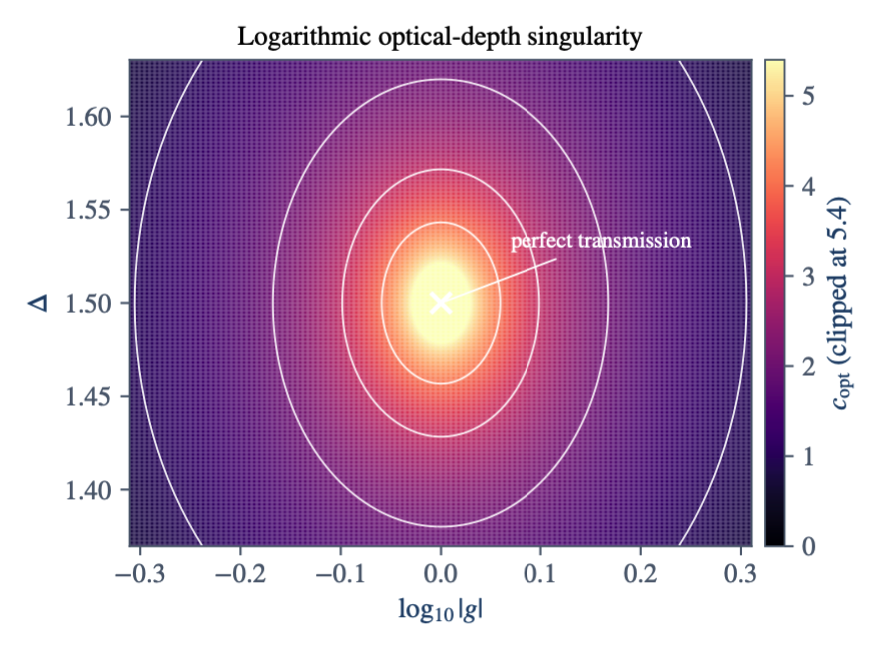}
 \caption{Optical depth near perfect transmission.  White curves are
 contours of constant optical depth, with $c_{\rm opt}=1,2,3,4$;
 the color scale is clipped at $5.4$ to display the neighborhood.
 The cross at $(\log_{10}|g|,\Delta)=(0,3/2)$ marks the divergent
 point}
 \label{fig:opt-critical}
\end{figure}

The optical depth law differs from its weak-coupling approximation
even when $T_{\rm UV}$ is moderately large, as shown in
Fig.~\ref{fig:opt-law}.  Its maximum for fixed $\Delta\ne3/2$
occurs at $|g|=1$ and is
$T_{\max}=\sin^2 A$,
$\Gamma_{\max}=-(2/\pi)\log|\cos A|$.
At $\Delta=3/2$ and $|g|=1$, the UV pole equation has a
perfect transmission singularity.
Indeed, putting $t=\log|g|$ gives
\begin{equation}
\begin{aligned}
 1-T_{\rm UV}&=\tanh^2t\\
  &\quad+\sin^2\!\bigl[\pi(\Delta-3/2)\bigr]
     \operatorname{sech}^2t\\
  &\simeq t^2+\pi^2(\Delta-3/2)^2,
\end{aligned}
 \label{eq:critical-optical-depth}
\end{equation}
which explains the logarithmic peak in Fig.~\ref{fig:opt-critical}.
\begin{figure}[H]
    \centering
    \includegraphics[width=1.1\linewidth]{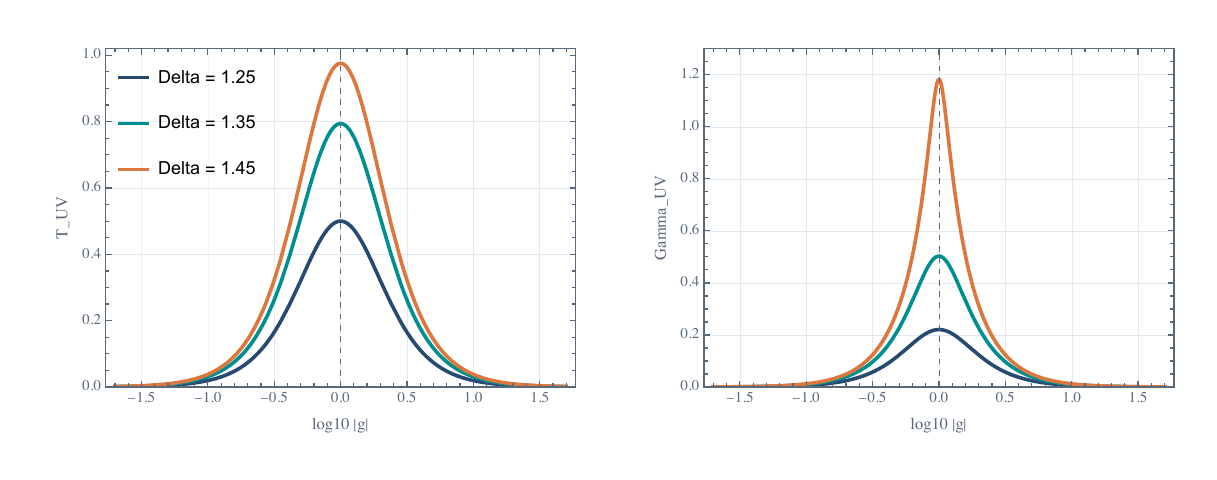}
    \caption{UV transmission parameter (left) and pole width (right)
 versus $\log_{10}|g|$ in the common boundary normalization.
 The three curves have $\Delta=1.25,1.35,1.45$; reflection about
 $|g|=1$ leaves both observables unchanged.  The physical damping rate
 is $\gamma_{\rm UV}=\Gamma_{\rm UV}/2$}
    \label{fig:opt-duality}
\end{figure}
\begin{figure}[H]
    \centering
    \includegraphics[width=\linewidth]{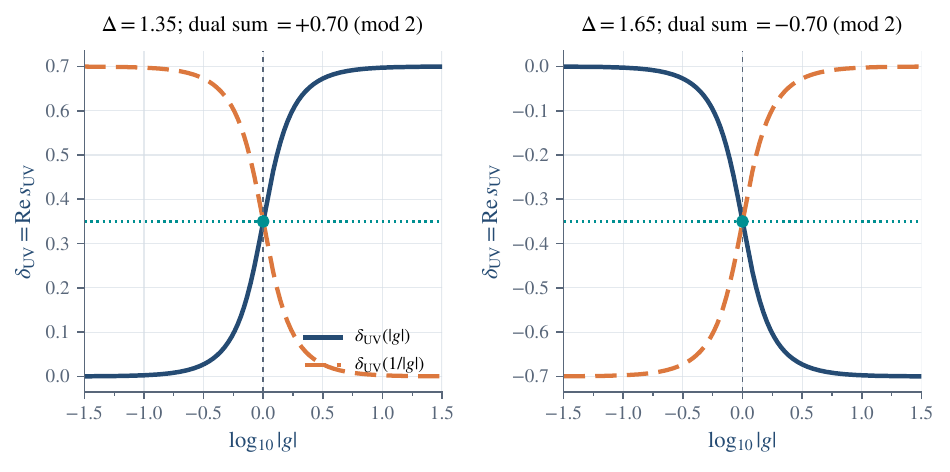}
    \caption{Real shift at $|g|$ (solid) and its reciprocal partner at $1/|g|$ (dashed). The two curves meet at the self-dual coupling but are otherwise different: their sum is $0.70$ for $\Delta=1.35$ and $-0.70$ for the principal branch at $\Delta=1.65$. The latter is equivalent to $2(\Delta-1)=1.30$ modulo the AdS level spacing of $2$. The dual sum is the vertical length between top and bottom.}
    \label{fig:placeholder}
\end{figure}
\begin{figure}[H]
 \centering
 \includegraphics[width=0.7\linewidth]{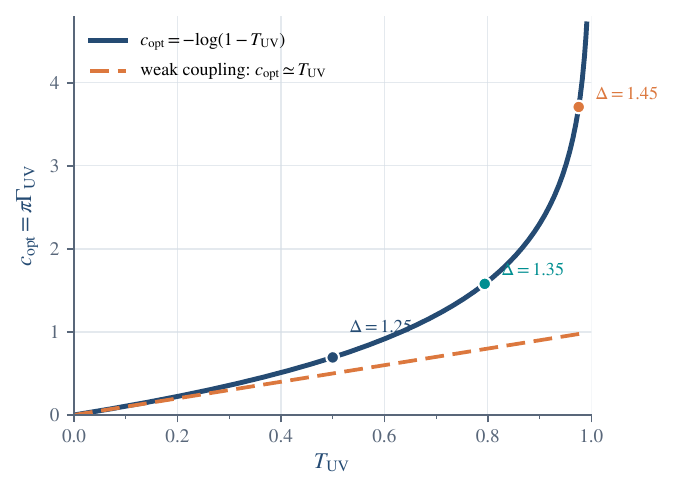}
 \caption{Optical depth $c_{\rm opt}=\pi\Gamma_{\rm UV}$ against
 $T_{\rm UV}$ (solid), compared with its weak-coupling tangent
 $c_{\rm opt}\simeq T_{\rm UV}$ (dashed).  Colored markers indicate
 the maximal transmission at $|g|=1$ for the same three dimensions
 as Fig.~\ref{fig:opt-duality}.}
 \label{fig:opt-law}
\end{figure}

\section{Summary and Outlook}
In this work, we studied a double-trace deformation coupling two CFTs and treated one sector as an open quantum field theory by tracing out the other. Using the Schwinger–Keldysh formalism, we integrated out the bath degrees of freedom and derived the retarded Green’s function of the resulting open system. Using the holographic expression for the retarded Green's functions, we computed the spectral densities which preserve the expected universal scaling of scale invariant open quantum systems. We then employed the holographic prescription for the retarded Green’s function to analyze its pole structure and compute the dissipative quantities presented in Section 4.4.4.

We investigated the gravitational signatures of the resulting open system dynamics through a perturbative analytic treatment of the quasinormal modes, deriving closed form expressions that also clarify the dissipation introduced due to the bath. We considered three configurations: (i) pure AdS coupled to a nonrotating BTZ black hole, (ii) two coupled nonrotating BTZ black holes, and (iii) a rotating BTZ black hole coupled to a empty AdS. For each configuration, we examined both possible system bath assignments by interchanging the roles of the two sectors. In the first two cases, our analytic expressions are in good agreement with the numerical results reported in \cite{Karch:2025hof} up to higher orders in the coupling.  

\textit{\textbf{Outlook}} Several interesting questions arise from studying deformations of CFTs in the context of studying open quantum systems \begin{itemize}
    \item In our analysis, the bath degrees of freedom are traced out, so that the remaining system is generically described by a mixed reduced density matrix. While the present work has focused on dissipation, it would be interesting to investigate the information-theoretic structure of this mixed state in the double-trace deformed theory. In particular, one could study how correlations between the system and bath, or between spatial subregions of the two theories, evolve as the coupling is varied. From the holographic perspective, quantities such as the reflected entropy and entanglement of purification are closely related to the entanglement-wedge cross section, providing a geometric probe of mixed-state correlations. It would be particularly interesting to determine whether the coupling dependence and weak/strong coupling structure found numerically in \cite{Karch:2025hof} for dissipation have counterparts in these mixed-state correlation measures.
    \item It would also be interesting to explore lower-dimensional realizations of the present framework, particularly in nearly $(\mathrm{AdS}_2)$ systems and coupled SYK models. Such models could provide a microscopic setting in which to study the interplay between system couplings, energy transfer, and dissipation, and to compare directly with the holographic description developed here.
    \item The Optical depth law for AdS-BTZ required taking the high-$n$ limit of the open pole equation, and simplifying both bath and system green's functions to obtain an algebraic equation and then find \ref{eq:GammaOptical} and \ref{eq:deltaUV} as the solution to such equation. This procedure could be generalized to other coupled gravitational systems such as the Janus-BTZ system discussed in \cite{Karch:2026spf}. This would allow a more general interpretation of the high-$n$ limit of Open Holography.
\end{itemize}

\acknowledgments
We would like to thank Andreas Karch for fruitful discussion and feedback. ChatGPT Pro was used for proofreading and improving the analytics. This work was supported in part by DOE grant DE-SC0022021.

\appendix

\section{Schwinger-Keldysh for Tracing out Holographic Bath}

\begin{equation}
e^{iS_{\mathrm{IF}}[O_{L,+},O_{L,-}]}
=\left\langle
T_C
\exp\left[
-ih\int_\mathcal{C} d^d x\,
\left(
O_{L}(x)O_{R}(x)
\right)
\right]
\right\rangle_R.
\end{equation}
\\
To see where the bath information goes, expand:
\[
e^{iS_{\mathrm{IF}}}
=
\left\langle
1
-ih\int_C d^d x\, O_L(x)O_R(x)
-\frac{h^2}{2}
\int_C d^d x\,d^d y\,
O_L(x)O_L(y)O_R(x)O_R(y)
+\cdots
\right\rangle_R .
\]
Taking the bath expectation value gives

\[
e^{iS_{\mathrm{IF}}}
=
1
-ih\int_C d^d x\, O_L(x)
\langle O_R(x)\rangle_R
-\frac{h^2}{2}
\int_C d^d x\,d^d y\,
O_L(x)O_L(y)
\left\langle
T_C O_R(x)O_R(y)
\right\rangle_R
+\cdots .
\]
Therefore, after tracing out the bath, the explicit bath operators \(O_R\)
are replaced by bath correlation functions such as
\[
\langle O_R(x)\rangle_R,
\qquad
\left\langle T_C O_R(x)O_R(y)\right\rangle_R,
\qquad
\left\langle T_C O_R(x_1)O_R(x_2)O_R(x_3)\right\rangle_R,
\]
and so on.

A compact way to write this is the cumulant expansion:

\[
iS_{\mathrm{IF}}[O_L]
=
\sum_{n=1}^{\infty}
\frac{(-ih)^n}{n!}
\int_C d^d x_1\cdots d^d x_n\,
O_L(x_1)\cdots O_L(x_n)
\left\langle
T_C O_R(x_1)\cdots O_R(x_n)
\right\rangle_{R,c}.
\]
Here the subscript \(c\) means connected bath correlator. This formula makes
the meaning of integrating out the bath explicit: the bath degrees of freedom
are removed as dynamical variables, but they leave behind a tower of connected
bath correlation functions.
\\\\
In our setup, we will perturb around the vacuum state where the solution $\phi = 0$ sets $\alpha_L=\beta_L = \alpha_R = \beta_R =0$. In other words,  
\[
\langle O_R(x)\rangle_R=0,
\]
so the leading contribution to the influence action is quadratic in the
coupling \(h\):
\[
iS_{\mathrm{IF}}^{(2)}
=
-\frac{h^2}{2}
\int_C d^d x\,d^d y\,
O_L(x)O_L(y)
\left\langle
T_C O_R(x)O_R(y)
\right\rangle_{R,c}.
\]
\[
iS_{\mathrm{IF}}^{(2)}
=
-\frac{h^2}{2}
\int_C d^d x\,d^d y\,
O_L(x)G^\mathcal{C}_R(x,y)O_L(y).
\]

\begin{equation}
  G_R^{\mathcal{C}}(x,x')
  =
  \langle T_{\mathcal{C}}O_R(x)O_R(x')\rangle_{R,c}.
\end{equation}
This term is important because the bath two-point function produces
the dissipative and noise kernels of the open system. In other words, this is the step where the reservoir
degrees of freedom \(O_R\) are traced out, leaving an effective non-unitary
dynamics for the system degrees of freedom \(O_L\).

\subsection{Changing from the \texorpdfstring{\(+,-\)}{+,-} left/right Basis to the $c/q$ (classical/quantum) Basis}

Starting from the influence functional, the leading nonzero contribution is quadratic in the system-bath coupling. We write it in the form
\begin{equation}
iS_{\rm IF}^{(2)}
=
-\frac{h^2}{2}
\int_C d^dx\,d^dx'\,
O_L(x)\,
G_B^C(x,x')\,
O_L(x'),
\label{eq:IF_contour_quadratic}
\end{equation}
where
\begin{equation}
G_B^C(x,x')
\equiv
\left\langle
T_C O_R(x)O_R(x')
\right\rangle_R .
\label{eq:contour_correlator_definition}
\end{equation}
\\\\
Splitting the contour into its forward and backward branches gives
\begin{equation}
iS_{\rm IF}^{(2)}
=
-\frac{h^2}{2}
\sum_{\sigma,\sigma'=\pm}
\sigma\sigma'
\int d^dx\,d^dx'\,
O_\sigma(x)\,
G_{\sigma\sigma'}(x,x')\,
O_{\sigma'}(x'),
\label{eq:IF_pm_basis}
\end{equation}
where \(+\equiv +1\), \(-\equiv -1\), and
\begin{equation}
G_{\sigma\sigma'}(x,x')
=
\left\langle
T_C O_{R,\sigma}(x)O_{R,\sigma'}(x')
\right\rangle_R .
\end{equation}
The four real-time components are
\begin{align}
G_{++}(x,x')
&=
\left\langle
T O_R(x)O_R(x')
\right\rangle_R,
\\
G_{--}(x,x')
&=
\left\langle
\widetilde T O_R(x)O_R(x')
\right\rangle_R,
\\
G_{-+}(x,x')
&=
\left\langle
O_R(x)O_R(x')
\right\rangle_R
\equiv G^>(x,x'),
\\
G_{+-}(x,x')
&=
\left\langle
O_R(x')O_R(x)
\right\rangle_R
\equiv G^<(x,x').
\end{align}
Equivalently,
\begin{align}
G_{++}(x,x')
&=
\theta(t-t')G^>(x,x')
+
\theta(t'-t)G^<(x,x'),
\\
G_{--}(x,x')
&=
\theta(t'-t)G^>(x,x')
+
\theta(t-t')G^<(x,x').
\end{align}
\\
Now define the average and difference fields / classical and quantum fields
\begin{equation}
O_c(x)
=
\frac{1}{2}\left(O_+(x)+O_-(x)\right),
\qquad
O_q(x)
=
O_+(x)-O_-(x).
\label{eq:ra_definition}
\end{equation}
Thus
\begin{equation}
O_+(x)=O_c(x)+\frac{1}{2}O_q(x),
\qquad
O_-(x)=O_c(x)-\frac{1}{2}O_q(x).
\label{eq:pm_in_terms_of_ra}
\end{equation}
Substituting \eqref{eq:pm_in_terms_of_ra} into \eqref{eq:IF_pm_basis}, one finds that the \(O_rO_r\) term cancels:
\begin{equation}
G_{++}+G_{--}-G_{+-}-G_{-+}=0.
\label{eq:rr_cancellation}
\end{equation}
This cancellation is a consequence of Schwinger--Keldysh normalization: if the two histories are identical, \(O_+=O_-\), then \(O_a=0\), and the influence functional cannot change the trace of the density matrix.
\\\\
The remaining terms can be written in terms of the retarded and Keldysh bath kernels, \cite{Haehl_2017}
\begin{equation}
G_B^R(x,x')
=
-i\theta(t-t')
\left\langle
[O_R(x),O_R(x')]
\right\rangle_R
\label{eq:bath_retarded_kernel}
\end{equation}
and
\begin{equation}
G_B^K(x,x')
=
\frac{1}{2}
\left\langle
\{O_R(x),O_R(x')\}
\right\rangle_R
\label{eq:bath_keldysh_kernel}
\end{equation}
In terms of \(G^>\) and \(G^<\), these are
\begin{equation}
G_B^R(x,x')
=
-i\theta(t-t')
\left(
G^>(x,x')-G^<(x,x')
\right),
\end{equation}
and
\begin{equation}
G_B^K(x,x')
=
\frac{1}{2}
\left(
G^>(x,x')+G^<(x,x')
\right).
\end{equation}

Therefore the quadratic influence action in the \(r/a\) basis is
\begin{equation}
iS_{\rm IF}^{(2)}
=
-ih^2
\int d^dx\,d^dx'\,
O_q(x)G_B^R(x,x')O_c(x')
-
\frac{h^2}{2}
\int d^dx\,d^dx'\,
O_q(x)G_B^K(x,x')O_q(x').
\label{eq:iSIF_ra_basis}
\end{equation}
Equivalently, dividing by \(i\),
\begin{equation}
S_{\rm IF}^{(2)}
=
-h^2
\int d^dx\,d^dx'\,
O_q(x)G_B^R(x,x')O_c(x')
+
\frac{ih^2}{2}
\int d^dx\,d^dx'\,
O_q(x)G_B^K(x,x')O_q(x').
\label{eq:SIF_ra_basis}
\end{equation}
\\\\
This is the standard separation of the influence action into response and noise. The term
\begin{equation}
-h^2 O_q G_B^R O_c
\end{equation}
is the causal \textbf{dissipative response of the bath}, while
\begin{equation}
\frac{ih^2}{2}O_qG_B^K O_q
\end{equation}
is the \textbf{fluctuation or noise term}. In the exponent \(e^{iS_{\rm IF}}\), the second term contributes
\begin{equation}
i\left(\frac{ih^2}{2}O_qG_B^K O_q\right)
=
-\frac{h^2}{2}O_qG_B^K O_q,
\end{equation}
so it suppresses configurations where the forward and backward histories differ strongly. This is the origin of decoherence in the reduced open-system description.

\section{Open Retarded Green's Function Relation and recovered Transparent Boundary Conditions} 
\label{appendix_G_open}
\begin{align}
Z_{\rm open}[J_+,J_-]
=
\int D\phi_+D\phi_-\,
\exp\left\{
iS_L[\phi_+]-iS_L[\phi_-]
+iS_{\rm IF}[O_+,O_-]
+i\int_x\left(J_+O_+-J_-O_-\right)
\right\}.
\end{align}
At quadratic order, write the left-system part in the original $+/-$ basis as
\begin{align}
S_L^{(2)}
=
\int_{x,y}
\begin{pmatrix}
O_+(x) & O_-(x)
\end{pmatrix}
\begin{pmatrix}
K_{++}(x,y) & K_{+-}(x,y)\\
K_{-+}(x,y) & K_{--}(x,y)
\end{pmatrix}
\begin{pmatrix}
O_+(y)\\
O_-(y)
\end{pmatrix}.
\end{align}
Here the signs coming from the opposite orientation of the two SK branches are already
included in the entries $K_{\sigma\sigma'}$. Now perform the Keldysh rotation
\begin{align}
O_c=\frac12(O_++O_-),
\qquad
O_q=O_+-O_-,
\end{align}
so that
\begin{align}
O_+=O_c+\frac12 O_q,
\qquad
O_-=O_c-\frac12 O_q.
\end{align}
Substituting this directly into the $+/-$ quadratic form gives
\begin{align}
S_L^{(2)}
=
\int_{x,y}
\Bigg[
&\left(O_c+\frac12O_q\right)_x
K_{++}(x,y)
\left(O_c+\frac12O_q\right)_y
+
\left(O_c+\frac12O_q\right)_x
K_{+-}(x,y)
\left(O_c-\frac12O_q\right)_y
\nonumber\\
&+
\left(O_c-\frac12O_q\right)_x
K_{-+}(x,y)
\left(O_c+\frac12O_q\right)_y
+
\left(O_c-\frac12O_q\right)_x
K_{--}(x,y)
\left(O_c-\frac12O_q\right)_y
\Bigg].
\end{align}
Collecting powers of $O_c$ and $O_q$,
\begin{align}
S_L^{(2)}
=
\int_{x,y}
\begin{pmatrix}
O_c(x) & O_q(x)
\end{pmatrix}
\begin{pmatrix}
K_{cc}(x,y) & K_{cq}(x,y)\\
K_{qc}(x,y) & K_{qq}(x,y)
\end{pmatrix}
\begin{pmatrix}
O_c(y)\\
O_q(y)
\end{pmatrix},
\end{align}
with
\begin{align}
K_{cc}
&=
K_{++}+K_{+-}+K_{-+}+K_{--},
\\
K_{cq}
&=
\frac12\left(K_{++}-K_{+-}+K_{-+}-K_{--}\right),
\\
K_{qc}
&=
\frac12\left(K_{++}+K_{+-}-K_{-+}-K_{--}\right),
\\
K_{qq}
&=
\frac14\left(K_{++}-K_{+-}-K_{-+}+K_{--}\right).
\end{align}
Equivalently, in matrix notation,
\begin{align}
\begin{pmatrix}
K_{cc} & K_{cq}\\
K_{qc} & K_{qq}
\end{pmatrix}
=
R^T
\begin{pmatrix}
K_{++} & K_{+-}\\
K_{-+} & K_{--}
\end{pmatrix}
R,
\qquad
R=
\begin{pmatrix}
1 & \frac12\\
1 & -\frac12
\end{pmatrix},
\end{align}
where
\begin{align}
\begin{pmatrix}
O_+\\
O_-
\end{pmatrix}
=
R
\begin{pmatrix}
O_c\\
O_q
\end{pmatrix}.
\end{align}
The Schwinger--Keldysh normalization condition says that if the two histories are
identified, then the closed-time-path evolution is trivial:
\begin{align}
O_+=O_-
\qquad\Longleftrightarrow\qquad
O_q=0,
\qquad
Z[J,J]=1.
\end{align}
Therefore the quadratic action must vanish for arbitrary $O_c$ when $O_q=0$:
\begin{align}
S_L^{(2)}[O_c,O_q=0]
=
\int_{x,y}O_c(x)K_{cc}(x,y)O_c(y)
=
0.
\end{align}
Since this must hold for arbitrary $O_c$, we require
\begin{align}
K_{cc}(x,y)=0.
\end{align}
Thus the quadratic action becomes
\begin{align}
S_L^{(2)}
=
\int_{x,y}
\begin{pmatrix}
O_c(x) & O_q(x)
\end{pmatrix}
\begin{pmatrix}
0 & K_{cq}(x,y)\\
K_{qc}(x,y) & K_{qq}(x,y)
\end{pmatrix}
\begin{pmatrix}
O_c(y)\\
O_q(y)
\end{pmatrix}.
\end{align}
Finally, identifying the $c/q$ components with the usual retarded, advanced, and
Keldysh kernels,
\begin{align}
K_{cq}\equiv K_L^A,
\qquad
K_{qc}\equiv K_L^R,
\qquad
K_{qq}\equiv K_L^K,
\end{align}
we obtain
\begin{align}
S_L^{(2)}
=
\int_{x,y}
\begin{pmatrix}
O_c(x) & O_q(x)
\end{pmatrix}
\begin{pmatrix}
0 & K_L^A(x,y)\\
K_L^R(x,y) & K_L^K(x,y)
\end{pmatrix}
\begin{pmatrix}
O_c(y)\\
O_q(y)
\end{pmatrix}.
\end{align}
This is Eq.~(4.1). Written out without matrix notation,
\begin{align}
S_L^{(2)}
=
\int_{x,y}
\left[
O_c(x)K_L^A(x,y)O_q(y)
+
O_q(x)K_L^R(x,y)O_c(y)
+
O_q(x)K_L^K(x,y)O_q(y)
\right].
\end{align}

\subsection{Open $G^R_{L,open}$ Green's function  relation after tracing out the bath}

This is the most general quadratic Schwinger--Keldysh action consistent with normalization \(S[O_c,O_q=0]=0\) has the form \cite{Sieberer:2015svu} (See \cite{Kamenev:2009jj} Sec 2.3)
The entries \(K^R_L\), \(K^A_L\) are the inverse of their respective green's functions (except $K^K$). 
\\\\
Define
\begin{equation}
\mathbb G(x,y)
=
\begin{pmatrix}
 G^K_L(x,y) & G^R_L(x,y) \\
G^A_L(x,y) & 0
\end{pmatrix},
\label{eq:green_matrix}
\end{equation}
\begin{equation}
\mathbb K(x,y)
=
\begin{pmatrix}
0 & K^A_L(x,y) \\
K_L^R(x,y) & K^K_L(x,y)
\end{pmatrix}.
\end{equation}
where \(G^R\) is the retarded Green's function, \(G^A\) is the advanced Green's function, and \(\ G^K\) is the Keldysh Green's function. For a Hermitian bosonic operator,
\begin{equation}
G^R_L(x,y)
= (G^{A}_L(x,y))^*=
-i\theta(t_x-t_y)\left\langle [O(x),O(y)]\right\rangle,
\end{equation}
\begin{equation}
G^K_L(x,y)
=
-i\left\langle \{O(x),O(y)\}\right\rangle.
\end{equation}
Therefore,
\begin{equation}
\mathbb K\circ \mathbb G=\mathbb I,
\label{eq:kernel_inverse_green_general}
\end{equation}
Here \(\circ\) denotes convolution:
\begin{equation}
(A\circ B)(x,y)
\equiv
\int_z A(x,z)B(z,y).
\end{equation}
Multiplying the matrices in 
\eqref{eq:kernel_inverse_green_general} gives
\begin{equation}
\begin{pmatrix}
K^A_L\circ G^A_L & 0 \\
K^R_L\circ G^K_L + K^K_L \circ G^A_L & K^R_L \circ G^R_L
\end{pmatrix}=\begin{pmatrix}
    1 & 0\\
    0 & 1
\end{pmatrix}
\end{equation}
\begin{equation}
K^R_L\circ G^R_L=1,
\qquad
K^A_L\circ G^A_L=1
\label{eq:retarded_inverse_relation}
\end{equation}
and
\begin{equation}
K^R_L\circ  G^K_L + K^K_L\circ G^A_L=0.
\end{equation}
Therefore, we derived expressions that should be true for any Kernel and Green's function matrix satisfying \eqref{eq:kernel_inverse_green_general} so we take the subscripts out
\begin{equation}
G^R=(K^R)^{-1},
\qquad
G^A=(K^A)^{-1},
\label{eq:retarded_kernel_inverse}
\end{equation}
and
\begin{equation}
 G^K
=
-G^R\circ K^K\circ G^A.
\label{eq:keldysh_green_from_noise_kernel}
\end{equation}
\\
For the isolated left theory, the quadratic retarded part of the effective action is therefore
\begin{equation}
S^{(2)}_{L,{\rm ret}}
=
\int_{x,y}
O_q(x)K_L^R(x,y)O_c(y),
\label{eq:left_retarded_kernel_action}
\end{equation}
Now include the bath influence functional to second order in coupling.
\begin{equation}
S_{\rm IF}^{(2)}
=
-h^2
\int_{x,y}
O_q(x)G_B^R(x,y)O_c(y)
+
\frac{i h^2}{2}
\int_{x,y}
O_q(x)G_B^K(x,y)O_q(y).
\label{eq:quadratic_influence_action}
\end{equation}
\\
Combining the isolated left action with the influence action gives
\begin{equation}
S^{(2)}_{L,\rm open,ret}
=
S^{(2)}_{L,\rm ret}+S^{(2)}_{\rm IF, ret} = \int_{x,y}
O_q(x)K_L^R(x,y)O_c(y) -h^2
\int_{x,y}
O_q(x)G_B^R(x,y)O_c(y)
\end{equation}
Combining the terms, the open-system retarded kernel is
\begin{equation}
K^R_{L,\rm open}
=
K_L^R-h^2G_B^R.
\label{eq:open_retarded_kernel}
\end{equation}
Using \(K_L^R=(G_L^R)^{-1}\), the open retarded Green's function is
\begin{equation}
G^R_{L,\rm open}
=
\left[
(G_L^R)^{-1}-h^2G_B^R
\right]^{-1}.
\label{eq:open_retarded_green}
\end{equation}
This is known in the literature as a Dyson equation: the bath retarded Green's function acts as a retarded self-energy for the left system. \cite{Kamenev:2009jj}

\begin{equation}
\Sigma_L^R=h^2G_B^R.
\end{equation}
The same logic gives the noise/Keldysh sector. 
\begin{equation}
    S_{L,open,K}^{(2)}=S_{\rm L,noise}^{(2)}+S_{\rm IF,noise}^{(2)}= \int_{x,y} O_q(x)(K^K_L+\frac{ih^2}{2}G_B^K)O_q(y)
\end{equation}
From \eqref{eq:quadratic_influence_action}, the \(O_qO_q\) term shifts the Keldysh inverse kernel:
\begin{equation}
K^K_{L, \rm open}
=
K_L^K+\frac{i h^2}{2}G_B^K.
\label{eq:open_noise_kernel}
\end{equation}
Here \(K_L^K\) represents any intrinsic Keldysh/noise kernel already present in the isolated left state. If the isolated left theory is treated as having no intrinsic noise, then \(K_L^K=0\), and the only noise comes from tracing out the bath.
\\\\
The full open-system Keldysh Green's function is obtained by inverting the full Keldysh matrix. (Notice how for the Keldysh Green's function, the Kernel is not equal necessarily to the inverse of its green funciton). Using \eqref{eq:keldysh_green_from_noise_kernel}, one obtains
\begin{equation}
 G^K_{\rm L,open}
=
-G^R_{L,\rm open}
\circ
K^K_{\rm L, open}
\circ
G^A_{\rm L,open}.
\label{eq:open_keldysh_green_general}
\end{equation}
Substituting the expression for the Open Keldysh Kernel,
\begin{equation}
 G^K_{L,\rm open}
=
-G^R_{L,\rm open}
\circ
\left(K_L^K+\frac{i h^2}{2}G_B^K\right)
\circ
G^A_{L,\rm open}
\label{eq:open_keldysh_green}
\end{equation}
If the isolated left system has no intrinsic Keldysh noise kernel, this reduces to
\begin{equation}
\
 G^K_{\rm L,open}
=
-i h^2\,
G^R_{\rm L,open}
\circ
G_B^K
\circ
G^A_{\rm L,open}.
\
\label{eq:open_keldysh_green_no_intrinsic_noise}
\end{equation}
\\
Convolutions become ordinary products in Fourier space, and using $G_{L,open}^A=G^A_L=(G^R_L)^*$. Then, our final expressions are,
\begin{equation}
G^R_{L,\rm open}(\omega,k)
=
\frac{1}
{
(G_L^R(\omega,k))^{-1}
-
h^2G_B^R(\omega,k)
}.
\label{eq:open_retarded_green_fourier}
\end{equation}
\begin{equation}
 G^K_{L,\rm open}(\omega,k)
=
-
G^R_{L,\rm open}(\omega,k)
\left[
K_L^K(\omega,k)+\frac{i h^2}{2}G_B^K(\omega,k)
\right]
(G^R_L)^*(\omega,k).
\label{eq:open_keldysh_green_fourier}
\end{equation}

\section{Simplification of the Empty-AdS Residue}
\label{app:residue-simplification}

For the $\ell=0$ empty-AdS system in alternate quantization,
\begin{equation}
    \Delta_S=2-\Delta,
    \qquad
    \Omega_{n,\pm}=\pm\Omega_n,
    \qquad
    \Omega_n=2-\Delta+2n,
    \qquad n=0,1,2,\ldots .
\end{equation}
The corresponding Green-function normalization is
$2\Delta_S-2=2-2\Delta$. At the positive-frequency pole,
\begin{equation}
    a_1(\Omega_n)=-n,
    \qquad
    a_2(\Omega_n)=n+2-\Delta,
    \qquad
    b_1(\Omega_n)=\Delta-1-n,
    \qquad
    b_2(\Omega_n)=n+1.
\end{equation}
Moreover, since $\epsilon=-\tfrac12(\omega-\Omega_n)$,
\begin{equation}
    \Gamma(-n+\epsilon)
    =\frac{(-1)^n}{n!\,\epsilon}+\mathcal{O}(1)
    =-\frac{2(-1)^n}{n!(\omega-\Omega_n)}+\mathcal{O}(1).
\end{equation}
The positive-frequency residue following from Eq.~(5.18) is therefore
\begin{equation}
    R_{n,+}
    =4(\Delta-1)
    \frac{\Gamma(\Delta-1)}{\Gamma(1-\Delta)}
    \frac{(-1)^n\Gamma(n+2-\Delta)}
    {\Gamma(\Delta-1-n)(n!)^2}.
    \label{eq:app-unsimplified-residue}
\end{equation}
Using
\begin{equation}
    \frac{(-1)^n}{\Gamma(\Delta-1-n)}
    =\frac{(2-\Delta)_n}{\Gamma(\Delta-1)},
    \qquad
    \Gamma(n+2-\Delta)=\Gamma(2-\Delta)(2-\Delta)_n,
\end{equation}
where $(a)_n\equiv\Gamma(a+n)/\Gamma(a)$, Eq.~\eqref{eq:app-unsimplified-residue}
becomes
\begin{align}
    R_{n,+}
    &=4(\Delta-1)
      \frac{\Gamma(2-\Delta)}{\Gamma(1-\Delta)}
      \left[\frac{(2-\Delta)_n}{n!}\right]^2 \notag\\
    &=-4(\Delta-1)^2
      \left[\frac{(2-\Delta)_n}{n!}\right]^2
      \equiv-Z_n,
\end{align}
where $\Gamma(2-\Delta)=(1-\Delta)\Gamma(1-\Delta)$ was used. Since
$G_{\mathrm{AdS}}^R(-\omega)=G_{\mathrm{AdS}}^R(\omega)$, the reflected pole has
the opposite residue. Thus
\begin{equation}
    R_{n,+}=-Z_n,
    \qquad
    R_{n,-}=+Z_n,
    \qquad
    Z_n=4(\Delta-1)^2
    \left[\frac{(2-\Delta)_n}{n!}\right]^2,
    \label{eq:app-simplified-residue}
\end{equation}

\bibliographystyle{JHEP}
\bibliography{main2}
\end{document}